\documentclass[pdflatex,sn-mathphys-num]{sn-jnl}
\usepackage{amsmath}
\usepackage{amsthm}
\usepackage{amsfonts}
\usepackage{amssymb,enumerate}
\usepackage{comment}
\usepackage{natbib}
\usepackage{setspace}
\usepackage{footnote}
\usepackage{algorithm}
\usepackage{algorithmic}
\usepackage{graphicx}
\usepackage{pstricks}

\usepackage{pst-plot}
\usepackage{pstricks-add} 
\usepackage{color,pstricks} 
\usepackage{colortbl} 
\usepackage{caption} 
\usepackage{subcaption}
\usepackage{epsf}
\usepackage{color}
\usepackage{booktabs}
\usepackage{multirow}
\usepackage{hyperref}
\usepackage{comment, tabularx, adjustbox}
\usepackage{float}
\usepackage[title]{appendix}
\usepackage{abstract}

\theoremstyle{thmstyleone}
\newtheorem{assumption}{Assumption}

\begin{document}

\title[Robustifying and Selecting Cohort-Appropriate Prognostic Models under Distributional Shifts]
{Robustifying and Selecting Cohort-Appropriate Prognostic Models under Distributional Shifts 
}
\author[1,2]{\fnm{Dimitris} \sur{Bertsimas}}
\author[1]{\fnm{Carol} \sur{Gao}}
\author[1]{\fnm{Angelos} \sur{Koulouras}}
\author*[3,4]{\fnm{Georgios} \sur{Antonios Margonis}}\email{margonig@mskcc.org}

\affil[1]{\orgdiv{Operations Research Center}, \orgname{Massachusetts Institute of Technology}, \orgaddress{\city{Cambridge}, \state{MA}, \postcode{02139}, \country{USA}}}
\affil[2]{\orgdiv{Sloan School of Management}, \orgname{Massachusetts Institute of Technology}, \orgaddress{\city{Cambridge}, \state{MA}, \postcode{02139}, \country{USA}}}
\affil[3]{\orgdiv{Department of Surgery}, \orgname{Memorial Sloan Kettering Cancer Center}, \orgaddress{\city{New York}, \state{NY}, \postcode{10065}, \country{USA}}}
\affil[4]{\orgname{Charité -- Universitätsmedizin Berlin}, \orgaddress{\city{Berlin}, \postcode{10117}, \country{Germany}}}

\maketitle

\begin{abstract}
Introduction: External validation is widely regarded as the gold standard for prognostic model evaluation. In this study, we challenge the assumption that successful external validation of calibration guarantees model generalizability and propose two complementary strategies to improve the transportability of prognostic models across cohorts.

Methods: Using six real-world surgical cohorts from tertiary academic centers, we first tested the hypothesis that successful external calibration does not necessarily guarantee model generalizability, but instead depends largely on similarity in covariates and outcomes between the development and validation cohorts. Cohort similarity was quantified using Kullback–Leibler (KL) divergence, and calibration was assessed using the Integrated Calibration Index (ICI). Second, from the model-developer's perspective, we aimed to train the “best-on-average” prognostic model by tuning models toward a meta-analysis-derived covariate and outcome distribution, treating this distribution as an approximation of the broader target population, and then evaluating whether this improved calibration. Third, from the end-user perspective, we proposed a simple measure of cohort outcome similarity to identify, among available published models, the prognostic model most suitable for a given target cohort with respect to both calibration and clinical utility, as assessed by Decision Curve Analysis (DCA). 

Results: External calibration worsened as distributional mismatch increased. Higher KL divergence was associated with higher ICI in both surgery-alone (Spearman $\rho=0.614, p=0.004$) and surgery + adjuvant chemotherapy cohorts (Spearman $\rho=0.738, p<0.001$). Meta-analysis-informed weighting improved calibration in most settings without materially affecting discrimination, with the clearest benefit observed when performance was evaluated against the aggregated external population (p=0.037), consistent with the goal of approximating a broader target population. From the end-user perspective, models developed in cohorts more similar to the target cohort achieved lower ICI in both surgery alone (Spearman $\rho=0.803, p<0.001$) and surgery + adjuvant chemotherapy cohorts (Spearman $\rho=0.737, p<0.001$), and also provided greater clinical utility on DCA.

Conclusion: Successful external calibration depends strongly on the similarity between development and validation cohorts and therefore does not, by itself, guarantee model generalizability to additional settings. We propose one strategy to train the “best-on-average” prognostic model and another to help users select the “best-for-their-cohort” prognostic model.
\end{abstract}


\section{Introduction}
Accurate prognostication is one of the pillars of clinical practice. Prognostic model outputs are commonly used to inform patients about what to expect after treatment, to guide physicians in deciding whether treatment escalation is warranted (for example, surgery plus chemotherapy vs surgery alone), and to support shared decision-making between physicians and patients. Accordingly, evaluating the performance of prognostic models is a central task in clinical outcome research.\\
\indent The current standard for prognostic model evaluation includes two complementary dimensions. The first is discrimination, that is, the ability of the model to correctly rank patients according to their risk. In survival analysis, this is typically assessed using metrics such as the concordance index (C-index) or time-dependent area under the curve (AUC). The second, and arguably even more clinically relevant, dimension is calibration: whether the absolute risk estimates assigned to individual patients, such as the predicted probability of 5-year survival, are accurate.\cite{alba2017discrimination} The most informative assessment of both discrimination and calibration occurs in an independent external cohort from another center, where good performance is generally interpreted as evidence that the model can generalize beyond the cohort used to develop it.\\
\indent Nonetheless, calibration is not purely an intrinsic property of the model. Rather, it depends in part on the cohort in which the model is evaluated. In clinical outcome research, external validation is often limited to one or, at most, two outside cohorts. However, successful validation in such a small number of datasets may be misleading: performance may appear favorable simply because the external cohort happens to share a similar underlying risk distribution with the training cohort, rather than because the model is truly generalizable. In practical terms, this challenges the credibility of many published prognostic tools, particularly those supported by only one or two external validations.\\
\indent This problem is especially important because, in most areas of medicine, the factors that determine patient outcomes are only incompletely understood. As a result, the risk distributions of different cohorts are not known a priori. Consequently, when a model performs well in an outside cohort, it often remains unclear whether this reflects genuine robustness or merely favorable alignment between the derivation and validation populations. This raises a fundamental question: is it truly feasible to develop a single prognostic model that generalizes well across all settings, or should model selection be tailored to the specific characteristics and needs of the target cohort?\\
\indent In this study, we address the problem both from the perspective of the model developer, who aims to build a robust and generalizable model across various cohorts, and from the perspective of the end user, who wishes to identify the most appropriate available model for their specific real-world cohort, even when that model is not the best on average. Our contributions are as follows: 
\begin{enumerate}
    \item We demonstrate empirically on 6 real-world cohorts that successful external validation does not necessarily imply generalization and provide a theoretical justification. 
    \item We develop a prognostic weighting approach that is more robust to distribution shifts. Specifically, we show across a range of external cohorts that the models weighted towards a meta-analysis distribution, on average, perform better in calibration than competing models. 
    \item We provide a structured framework for selecting, among the models available in the literature, the one that best fits the prognostic needs of a given external cohort. 
\end{enumerate}

\section{Methods}
The Methods are organized into five parts. First, we describe the study cohorts used for empirical evaluation. Second, we describe and formulate the problem of distribution shifts we attempt to solve. Third, from the perspective of the model developer, we present a framework to train prognostic models that are more robust to distributional shifts across cohorts. Fourth, from the perspective of the end user, we describe a framework for selecting the prognostic model that best fits a specific target cohort. Lastly, we summarize the statistical analyses used to evaluate model performance and clinical utility.

\subsection{Study cohorts}
As noted above, the aim of this study was not to address a specific clinical problem, such as prognostication in a single disease setting, but rather to propose: (1) a structured approach for training the best average prognostic model from the perspective of the model developer, and (2) a framework for selecting, from the perspective of the end user, the prognostic model from the literature that best fits the needs of a given target cohort.

To demonstrate the proposed methodology, we performed empirical analyses using high-quality data from tertiary institutions across distant geographic regions, including centers in the United States, Europe, and Japan. All cohorts consisted of patients undergoing surgery for colorectal liver metastases (CRLM), thereby providing a clinically coherent setting in which to evaluate model transportability across different populations.

Specifically, we included patients who underwent surgical treatment for colorectal liver metastases at six academic institutions: Johns Hopkins Hospital (JHH, USA), Cleveland Clinic Foundation (CCF, USA), Erasmus University of Rotterdam (UOR, the Netherlands), Yokohama City University (YCU, Japan), CHU Clermont-Ferrand (CHU, France), and the University of Graz (Austria). Eligible patients were aged 18 years or older and had undergone complete (R0 or R1) resection of both the primary colorectal tumor and all colorectal liver metastases.

\subsection{Problem formulation: distribution shifts in real-world medical data}
Consider a dataset of CRLM patients, where each patient $i$ is characterized by a set of prognostic covariates $\mathbf{x}_i$, a treatment status $t_i \in \{0,1\}$, where $t_i$ = 1 denotes receipt of adjuvant chemotherapy and a $t_i$ = 0 denotes no adjuvant chemotherapy and a recurrence outcome $y_i \in \{0,1\}$.

A central challenge in prognostic modeling is that a model trained in one real-world cohort may later be applied to external cohorts whose underlying data distributions differ. In this study, we consider two types of distribution shift that may affect model transportability across cohorts.

The first is covariate shift, which refers to differences in the distribution of observed prognostic factors across cohorts. These differences may arise from variation in patient selection criteria and differences in the “pool” of candidate patients across countries and centers.

The second is concept shift, where the conditional outcome distribution differs across cohorts: $p_{\text{train}}(y \mid \mathbf{x}) \neq p_{\text{test}}(y \mid \mathbf{x})$. Concept shifts may occur due to differences in standards of care between centers or because of varying prevalence of unobserved prognostic factors such as genomic alterations, radiomics features, or other latent clinical variables not captured in the dataset.

Formally, given $n_{train}$ patients each with covariates $\mathbf x_i$ and outcome $y_i$ in a training cohort, model training involves minimizing the loss function 
\begin{align}
    \mathcal{L}_{train} & = \int \ell(f(\mathbf x),y) p_{train}(\mathbf x,y) d \mathbf x dy.
\end{align}
When the joint distribution in the test cohort is similar to that of the training cohort, that is, $p_{train}(\mathbf x,y) \approx p_{test}(\mathbf x,y)$, the test loss is expected to be similar:
\begin{align}
     \mathcal{L}_{train} = \int \ell(f(\mathbf x),y) p_{train}(\mathbf x,y) d \mathbf x dy \approx \int \ell(f(\mathbf x),y) p_{test}(\mathbf x,y) d \mathbf x dy =  \mathcal{L}_{test}.
\end{align}

Accordingly, the methodological problem addressed in this study is how to improve prognostic modeling when the training and target cohorts differ because of covariate shift, concept shift, or both. In the following sections, we address this problem from two complementary perspectives: first, from the perspective of the model developer, who seeks to train a model that is more robust across heterogeneous external cohorts, and second, from the perspective of the end user, who seeks to identify the model most appropriate for a specific target cohort.

\subsection{Part I. Proposed solution from the model-developer perspective: training a more generalizable prognostic model}\label{sec:weights_method}

Having formalized the problem of distributional shift across cohorts, we next address it from the perspective of the model developer. Specifically, our goal is to train a prognostic model that is less dependent on the idiosyncrasies of a single training cohort and instead better aligned with the broader population in which the model may eventually be applied. To this end, we propose an importance-weighting framework based on internal prognostic risk distribution that shifts the effective training distribution toward a meta-analysis-derived target distribution.

Ideally, if the target test distribution $p_{test}({\bf x}, y)$ is known, we can introduce \emph{sample weights}
        $$
        w({\bf x},y) = \frac{p_{test}({\bf x},y)}{p_{train}({\bf x},y)}
        $$
        to recover the true test loss from training data, namely
        \begin{align}
            \mathcal{L}_{train}^w &= \int  w({\bf x},y)\ell(f(\mathbf x),y) p_{train}(\mathbf x,y) d \mathbf x dy  \\&= \int  \frac{p_{test}(\mathbf x,y)}{p_{train}(\mathbf x,y)}\ell(f(\mathbf x),y) p_{train}(\mathbf x,y) d \mathbf x dy \label{loss_function}  \\
            &= \int \ell(f(\mathbf x),y) p_{test}(\mathbf x,y) d \mathbf x dy  \\
            &= \mathcal{L}_{test}.
        \end{align}
        

In practice, however, the target test distribution is unknown when the prognostic model is trained. Moreover, there is no single universal target distribution, because each cohort in which the model may eventually be used may have a different distribution. Therefore, from the perspective of model development, it is desirable to train a prognostic model that is more robust to distributional shifts.

We aim to improve generalizability by adjusting the training process toward the distribution of the broader population of CRLM patients. To this end, we use summary statistics from a meta-analysis as a proxy for the general population distribution, under the rationale that a meta-analysis samples multiple studies across regions and is therefore more likely to reflect the broader distribution encountered in practice.

Empirically, in a training set, the training process is given by minimizing the empirical loss
\begin{align}
     \hat{\mathcal L}_{train} = \frac{1}{n_{train}} \sum_{i=1}^{n_{train}} w_i \ell(f({\bf x}_i),y_i),
\end{align}
where $w_i$ is the sample weight on observation $i$

We first consider the scenario where the data only suffer from concept shift. We propose a method that computes the concept weight.
\begin{align}
    w^{conc}_i = \frac{\hat p_{meta}(y_i \mid \mathbf x_i) }{\hat p_{train}(y_i \mid \mathbf x_i) },
\end{align}
where $\hat p_{train}(y_i \mid {\bf x}_i)$ can be directly estimated by fitting a survival model $S_i$ in the training cohort and making inference  $S_i({\bf x}_i)$ for patient i.  $\hat p_{meta}(y_i \mid {\bf x}_i)$ is obtained from a simulated distribution with aggregated statistics from the meta-analysis, namely the mean and standard deviation of recurrence.

This approach fully addresses distribution shift when the shift is purely due to differences in the conditional outcome distribution and there is no covariate shift. However, concept-based weighting alone is not generally sufficient to correct pure covariate shift, because in that setting the discrepancy arises through $p(\mathbf {\bf x}_i)$ rather than $p(y_i \mid \mathbf {\bf x}_i)$. Still, because differences in covariate distributions often manifest through differences in predicted-risk strata, the proposed approach that relies on an internal risk prognosis may partially reduce some forms of covariate mismatch in practice. 

The model used to compute weights must be trained on the training cohort. Since the weights are applied during training of the final prognostic model in that same cohort, in order to recover the general distribution $\hat p_{meta} (y_i \mid \mathbf {\bf x}_i)$, it is necessary that $\hat p_{train}(y_i \mid  \mathbf  {\bf x}_i)$ used in the weights is an accurate estimate of the $\hat p_{train}(y_i \mid {\bf x}_i)$ in the final model, as shown in equation \eqref{loss_function}.

Let $n_{train}$ denote the number of observations in the training set and $m$ the number of risk strata, such that 
$$
s(i) = k \quad \text{ if } \ell_k < \hat p_{train}(y_i \mid \mathbf x_i) \le u_k \quad \forall  k = 1,2,\hdots, m,
$$
where $\ell_k, u_k$ are the lower and upper thresholds of stratum $k$ respectively. Then, the importance of stratum $k$ in the training cohort is given by 
$$
{\hat{q}_{train}^k} = \frac{n_k}{n_{train}} =  \frac{\text{number of patients in bucket } k}{\text{total number of patients}}.
$$
Specifically, $\hat{q_k} = \mathbb{P}( \hat p_{train}(y \mid {\bf x}) \in \text{bucket } k)$.

We then generate a simulated cohort with synthetic recurrence risk as a normally distributed random variable $R \sim \mathcal{N}(\mu_y, \sigma_y)$ where $\mu_y, \sigma_y$ are the reported mean and standard deviation of recurrence in the meta-analysis. 
We then compute the importance of each stratum $k = 1,2,\hdots, m$ as 
\begin{align}
    \hat q_{meta}^k = \mathbb{P}(R \le u_k) - \mathbb{P}(R \le \ell_k) &\quad  k = 1, 2,3,\hdots, m
\end{align}
The weights are then computed for each stratum $k$ as 
\begin{align}
    w^{conc}_{k} = \frac{\hat q_{meta}^{k}}{\hat q_{train}^{k}},
\end{align}
or equivalently for each patient $i$ in the training set as 
$$
w^{conc}_{i} = \frac{\hat q_{meta}^{s(i)}}{\hat q_{train}^{s(i)}} = \frac{\text{importance of its bucket in general population}}{\text{importance of its bucket in training data}}.
$$

Next, observe that by conditional probability, 
\begin{align*}
        \frac{\hat p_{meta}({\bf x}_i, y_i)}{\hat p_{train}({\bf x}_i, y_i)} &= 
        \underbrace{\frac{\hat p_{meta}(y_i \mid {\bf x}_i) }{\hat p_{train}(y_i \mid {\bf x}_i)}}_{\text{concept weight}} \cdot \underbrace{\frac{\hat p_{meta}({\bf x}_i)}{\hat p_{train}({\bf x}_i)}}_{\text{covariate weight}} 
\end{align*}
To adjust when there exist both concept shifts and covariate shifts, we introduce the joint weight given by 
\begin{align}
    w^{joint}_i = w^{conc}_i \frac{\hat p_{meta}( {\bf x}_i)}{\hat p_{train}({\bf x}_i)},  
\end{align}
where $\frac{\hat p_{meta}({\bf x}_i)}{\hat p_{train}({\bf x}_i)} $ accounts for the covariate shift. 

\begin{assumption}[]\label{assum:constant_cov}
    The covariance between each pair of covariates is constant across cohorts.   
\end{assumption}
Let $\mathbf \Sigma^{train} \in \mathbb{R}^{d\times d}$ denote the covariance matrix of covariates $\mathbf{X}^{train}$ in the training data.
By Assumption \ref{assum:constant_cov}, we obtain the covariance matrix of the meta-analysis population $\boldsymbol{\Sigma}^{meta}$ by the following:
\begin{align*}
    &\boldsymbol{\Sigma}^{meta}_{j,j} = (\sigma^{meta}_j)^2, \\
    & \boldsymbol{\Sigma}^{meta}_{j,l} = \boldsymbol{\Sigma}^{train}_{j,l} \quad \text{for } j \neq l,
\end{align*}
where $(\sigma^{meta}_j)^2$ is the reported variance of feature $j$ in the meta-analysis. 
We can then simulate a random vector $\mathbf{X}^{meta} \sim \mathcal{N}(\boldsymbol{\mu}^{meta}, \boldsymbol{\Sigma}^{meta})$ where $\boldsymbol{\mu}^{meta}$ are the reported means of the features in the meta-analysis. 
After obtaining the simulated general population feature space, we then follow the literature and use Kernel Density Estimation (KDE) with a Gaussian kernel to obtain the probability density function of $\mathbf{X}^{meta}$ and $\mathbf{X}^{train}$ \citep{nair2019covariate}. Specifically, 
\begin{align*}
    \hat p_{train}(\mathbf{x}_i) &= \frac{1}{n_{train}} \sum_{n=1}^{n_{train}} \frac{1}{b^d}K(\frac{\mathbf{x}_i - \mathbf{x}_n}{b}), \\ 
    \hat p_{meta}(\mathbf{x}_i) &= \frac{1}{n_{meta}} \sum_{n=1}^{n_{meta}} \frac{1}{b^d}K(\frac{\mathbf{x}_i - \mathbf{x}_n}{b}),
\end{align*}
where $b$ is a chosen bandwidth and $K$ is the Multivariate Gaussian kernel.

\begin{algorithm}[H] 
\begin{algorithmic}
\STATE {\textbf{Input}: Training dataset $(\bf X_{train}, t_{train}, y_{train})$}, reported outcome mean and variance in a meta-analysis $\mu_y, \sigma^2_y$. 
\STATE {\textbf{Output}: Predicted 5-year recurrence risks on testing cohort patients  $\bf \hat y_{test}$}

\STATE \textit{Step 1: Risk prediction and stratification on training cohort}
\STATE{Obtain $\hat p_{train}(y_i|{\mathbf x_i})$ for each training patient $i$.}
\STATE{Stratify training patients into $m$ strata where $s(i) = k$ if $\ell_k \le \hat p_{train}(y_i \mid \mathbf x_i) \le u_k$ for $k = 1,2,\hdots, m$.}
\STATE{Obtain stratum density $\hat q_{train}^k = \frac{n_k}{n_{train}}$.}

{\STATE \textit{Step 2: Risk distribution approximation through meta-analysis}}
\STATE{Simulate recurrence risk in the general patient population $R \sim \mathcal{N}(\mu_y, \sigma_y)$ where $\mu_y, \sigma_y$ are the reported recurrence mean and standard deviation in meta-analysis.}
\STATE{Obtain stratum density $\hat q_{meta}^k = \mathbb{P}(r \le u_k) - \mathbb{P}(r \le \ell_k)$. }

{\STATE \textit{Step 3: Concept weight computation}}
\STATE Compute sample weight for each training patient $i$ to be $w_i = \frac{\hat q_{meta}^{s(i)}}{\hat q_{train}^{s(i)}}$ where $s(i)$ is the corresponding stratum of patient $i$.

\STATE \textit{Step 4: Train weighted prognostic model}
\end{algorithmic}
\caption{Weighted Prognostic Model Adjusting for Concept Shift}
\label{alg:weights}
\end{algorithm}

\normalfont 
\subsection{Part II. Proposed solution from the end-user perspective: selecting the best model for a target cohort}
Even if a model is optimized to perform well on average across heterogeneous populations, it will not necessarily be the best-performing model for every specific cohort. From the perspective of the end user, the relevant question is therefore different: rather than asking which model performs best on average, the goal is to identify which available model is most likely to be well calibrated and clinically useful in the user’s own patient population.

We therefore propose a cohort-specific model-selection strategy based on similarity between the target cohort and the original development cohorts of candidate models. Specifically, we hypothesized that models developed in cohorts whose outcome profile more closely resembles that of the target cohort are more likely to show better calibration when externally applied to that cohort.

Let $\mu^{test}_y$ and $\mu^{train}_y,$ denote the mean Kaplan Meier estimates at the time point of interest in the target cohort and a model-development cohort, respectively. We quantified the distance between the two cohorts using the Euclidean norm of the means, given by
$$
dist_{train, test} = \| \mu^{train}_y, -\mu^{test}_y\|_2.
$$
Smaller values indicate greater similarity between the development and target cohorts.

\subsection{Statistical analysis}
To quantify the training and testing distribution similarity, we compute the Kullback–Leibler (KL) divergence between the two cohorts. 
To capture both concept shift and covariate shift, we define
$$
\mathbf{z}_i = [\mathbf{x}_i, y_i]
$$
as the concatenated vector of the covariates and outcome. The probability density function is obtained similarly using KDE with a Gaussian kernel. Namely, for a pair of training and testing sets, 
\begin{align*}
    \hat q_{train}(\mathbf{z}_i) &= \frac{1}{n_{train}} \sum_{n=1}^{n_{train}} \frac{1}{b^d}K(\frac{\mathbf{z}_i - \mathbf{z}_n}{h}), \\ 
    \hat q_{test}(\mathbf{z}_i) &= \frac{1}{n_{test}} \sum_{n=1}^{n_{test}} \frac{1}{b^d}K(\frac{\mathbf{z}_i - \mathbf{z}_n}{h}).
\end{align*}
Then, the KL divergence between the two cohorts is given by 
\begin{align*}
    \text{KL}(\hat{q}_{train}|\hat{q}_{test}) = \mathbb{E}_{\mathbf{z} \sim \hat{q}_{train}}\left[\log \left (\frac{\hat{p}_{train}(\mathbf{z})}{\hat{p}_{test}(\mathbf{z})} \right ) \right ].
\end{align*}
A smaller KL divergence indicates that the training and testing cohort are more similar in distribution. 

Three prognostic model classes were evaluated: Cox Proportional Hazards, Random Survival Forests (RSF), representing the ensemble-tree class, and Optimal Survival Trees (OST), representing the single-tree class \cite{ishwaran2008random, bertsimas2022optimal}. Cox models were implemented in R using the survival package. RSF models were implemented in Python using the scikit-survival package, and OST models were implemented in R using the Interpretable AI modules (version 1.10.2).

Model performance was assessed along two complementary dimensions: discrimination and calibration. Discrimination was evaluated using Harrell’s concordance index. Calibration was assessed using the Integrated Calibration Index (ICI), which summarizes the average absolute difference between predicted and observed recurrence probabilities across the risk spectrum and provides a single interpretable measure that enables direct comparison across models. Formally, ICI on a test set is given by
\begin{align}
    \text{ICI}_{train, test} = \frac{\sum_{i=1}^{n_{test}} |\hat{p}_{train}(y_i \mid \mathbf{x}_i) - \mathbb{P}(y_i =1 \mid \hat{p}_{train}(y_i \mid \mathbf{x}_i)) |}{n_{test}} \in [0,1]. 
\end{align}

To formally evaluate the correlation between distribution similarity and calibration, we compute the Spearman correlation coefficient, given by 
\begin{align}
    \rho = 1 - \frac{6\displaystyle \sum_{train} \sum_{test \neq train} (R[\text{ICI}_{train,test}] -R[ \text{KL}(\hat{q}_{train}|\hat{q}_{test})])^2}{n(n^2-1)} \in [-1,1],
\end{align}
where $R[\text{ICI}_{train,test}]$ is the ranking of $\text{ICI}_{train,test}$ and $R[\text{KL}(\hat{q}_{train}|\hat{q}_{test})]$ is the ranking of $\text{KL}(\hat{q}_{train}|\hat{q}_{test})$ among all pairs of cohorts considered.
A large positive $\rho$ would indicate that calibration and cohort distribution similarity are highly positively correlated. 

Next, we want to formally evaluate whether weighting on average improves calibration. To this end, we compute the Wilcoxon signed-rank test that tests the hypothesis that whether the median ICI among the weighted models is statistically different from the median ICI among the unweighted models \citep{woolson2007wilcoxon}.

Lastly, to assess potential clinical utility in guiding treatment decisions, we also performed decision curve analysis (DCA). In the empirical evaluation of our framework, the treatment of interest was adjuvant chemotherapy. Therefore, DCA was restricted to patients treated with surgery alone, such that observed recurrence served as a pragmatic approximation of baseline risk. Net benefit was calculated across a range of threshold probabilities and compared with the default “treat all” and “treat none” strategies. For figure clarity, we emphasized the threshold range from 0.10 to 0.70, which broadly reflects clinically relevant risk-benefit tradeoffs. For each model, we recorded the maximum net benefit and the range of thresholds over which the model outperformed both default strategies.

All analyses were performed using Python and R.

\section{Results}

The baseline characteristics, sample sizes, and long-term outcomes of the six real-world CRLM cohorts are summarized in Table \ref{tab:sample_desc}. The 5-year RFS reported correspond to the mean 5- year Kaplan-Meier estimates. Marked heterogeneity was observed across cohorts. Among surgery-alone cohorts, the mean 5-year mean recurrence free survival ranged from 22\% to 62\%; among surgery + adjuvant chemotherapy cohorts, it ranged from 10\% to 60\%. Covariates used to train the prognostic models were selected on the basis of relevant surgical literature and included metastatic lymph node status, number of liver metastases, maximum metastasis size, laterality of liver metastases, carcinoembryonic antigen (CEA), extrahepatic disease, and resection margin status.

\begin{table}[h!]
    \centering
    \begin{tabular}{l}
    \includegraphics[width=\linewidth]{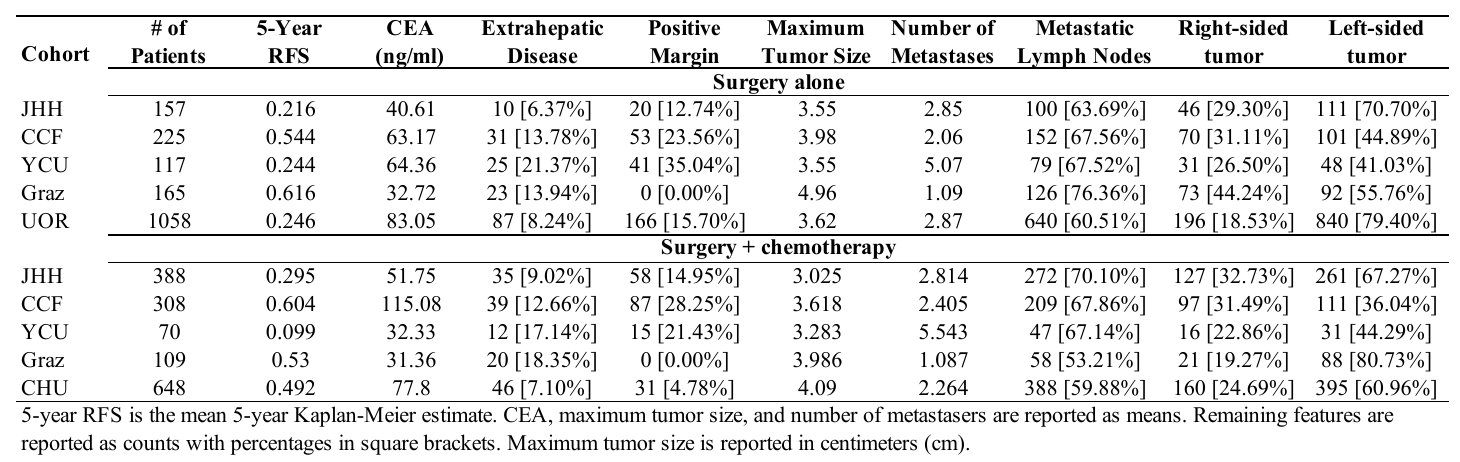}
    \end{tabular}
    \caption{Characteristics of the CRLM cohorts}
    \label{tab:sample_desc}
\end{table}

\subsection{Distributional similarity and external calibration}
To first establish the practical importance of distributional shift, we examined whether similarity between training and target cohorts was associated with external calibration. For each surgery-alone training cohort, we trained an unweighted prognostic model and evaluated it across each of the other available surgery alone cohorts. We then compared pairwise cohort similarity, quantified by KL divergence --- which reflects combined mismatch arising from both covariate shift and concept shift --- with calibration in the target cohort, quantified by ICI. The same analysis was repeated for surgery + chemotherapy cohorts.

\begin{figure}[ht!]
    \begin{subfigure}{0.3\textwidth}
        \centering
         \includegraphics[width=\textwidth]{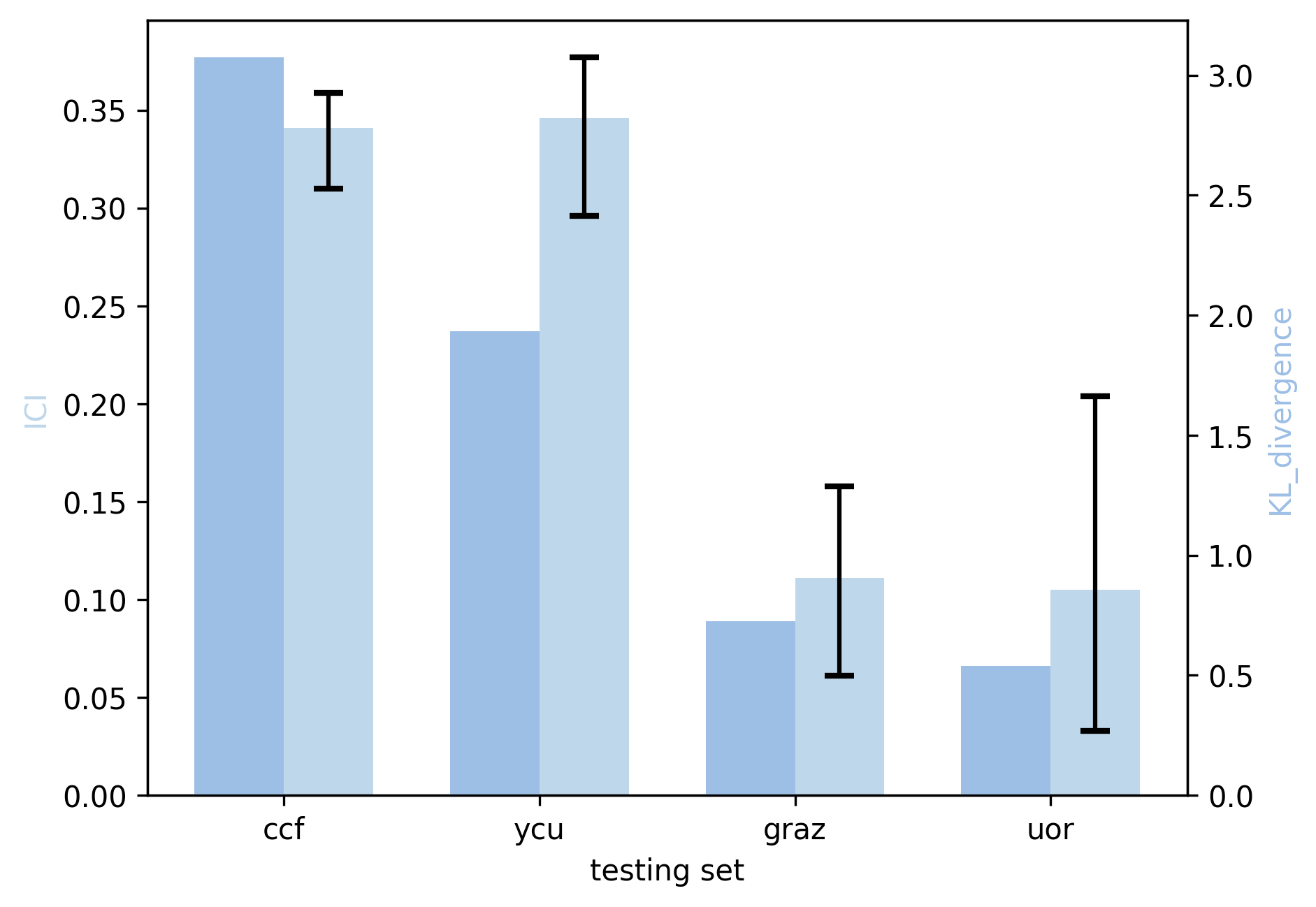}
         \caption*{\footnotesize A. Trained on JHH}
    \end{subfigure}
    \begin{subfigure}{0.3\textwidth}
        \centering
         \includegraphics[width=\textwidth]{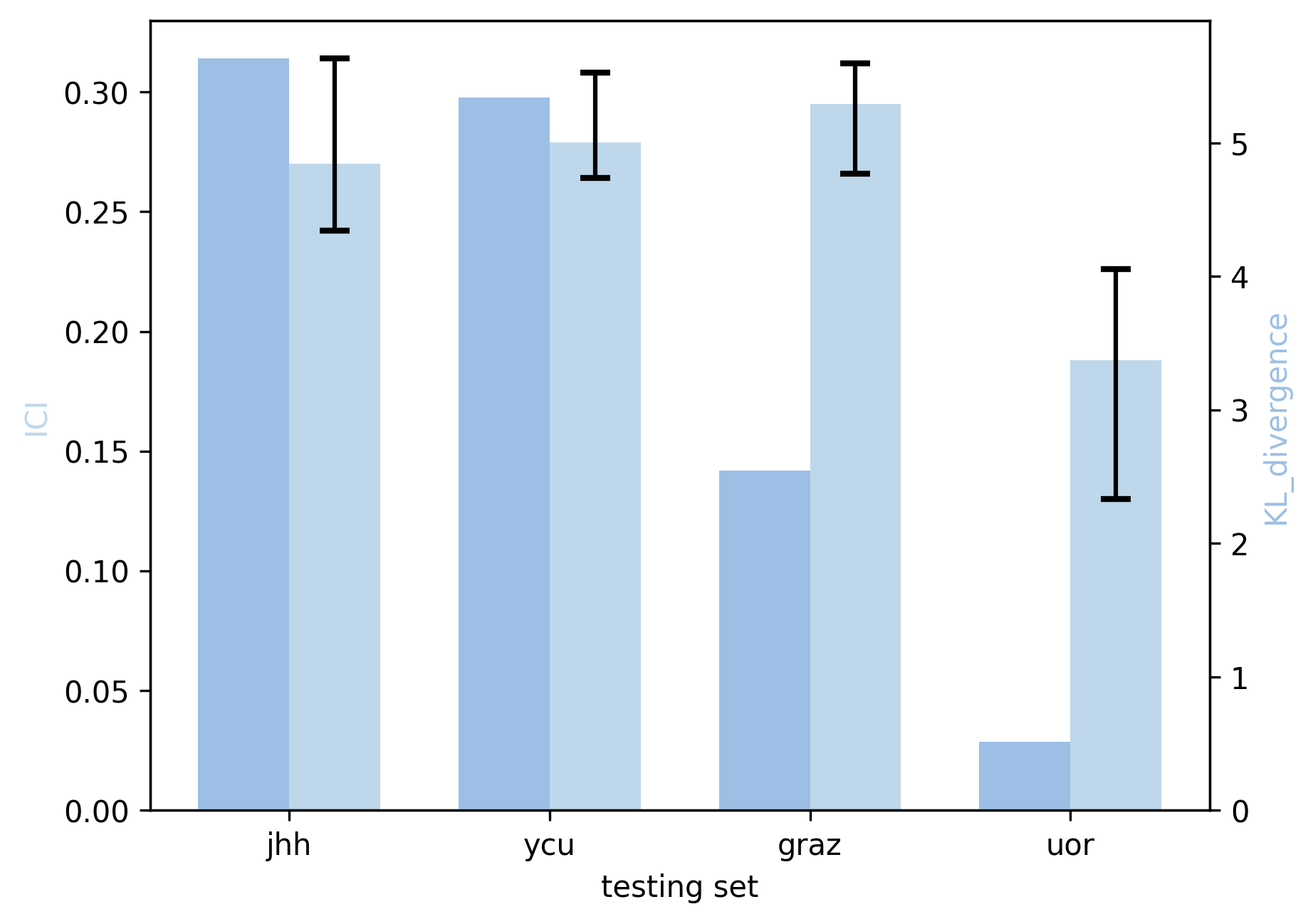}
        \caption*{\footnotesize B. Trained on CCF}
    \end{subfigure}
        \begin{subfigure}{0.3\textwidth}
        \centering
         \includegraphics[width=\textwidth]{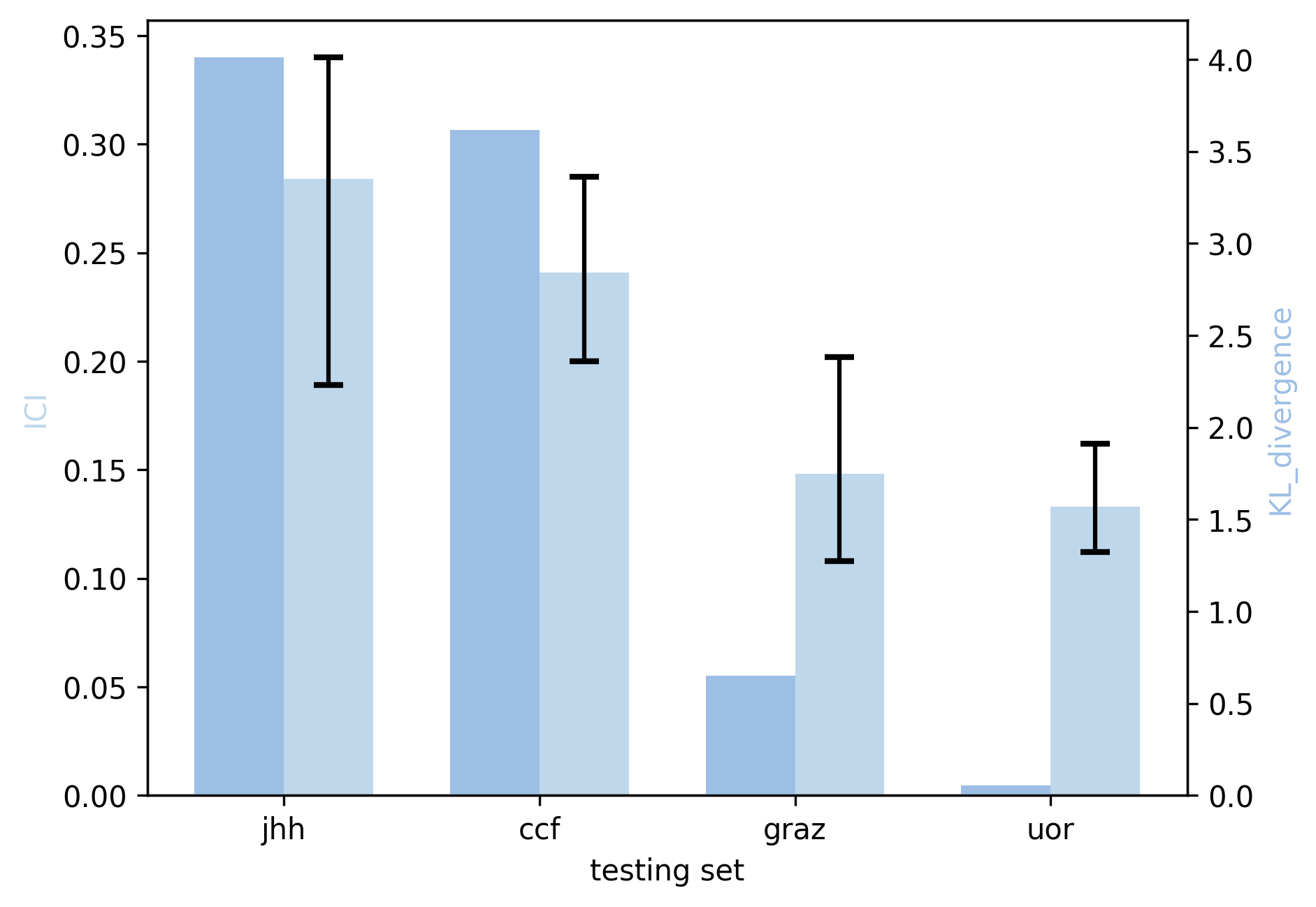}
        \caption*{\footnotesize C. Trained on YCU}
    \end{subfigure}\\
    \begin{subfigure}{0.3\textwidth}
        \centering
         \includegraphics[width=\textwidth]{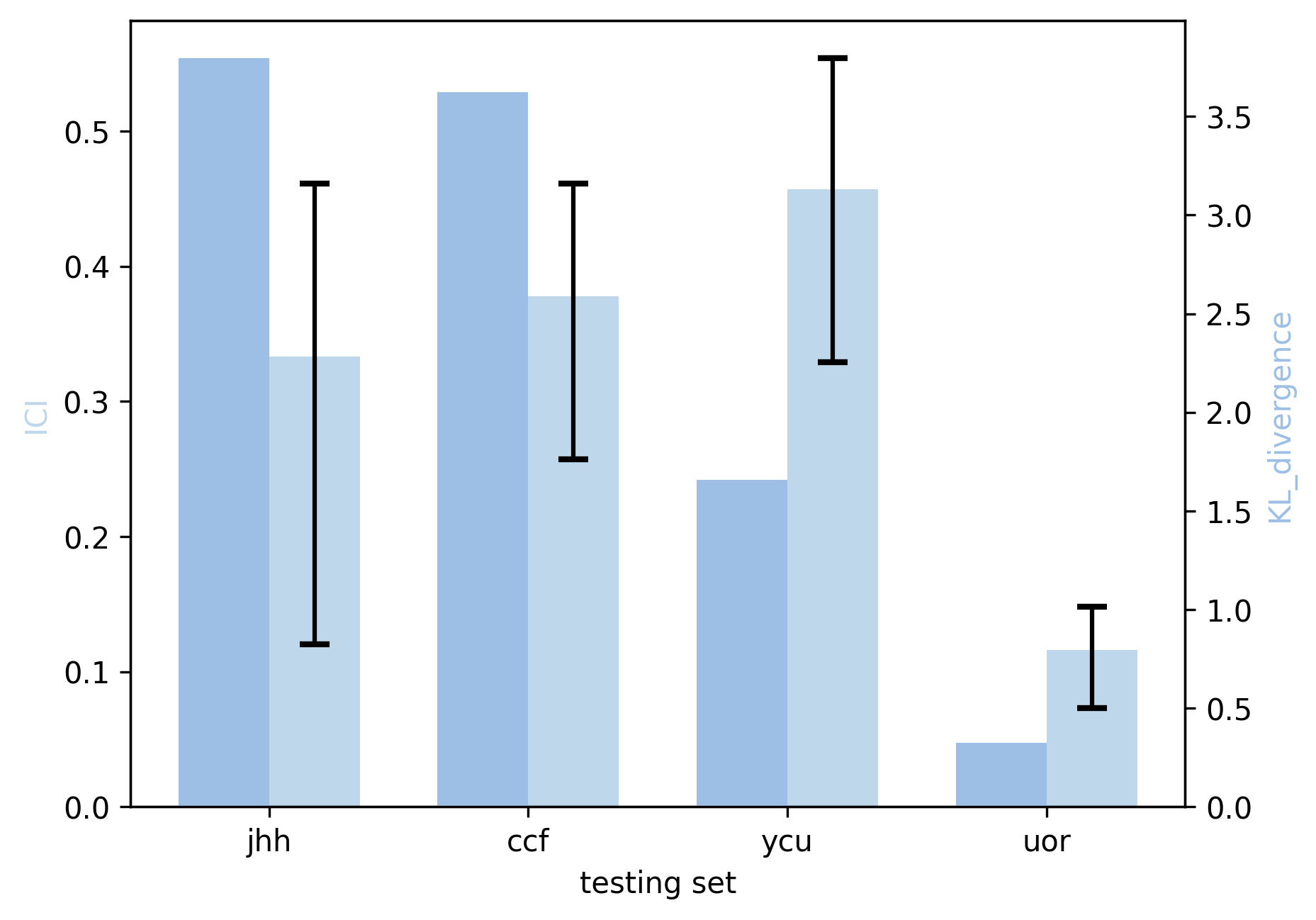}
        \caption*{\footnotesize D. Trained on Graz}
    \end{subfigure}
    \begin{subfigure}{0.3\textwidth}
        \centering
         \includegraphics[width=\textwidth]{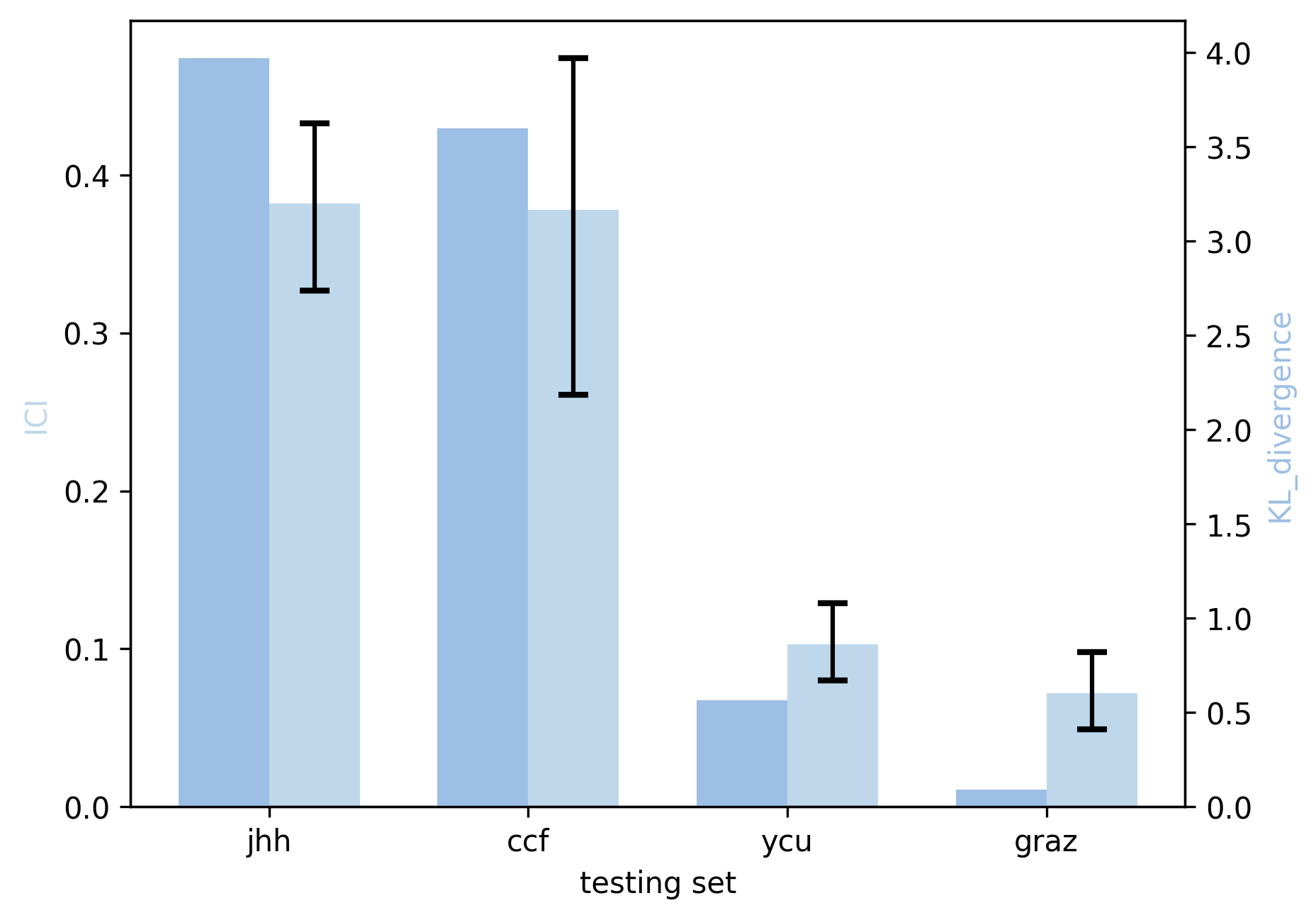}
        \caption*{\footnotesize  E. Trained on UOR}
    \end{subfigure}
    \caption{Pairwise KL divergence between training and test cohorts and the corresponding ICI for models trained in each surgery-alone cohort. ICIs are averaged across three models --- Cox, Optimal Survival Trees (OST), and Random Survival Forest (RSF) --- and the error bars represent the minimum and maximum ICIs across the three models. KL divergence is computed deterministically with fixed bandwidth. Each panel represents one training cohort, and the bars within that panel show results for all four test cohorts. Panels A–E correspond to models trained in JHH, CCF, YCU, Graz, and UOR, respectively.}
    \label{fig:untreated_part1}
\end{figure}

\begin{figure}[ht!]
    \begin{subfigure}[b]{0.3\textwidth}
        \centering
         \includegraphics[width=\textwidth]{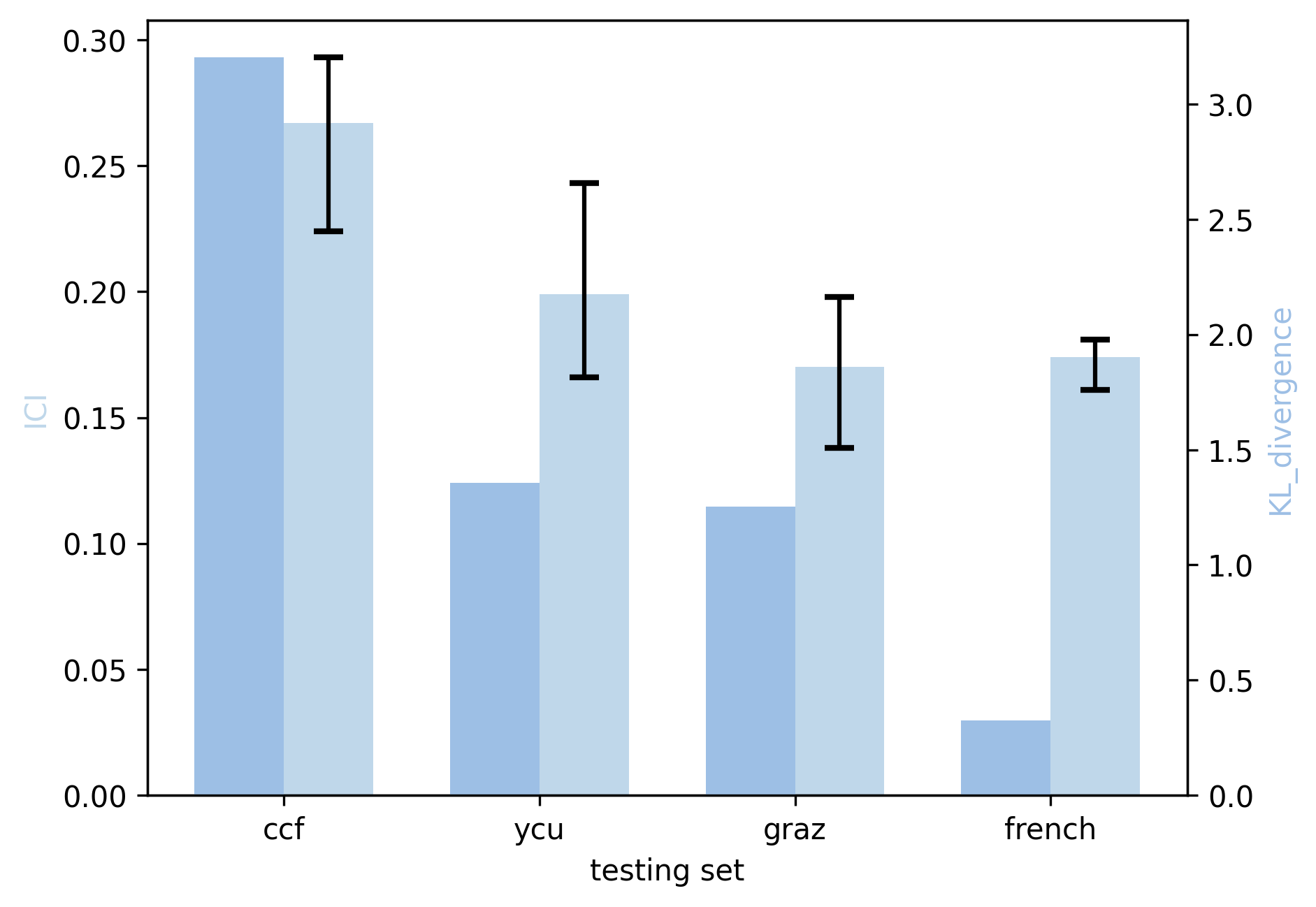}
        \caption*{\footnotesize A. Trained on JHH}
    \end{subfigure}
    \begin{subfigure}[b]{0.3\textwidth}
        \centering
         \includegraphics[width=\textwidth]{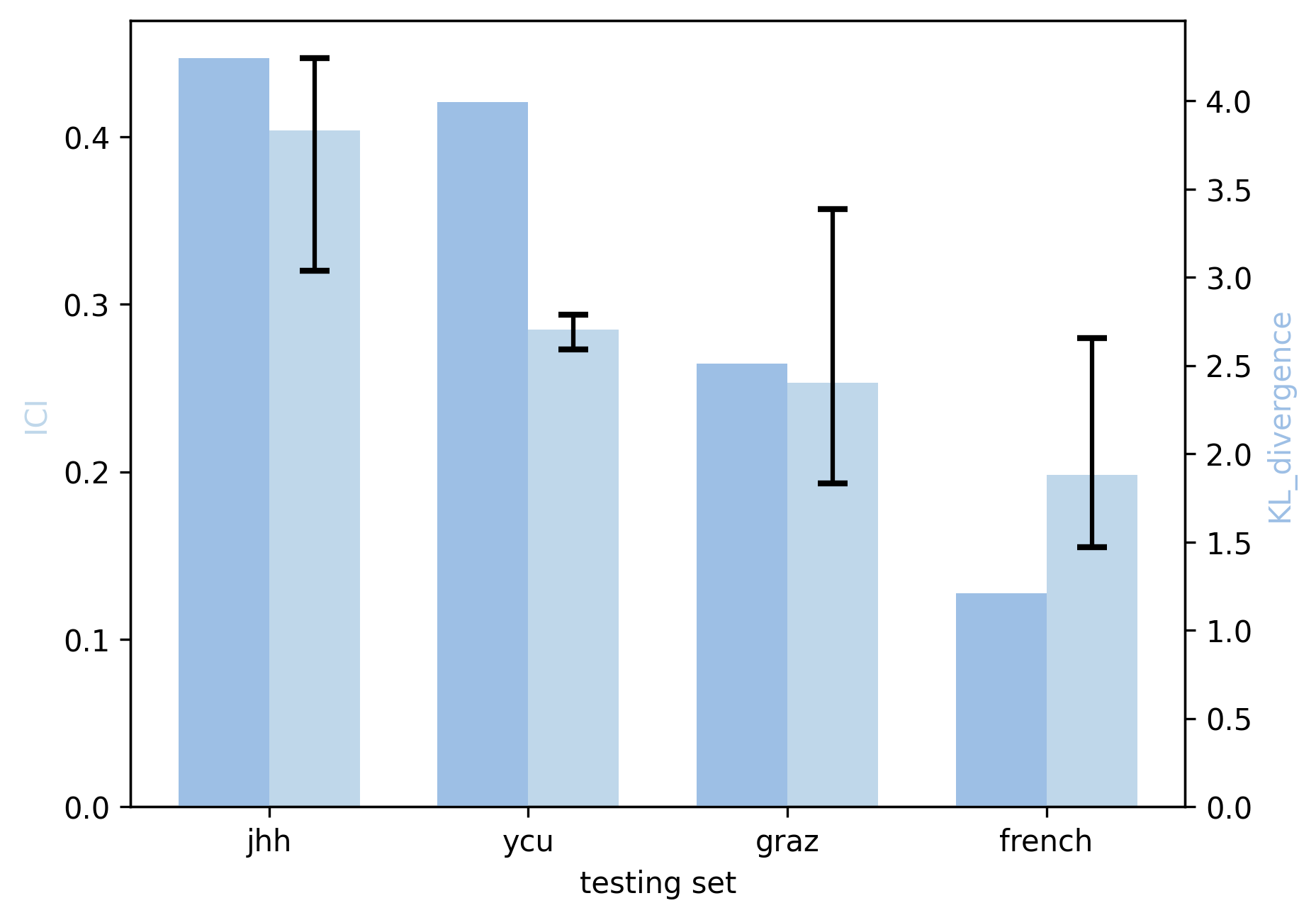}
         \caption*{\footnotesize B. Trained on CCF}
    \end{subfigure}
        \begin{subfigure}[b]{0.3\textwidth}
        \centering
         \includegraphics[width=\textwidth]{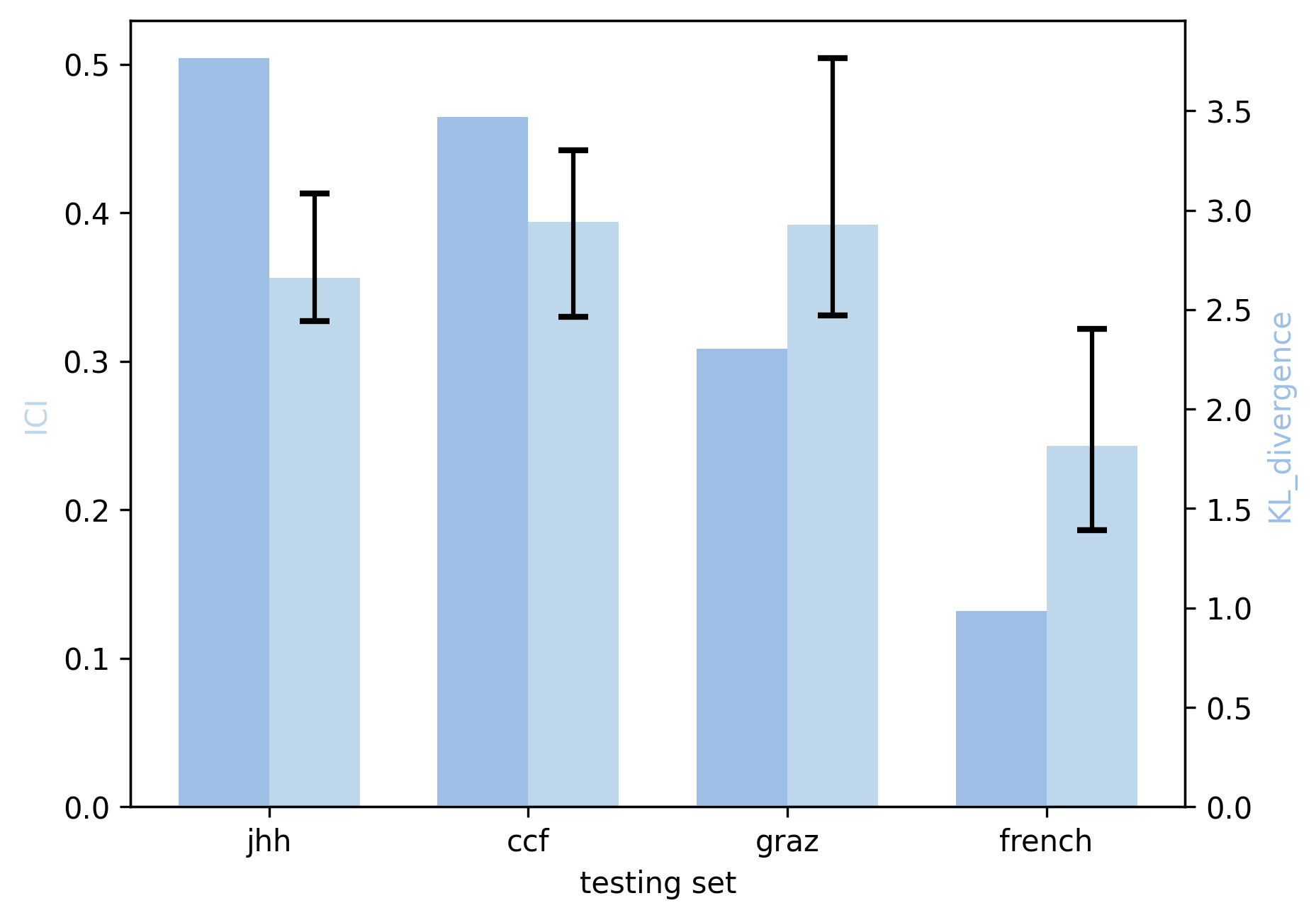}
            \caption*{\footnotesize C. Trained on YCU}
    \end{subfigure} \\
    \begin{subfigure}[b]{0.3\textwidth}
        \centering
         \includegraphics[width=\textwidth]{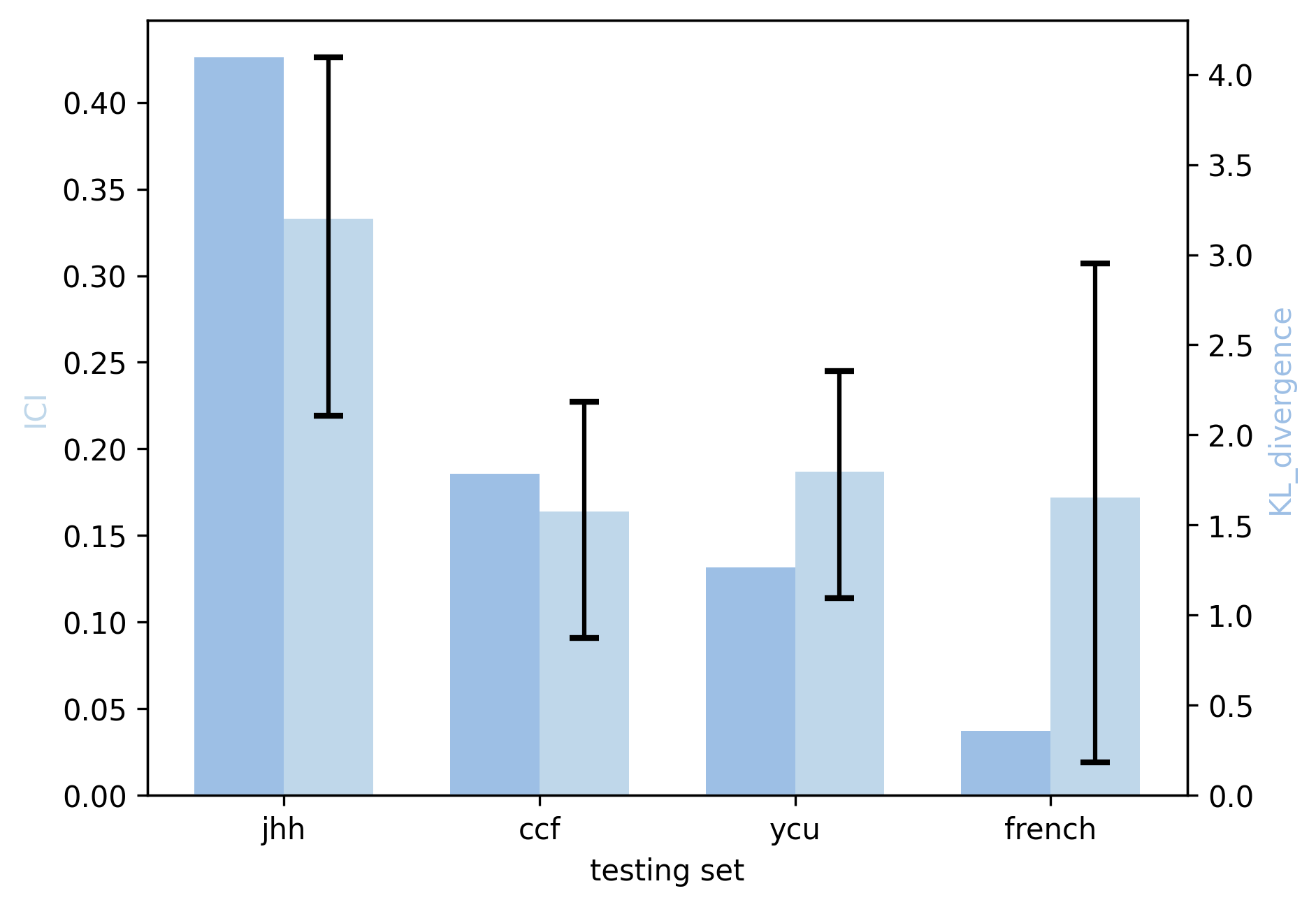}
         \caption*{\footnotesize D. Trained on Graz}
    \end{subfigure}
        \begin{subfigure}[b]{0.3\textwidth}
        \centering
         \includegraphics[width=\textwidth]{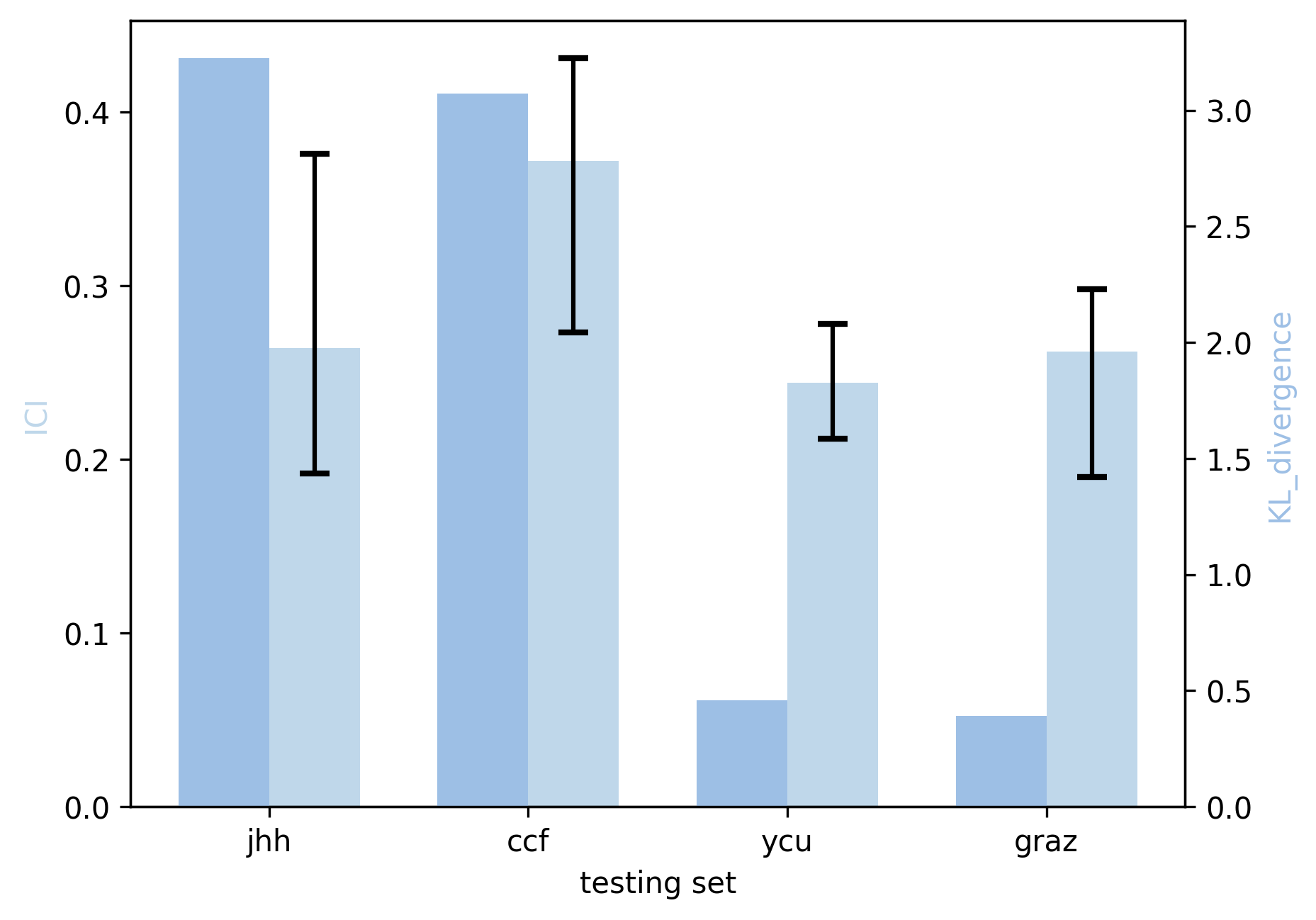}
        \caption*{\footnotesize  E. Trained on CHU}
    \end{subfigure}
    \caption{Pairwise KL divergence between training and test cohorts and the corresponding test-set ICI for models trained in each surgery + adjuvant chemotherapy cohort. ICIs are averaged across three models --- Cox, Optimal Survival Trees (OST), and Random Survival Forest (RSF) --- and the error bars represent the minimum and maximum ICIs across the three models. KL divergence is computed deterministically with fixed bandwidth. 
    Each panel represents one training cohort, and the bars within that panel show results for all four test cohorts. Panels A–E correspond to models trained in JHH, CCF, YCU, Graz, and CHU, respectively. }
    \label{fig:treated_part1}
\end{figure}

Across the analyses of patients who underwent surgery alone, models consistently achieved their best external calibration in target cohorts that were most similar in distribution to the corresponding training cohort. In other words, the smaller the distance between cohorts, the greater the distributional similarity and the lower the ICI, indicating better calibration (Figure \ref{fig:untreated_part1}).  The same qualitative pattern was observed in the analyses on patients who received adjuvant chemotherapy (Figure \ref{fig:treated_part1}). Overall, these results suggest that similarity in the joint distribution of covariates and outcomes is associated with better external calibration.

\begin{figure}
    \centering
    \includegraphics[width=0.75\linewidth]{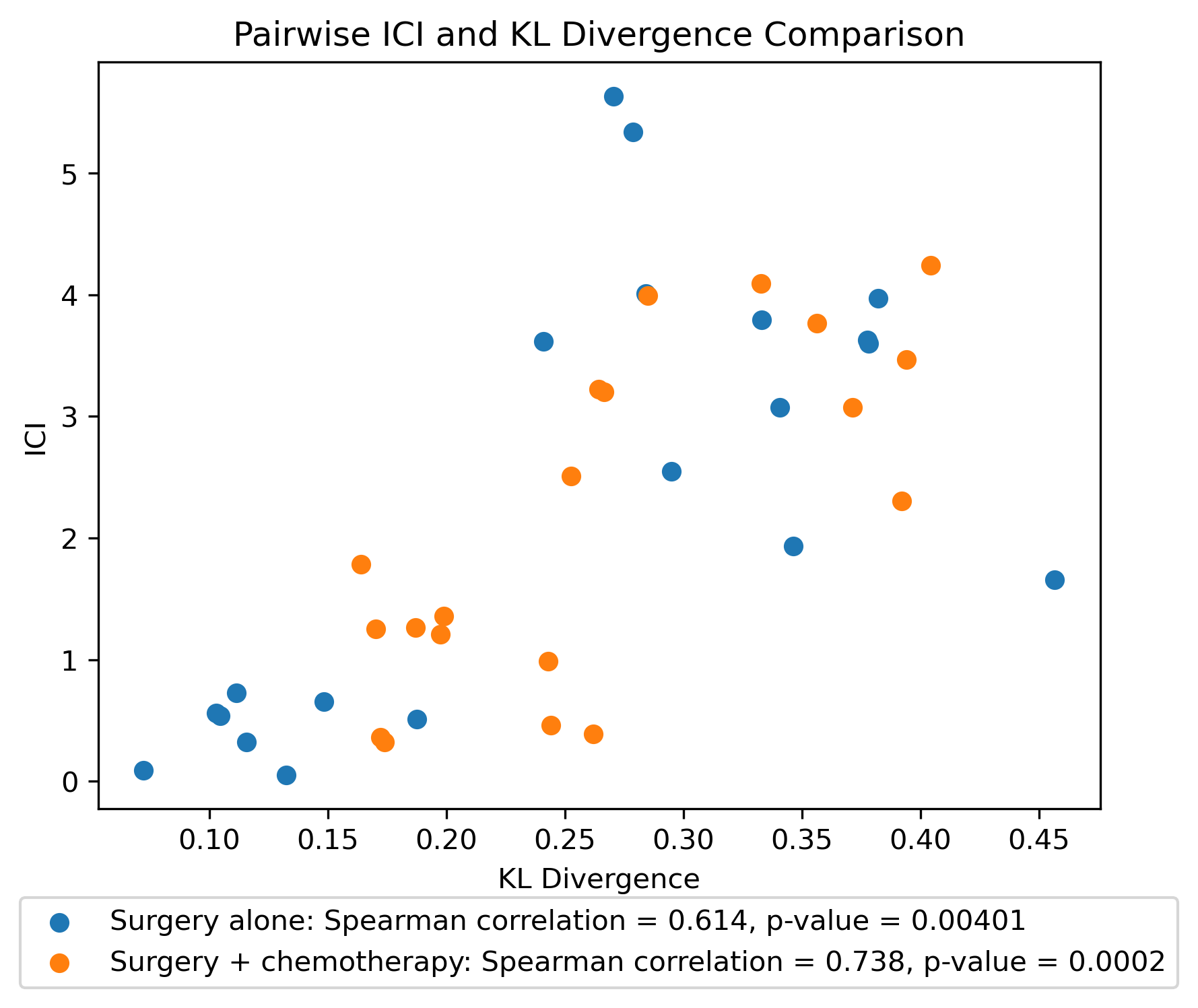}
    \caption{Correlation between ICI and KL Divergence}
    \label{fig:scatter_part1}
\end{figure}

The magnitude of this effect was substantial. For a given model trained on a single cohort, ICI ranged from values below 0.10 in distributionally similar target cohorts to values approaching 0.30 or higher in dissimilar target cohorts. For example, the model trained on surgery-alone patients from JHH, a cohort with high recurrence risk, achieved ICI values below 0.10 when applied to the UOR and YCU cohorts, which had similarly high recurrence risk, but its ICI increased to nearly 0.35 when applied to Graz, a substantially lower-risk cohort. Had external validation been performed only in the more similar cohorts, the model might have appeared broadly generalizable despite poor performance in distributionally distinct settings.
Moreover, the surgery-alone cohorts and surgery + chemotherapy cohorts yield statistically significant positive correlations of 0.614 (with p-value $= 0.004$) and 0.738 (with p-value $<0.001$) respectively between KL divergence and ICI. Appendix Figure \ref{fig:scatter_part1} plots these correlations in a scatterplot. These results further formally test the hypothesis that cohorts that are similar in distributions tend to have better calibration.  

These findings, which underscore the limited generalizability of prognostic models in the presence of covariate and/or concept shift, motivate the first methodological question addressed in this study: whether concept weighting or joint weighting can be used to train a prognostic model that generalizes better across heterogeneous external cohorts.

$$
$$

\subsection{Part I. Model-developer perspective: meta-analysis-informed weighting improves average external calibration}

We next evaluated whether the importance-weighting strategies described in Section \ref{sec:weights_method} improved average external calibration across cohorts. We applied the weighting framework using statistics derived from the meta-analysis by Bosma and colleagues \cite{bosma2021efficacy}.\footnote{The individual studies included in the meta-analysis are \cite{portier2006multicenter, nordlinger2013perioperative, hasegawa2016adjuvant, kanemitsu2021hepatectomy}. When 5-year recurrence was not explicitly reported in \cite{bosma2021efficacy}, it was extracted from the individual studies.}

For each training cohort, we estimated predicted 5-year recurrence risk and stratified patients into eight risk strata. We then trained unweighted models, models weighted by concept weights $w^{conc}$, and models weighted by joint weights $w^{joint}$, and evaluated each model separately in the remaining external cohorts. For each training cohort, the reported ICI represents the average across the corresponding external testing cohorts.

\begin{figure}[ht!]
    \begin{subfigure}[b]{0.45\textwidth}
        \centering
         \includegraphics[width=\textwidth]{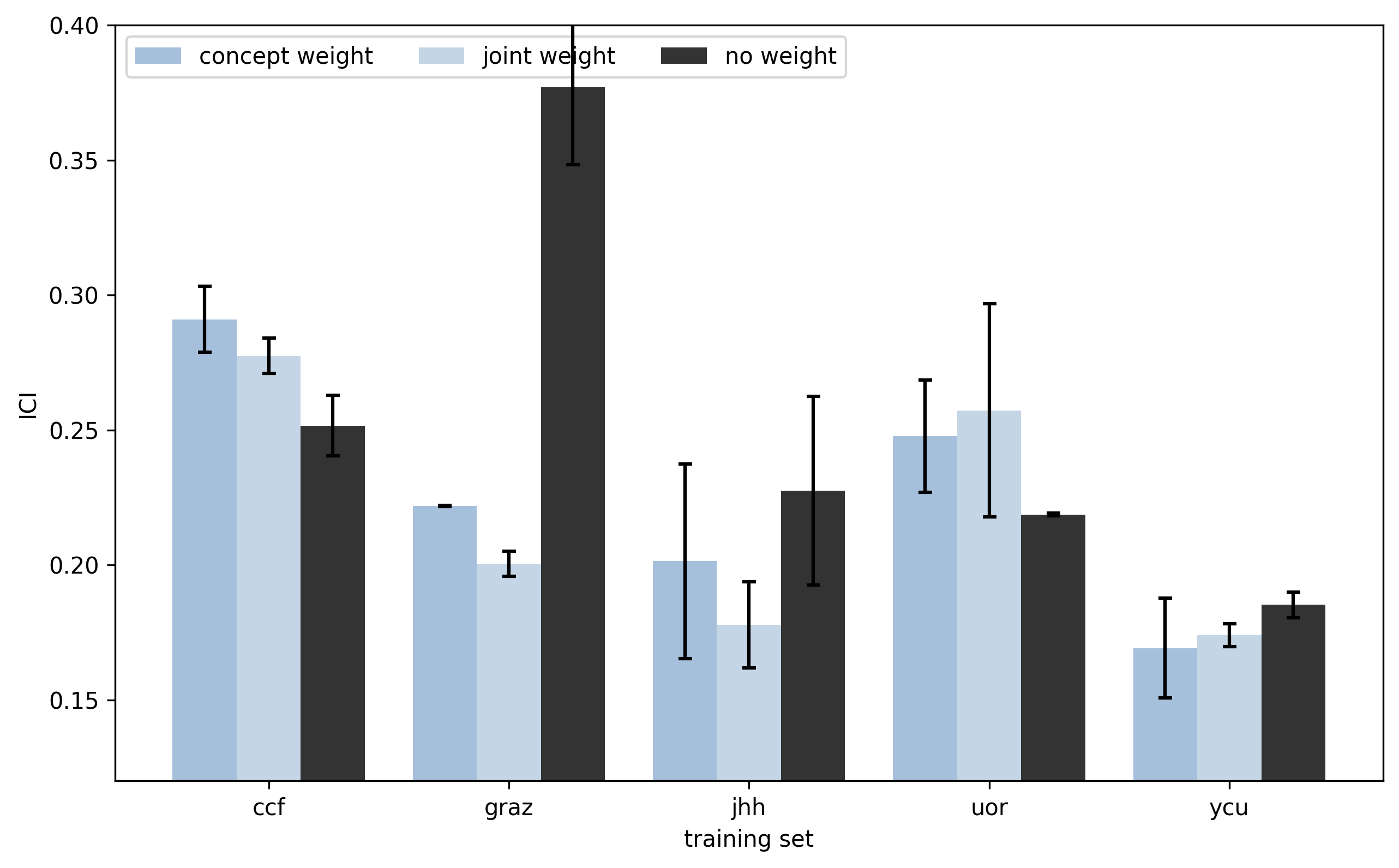}
        \caption*{\small A. Surgery-alone}
    \end{subfigure}
    \hspace{5mm}
    \begin{subfigure}[b]{0.45\textwidth}
        \centering
         \includegraphics[width=\textwidth]{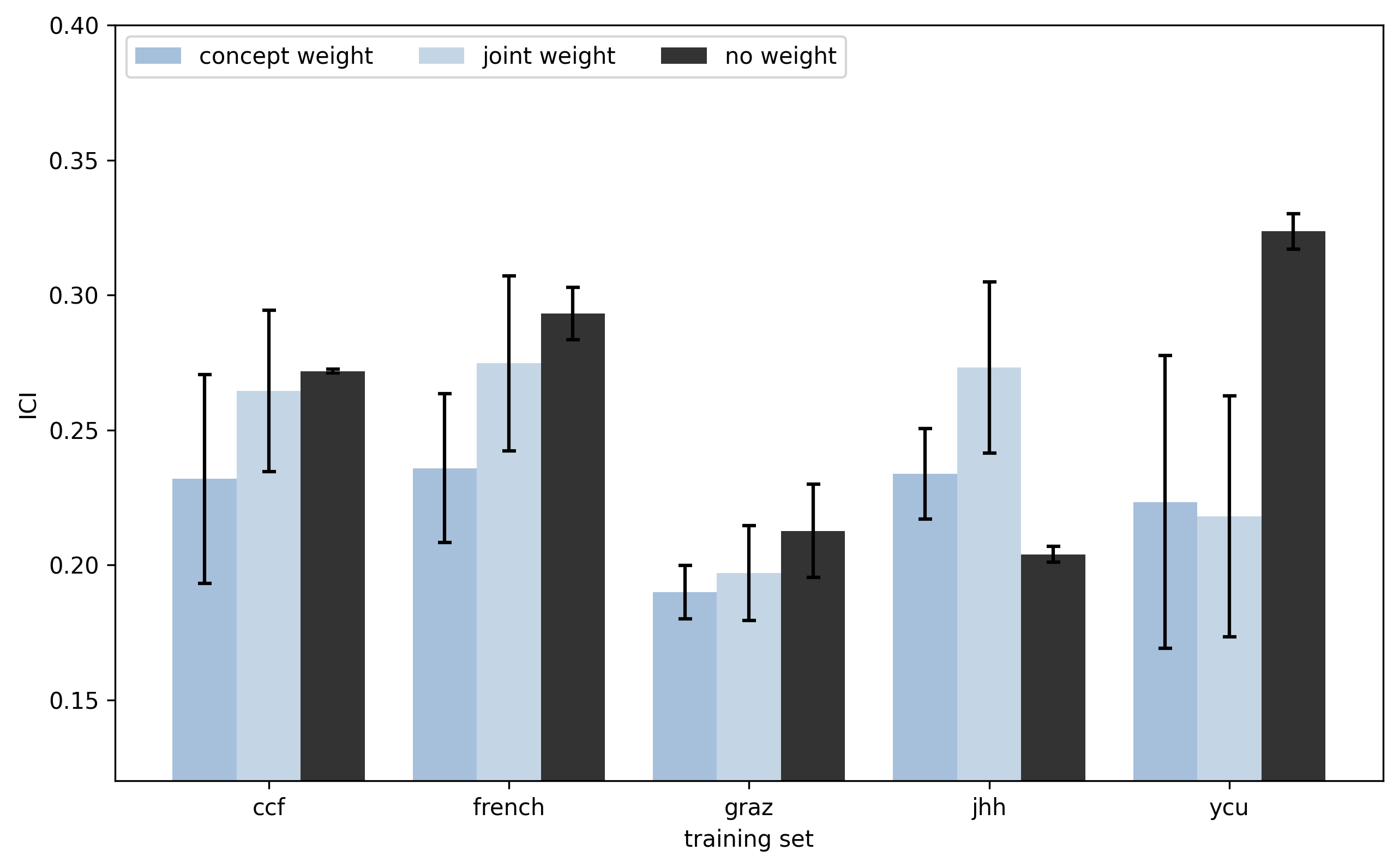}
        \caption*{\small B. Surgery + chemotherapy}
    \end{subfigure}
    \caption{Average test-set ICI by weighting strategy. For each training cohort, the reported ICI represents the average across the corresponding external test cohorts and across the models (OST and RSF). Error bars represent the minimum and maximum model-specific average ICIs. Panel A: surgery-alone cohorts. Panel B: surgery-plus-chemotherapy cohorts.}
    \label{fig:ici_part2}
\end{figure}

Among surgery-alone cohorts, either concept weighting or joint weighting improved average external calibration for 3 of the 5 training cohorts (Figure \ref{fig:ici_part2}, Panel A). 
Among the surgery-plus-chemotherapy cohorts, concept weighting improved average external calibration for 4 of the 5 training cohorts (Figure \ref{fig:ici_part2}, Panel B). The magnitude of improvement varied across settings, ranging from approximately 5\% in the Graz surgery-plus-chemotherapy cohort to more than 50\% in the YCU surgery-plus-chemotherapy cohort.
Concept weight tends to outperform joint weight, likely due to data limitations and covariate noise. Meta-analytic studies vary in their reporting practices — not all provide means and standard deviations for every model feature, and some report continuous means while others report binary counts relative to a threshold. This inconsistency produces noisy estimates of the proxy meta-analysis distribution. Our results suggest that correcting for concept shift alone is generally sufficient to improve model robustness, even when the covariate distribution of the general population is not directly estimable.

We further compare calibration on an aggregated cohort excluding the training cohort between weighted and unweighted models. This yields a Wilcoxon stat of 7.0 with p-value of 0.037. This suggests that weighted models validated on the remaining cohorts aggregated have significantly lower  median ICI than the unweighted models and therefore better calibration. 

Notably, the gains from weighting were largest when the baseline external calibration of the unweighted model was poorest. This pattern is consistent with the underlying rationale of the approach: when a training cohort is distributionally distinct from the external cohorts in which the model is likely to be used, shifting the effective training distribution toward a broader meta-analysis-derived target can yield more substantial improvements in average transportability.

\begin{figure}[ht!]
    \begin{subfigure}[b]{0.45\textwidth}
        \centering
         \includegraphics[width=\textwidth]{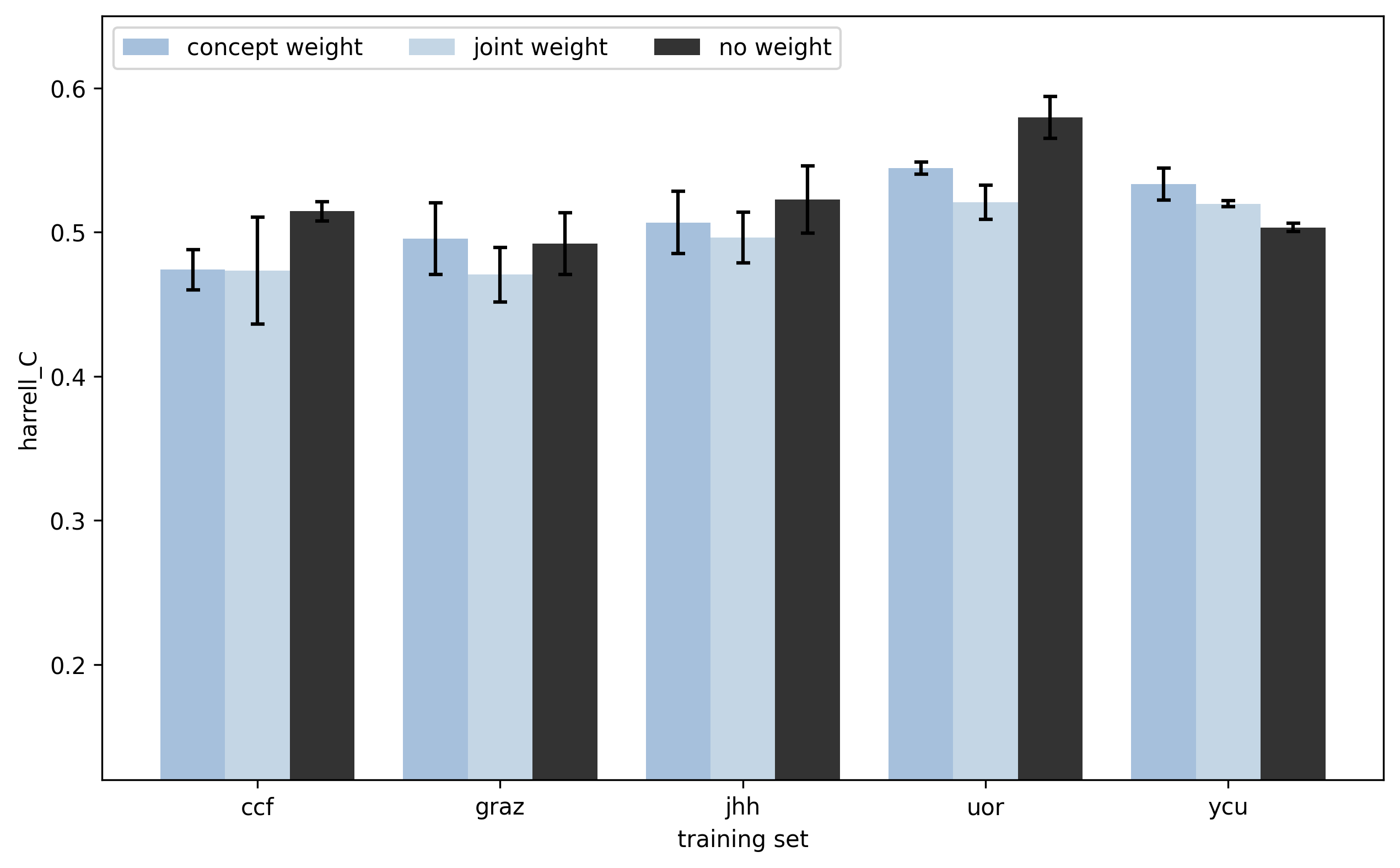}
        \caption*{\small A. Surgery alone}
    \end{subfigure}
    \hspace{5mm}
    \begin{subfigure}[b]{0.45\textwidth}
        \centering
         \includegraphics[width=\textwidth]{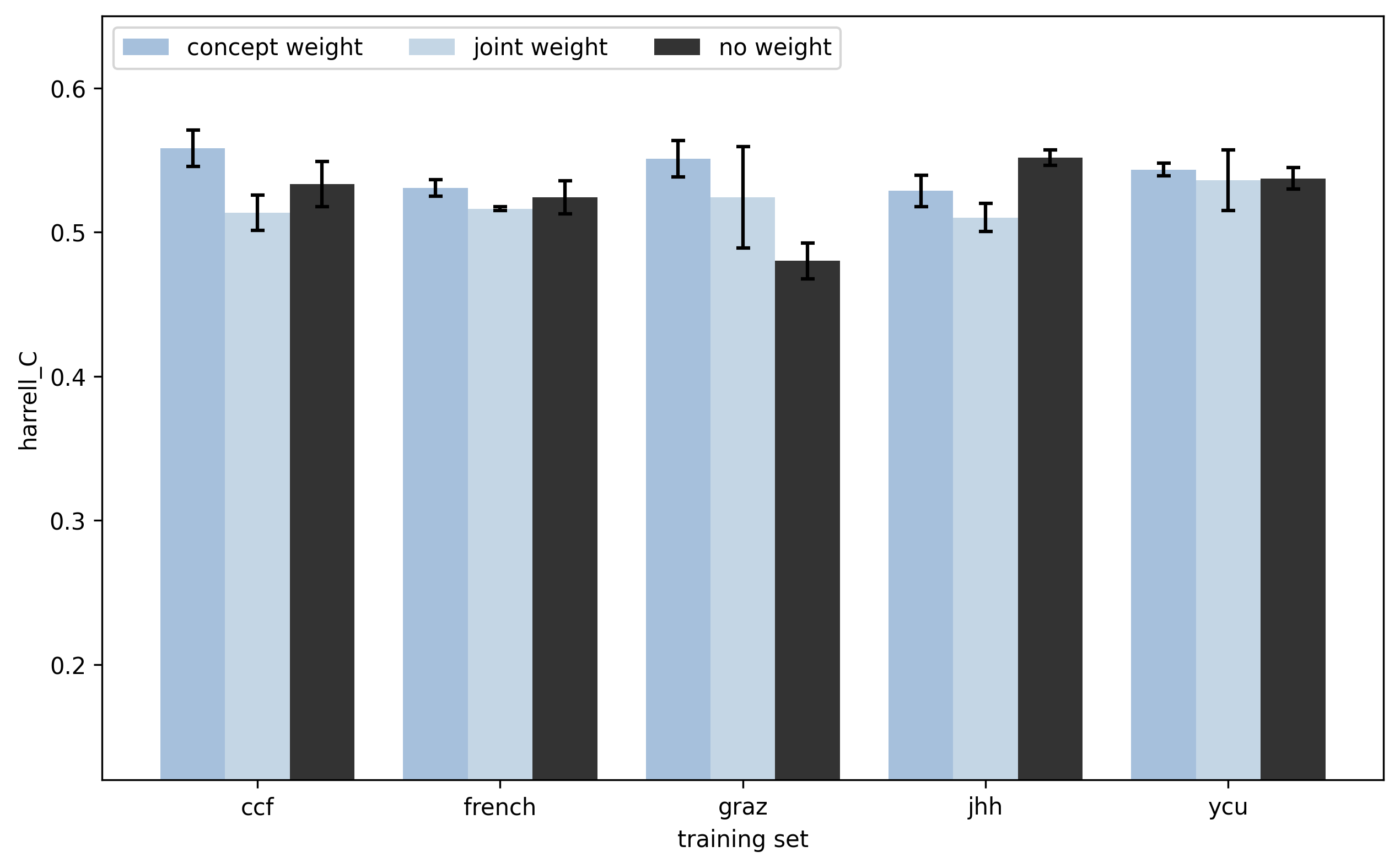}
        \caption*{\small B. Surgery + chemotherapy}
    \end{subfigure}
    \caption{Average test-set Harrell's C-index by weighting strategy. For each training cohort, the reported C-index represents the average across the corresponding external test cohorts and across the models (OST and RSF). Error bars represent the minimum and maximum model-specific average ICIs. Panel A: surgery-alone cohorts. Panel B: surgery-plus-chemotherapy cohorts.}
    \label{fig:c_index_part2}
\end{figure}

By contrast, weighting had little effect on discrimination. As shown in Figure \ref{fig:c_index_part2}, Harrell’s C-index remained largely unchanged after applying either concept weights or joint weights. Taken together, these findings suggest that the proposed weighting framework primarily improves calibration, rather than rank ordering, across heterogeneous external cohorts.

\subsection{Part II. End-user perspective: cohort-specific model selection improves calibration and clinical utility}
We next turned to the second perspective of the framework: the perspective of the end user, for whom the most broadly transportable model may not necessarily be the most appropriate model for a specific target cohort. We therefore examined whether simple cohort-level similarity metrics could guide selection of the model most likely to be well calibrated in a given external cohort.

\begin{figure}[ht!]
    \begin{subfigure}[b]{0.31\textwidth}
        \centering
         \includegraphics[width=\textwidth]{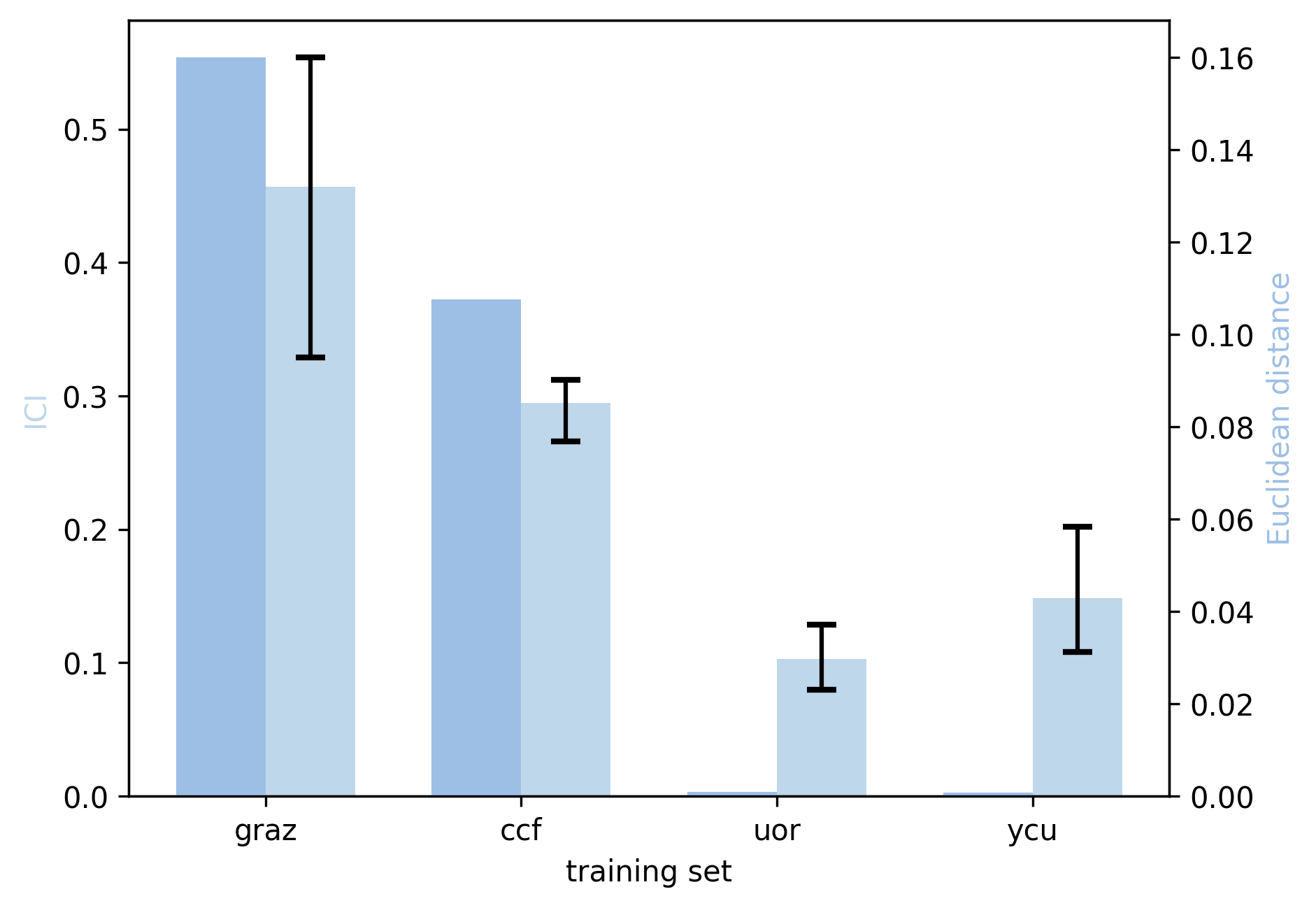}
        \caption*{\footnotesize A. Tested on JHH}
    \end{subfigure}
    \begin{subfigure}[b]{0.31\textwidth}
        \centering
         \includegraphics[width=\textwidth]{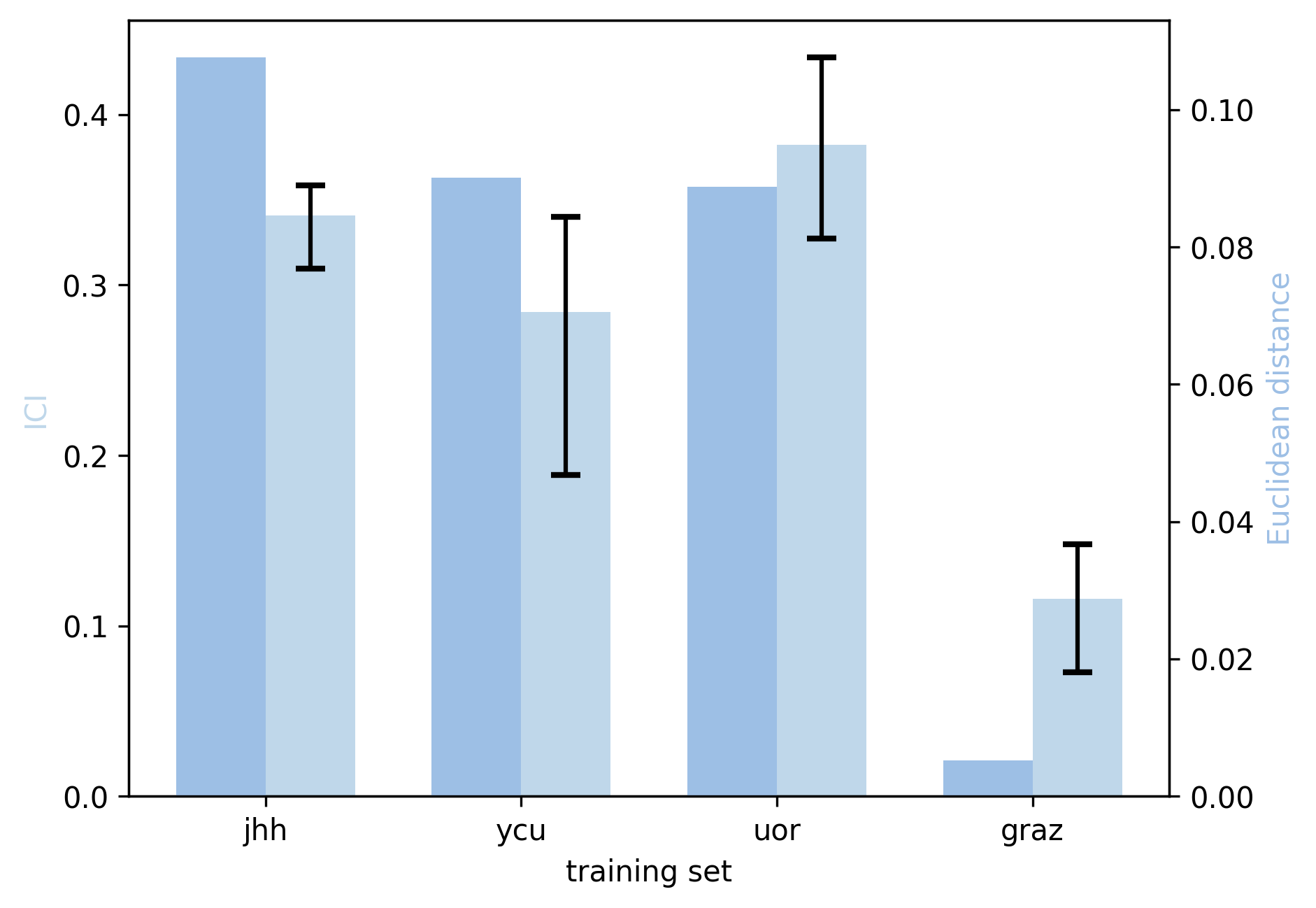}
        \caption*{\footnotesize B. Tested on CCF}
    \end{subfigure}
        \begin{subfigure}[b]{0.31\textwidth}
        \centering
         \includegraphics[width=\textwidth]{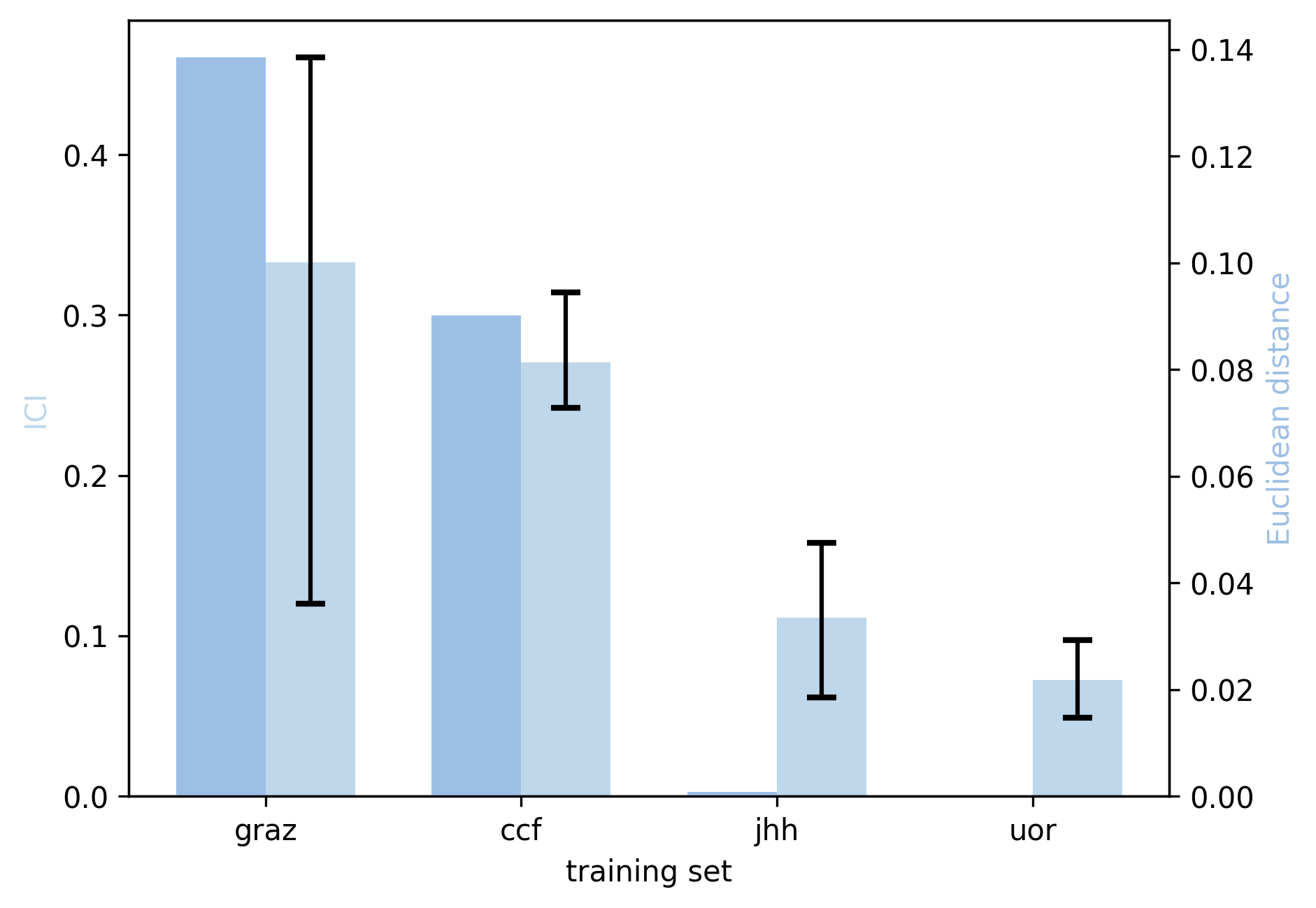}
        \caption*{\footnotesize C. Tested on YCU}
    \end{subfigure} \\
    \begin{subfigure}[b]{0.31\textwidth}
        \centering
         \includegraphics[width=\textwidth]{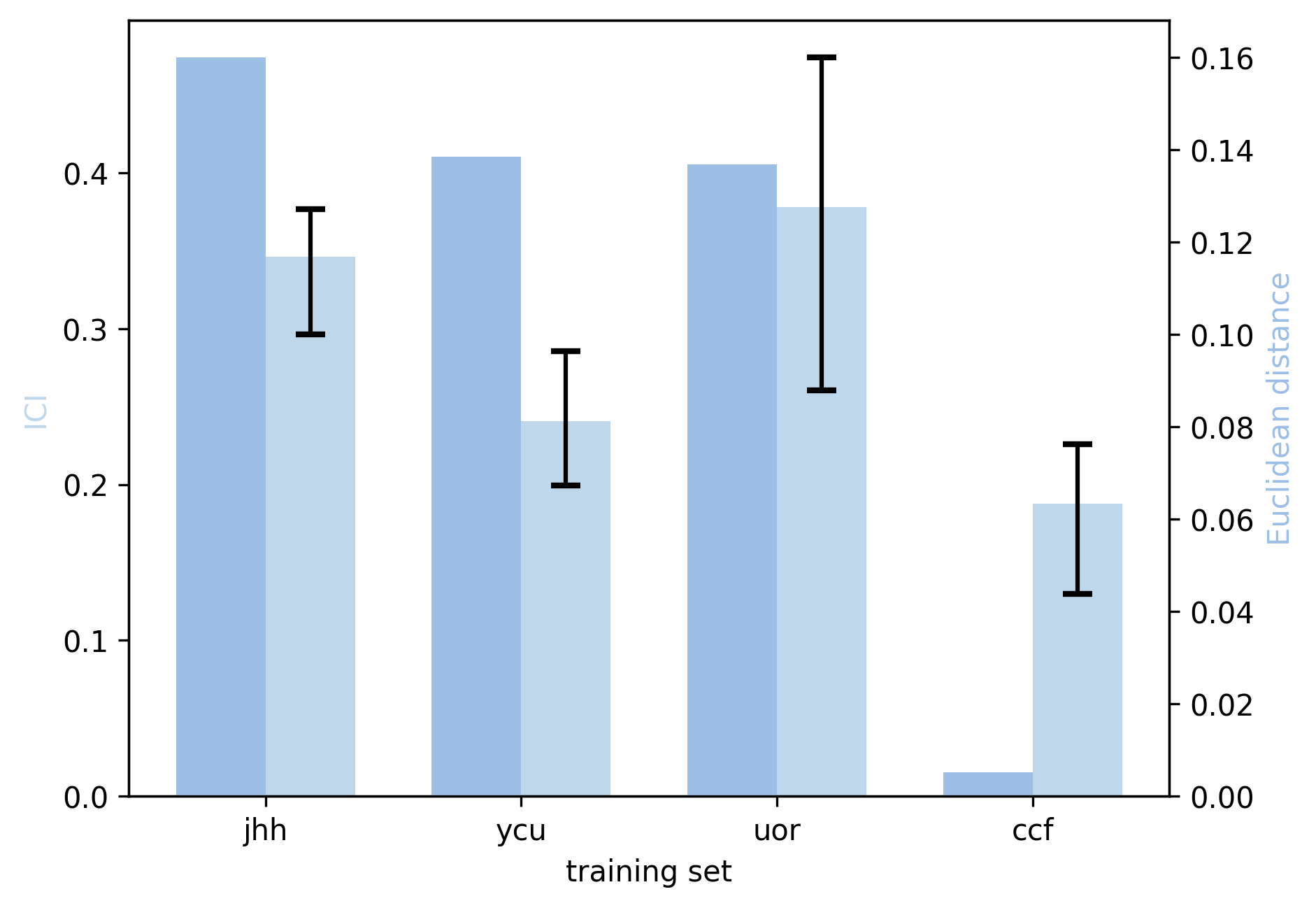}
        \caption*{\footnotesize D. Tested on Graz}
    \end{subfigure}
        \begin{subfigure}[b]{0.31\textwidth}
        \centering
         \includegraphics[width=\textwidth]{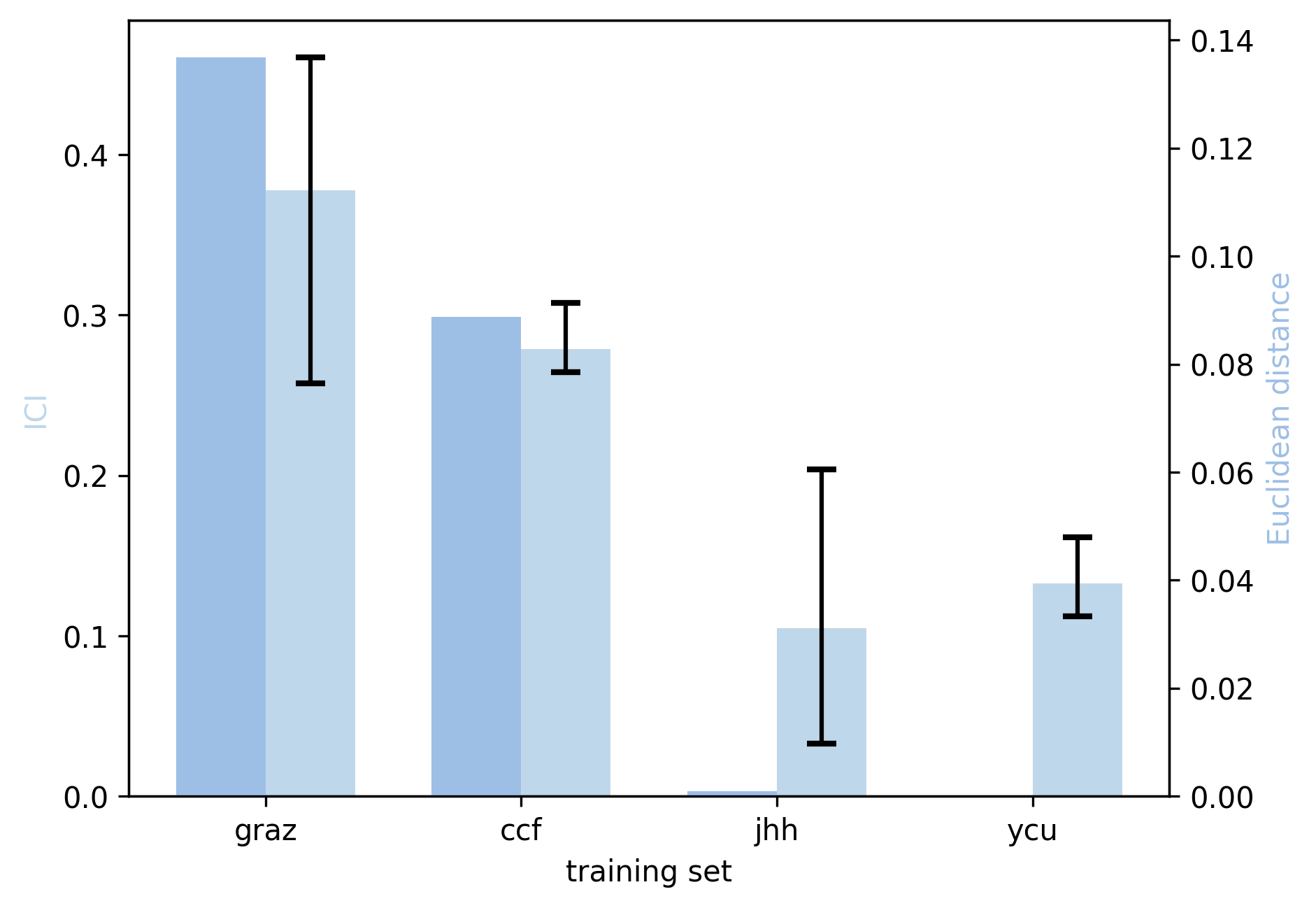}
        \caption*{\footnotesize  E. Tested on UOR}
    \end{subfigure}
    \caption{Pairwise cohort distance between training and test cohorts and the corresponding test-set ICI for models trained in each surgery-alone cohort. ICIs are averaged across three models --- Cox, Optimal Survival Trees (OST), and Random Survival Forest (RSF) --- and the error bars represent the minimum and maximum ICIs across the three models. Cohort distance is computed deterministically as the Euclidean distance between the cohorts' 5-year Kaplan Meier estimates. Each panel represents one training cohort, and the bars within that panel show results for all four test cohorts. Panels A–E correspond to models tested in JHH, CCF, YCU, Graz, and UOR, respectively.}
    \label{fig:untreated_part3}
\end{figure}

\begin{figure}[ht!]
    \begin{subfigure}[b]{0.31\textwidth}
        \centering
         \includegraphics[width=\textwidth]{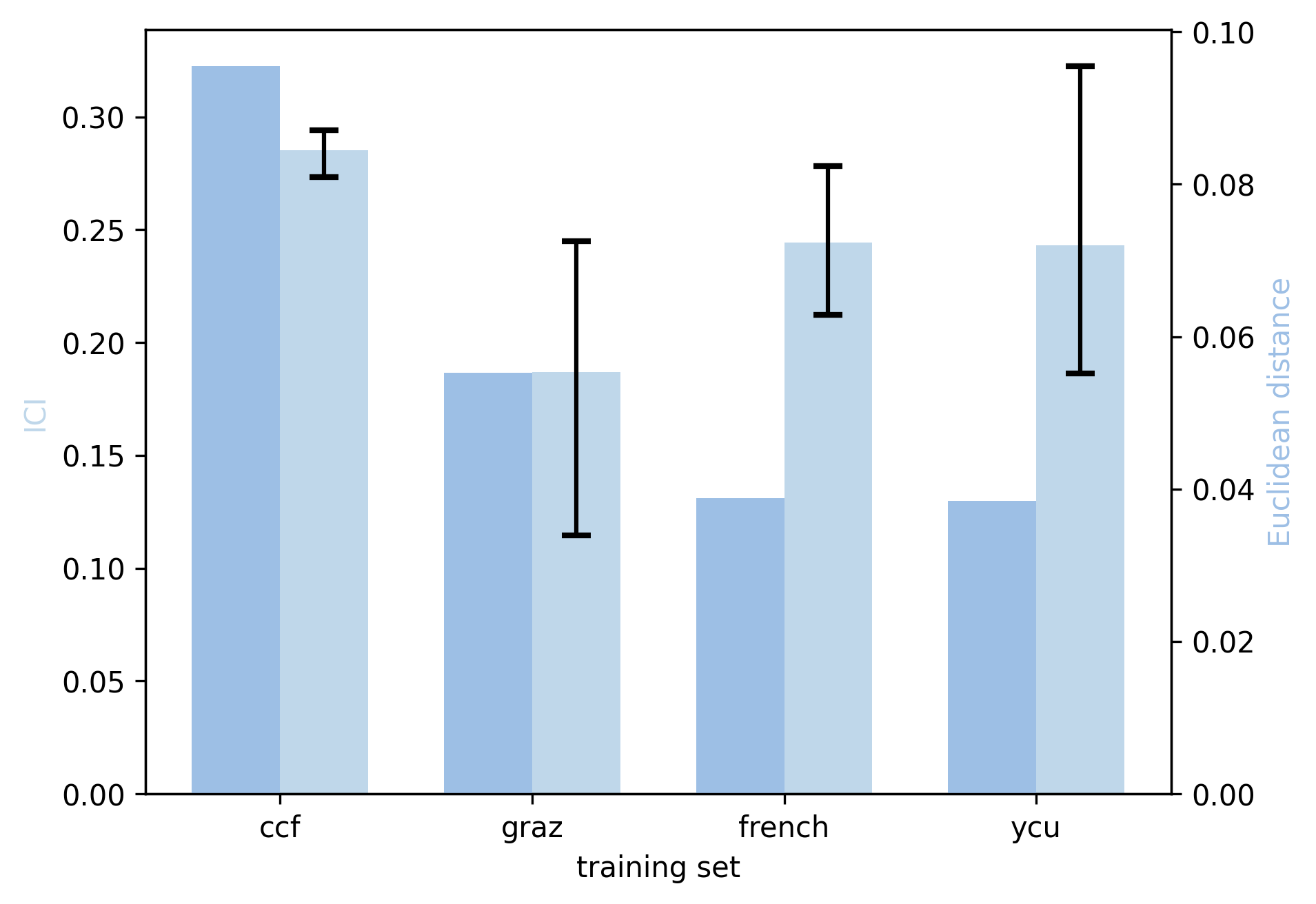}
        \caption*{\footnotesize A. Tested on JHH}
    \end{subfigure}
    \begin{subfigure}[b]{0.31\textwidth}
        \centering
         \includegraphics[width=\textwidth]{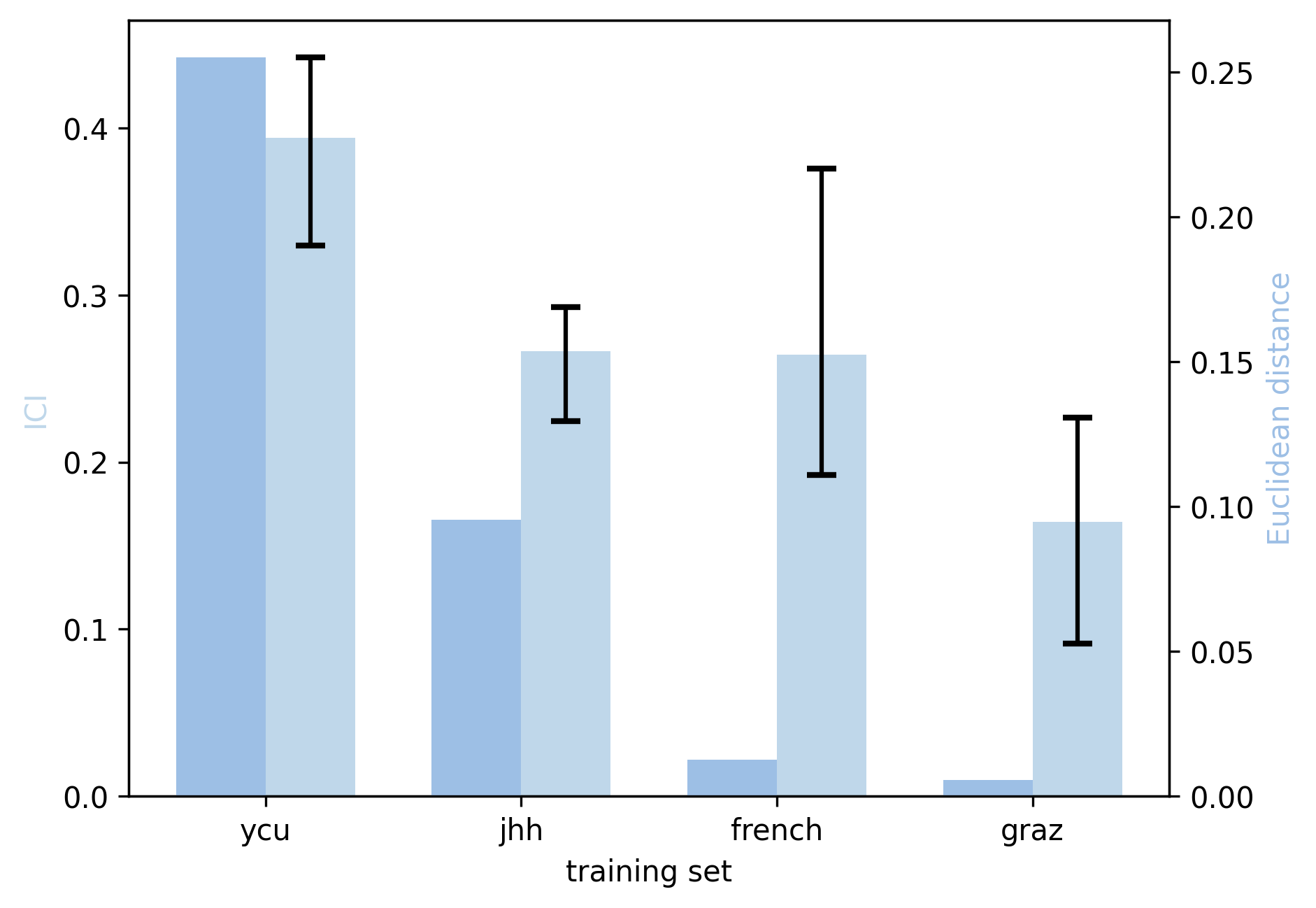}
        \caption*{\footnotesize B. Tested on CCF}
    \end{subfigure}
        \begin{subfigure}[b]{0.31\textwidth}
        \centering
         \includegraphics[width=\textwidth]{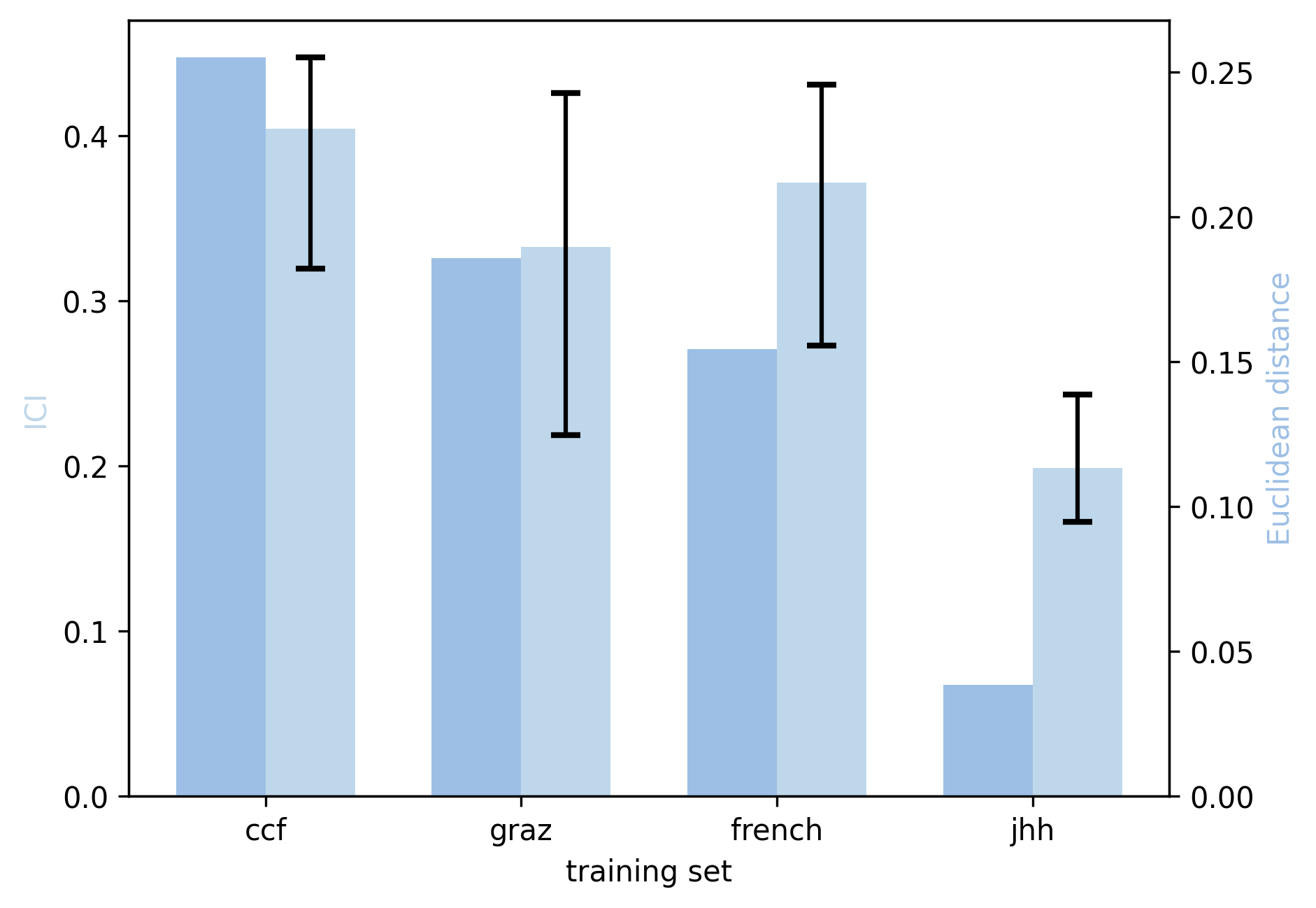}
        \caption*{\footnotesize C. Tested on YCU}
    \end{subfigure} \\
    \begin{subfigure}[b]{0.31\textwidth}
        \centering
         \includegraphics[width=\textwidth]{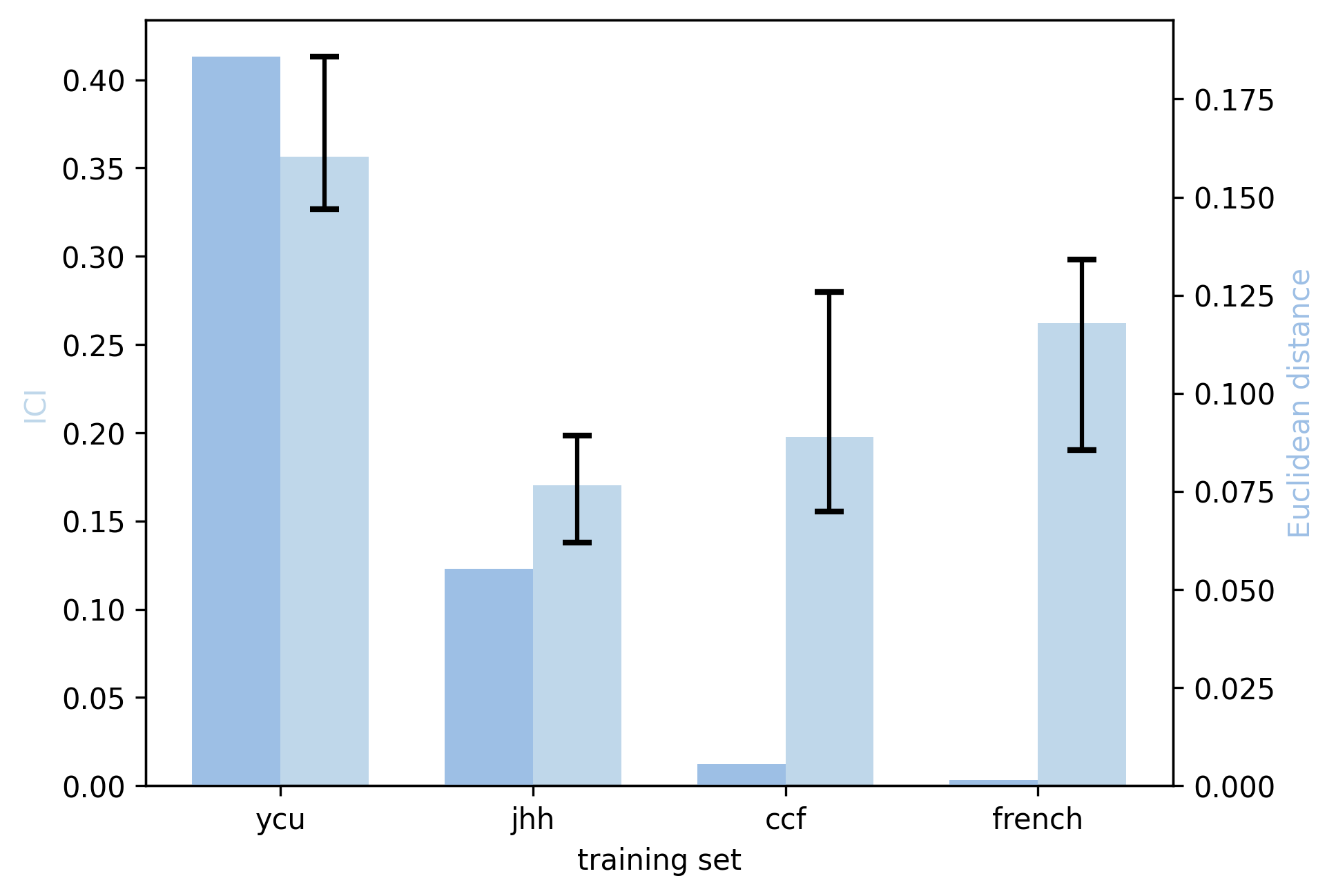}
        \caption*{\footnotesize D. Tested on Graz}
    \end{subfigure}
        \begin{subfigure}[b]{0.31\textwidth}
        \centering
         \includegraphics[width=\textwidth]{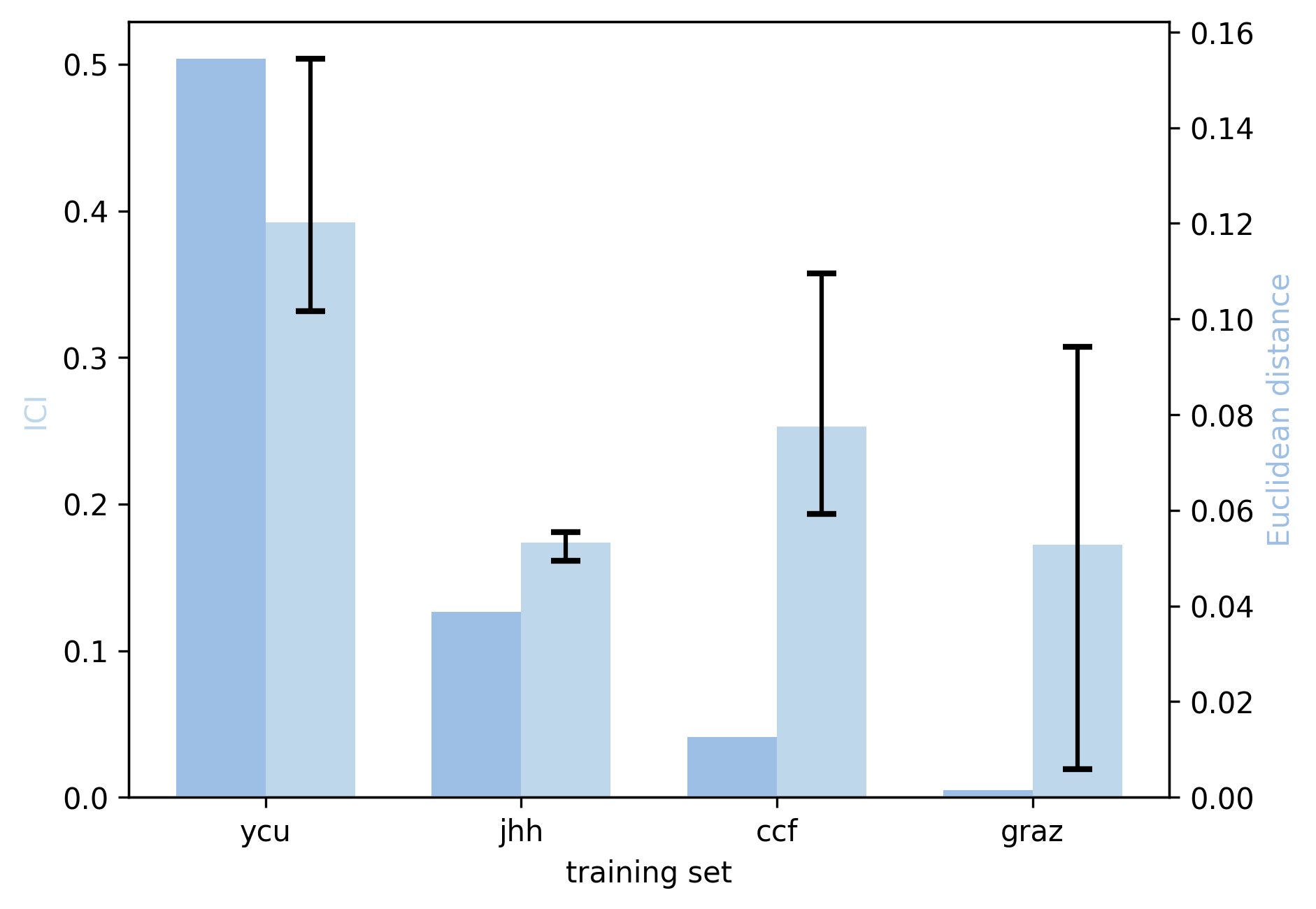}
         \caption*{\footnotesize  E. Tested on CHU}
    \end{subfigure}    
    \caption{Pairwise KL divergence between training and test cohorts and the corresponding test-set ICI for models trained in each surgery + adjuvant chemotherapy cohort. ICIs are averaged across three models --- Cox, Optimal Survival Trees (OST), and Random Survival Forest (RSF) --- and the error bars represent the minimum and maximum ICIs across the three models. Cohort distance is computed deterministically as the Euclidean distance between the cohorts' 5-year Kaplan Meier estimates. 
    Each panel represents one training cohort, and the bars within that panel show results for all four test cohorts. Panels A–E correspond to models tested in JHH, CCF, YCU, Graz, and CHU, respectively.}
    \label{fig:treated_part3}
\end{figure}

For each target cohort, we computed the distance $dist_{train, test}$ between that cohort and each candidate training cohort, using the cohort-level mean 5-year recurrence probability obtained from Kaplan-Meier estimates described in the Methods. We then compared this distance with the observed calibration of each externally applied model.
Among surgery-alone cohorts, in all of the target cohorts, the model trained on the most similar cohort also yielded either the lowest or second lowest ICI in the target cohort (Figure \ref{fig:untreated_part3}). In the surgery-alone UOR and JHH cohorts where the most similar cohort yields the second lowest ICI, the difference in both ICI and cohort similarity between the lowest and second lowest cohorts is marginal.
Similar patterns were observed in the surgery-plus-chemotherapy cohorts (Figure \ref{fig:treated_part3}). In addition, a formal examination of the correlation yields a correlation index of 0.803 between cohort similarity and ICI among the surgery-alone cohorts and 0.737 among the surgery-plus-chemotherapy cohorts, both with p-values $< 0.001$. This correlation is shown in Appendix Figure \ref{fig:scatter_part3}. These findings suggest that simple cohort-level outcome summaries may provide a useful and pragmatic guide for selecting among existing prognostic models when direct recalibration or local model development is not feasible.

\begin{figure}[ht!]
    \begin{subfigure}[b]{0.45\textwidth}
        \centering
        \vspace{-5mm}
         \includegraphics[width=\textwidth]{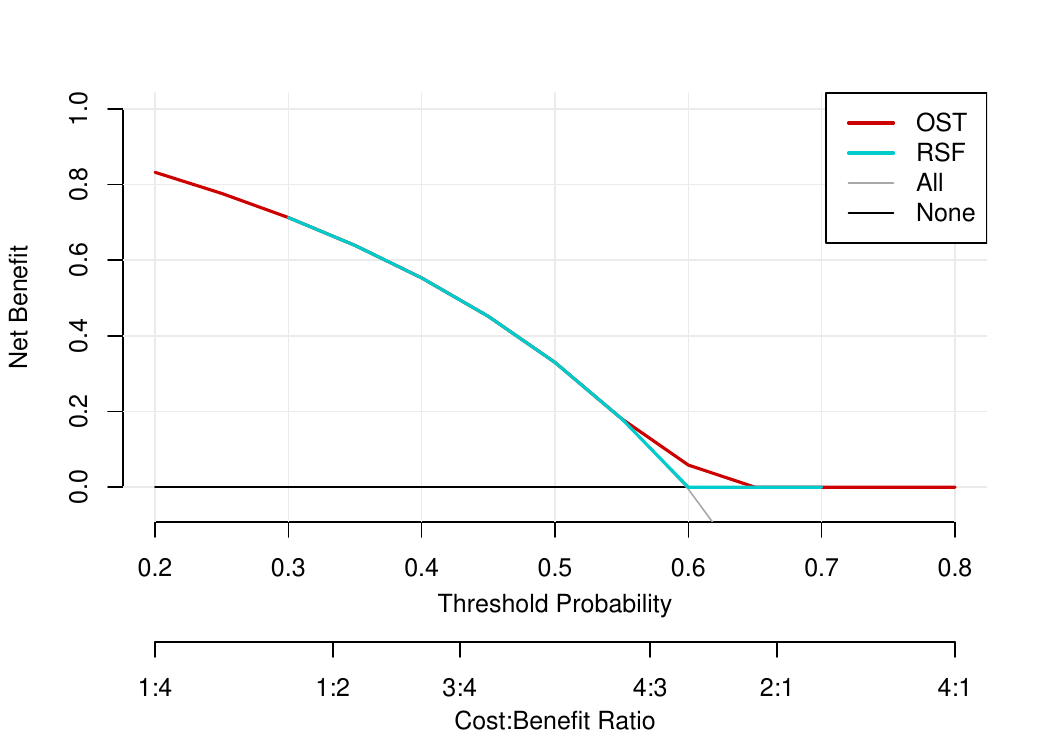}
        \caption*{\small A. Trained on CCF}
    \end{subfigure}
    \begin{subfigure}[b]{0.45\textwidth}
        \centering
          \vspace{-5mm}
         \includegraphics[width=\textwidth]{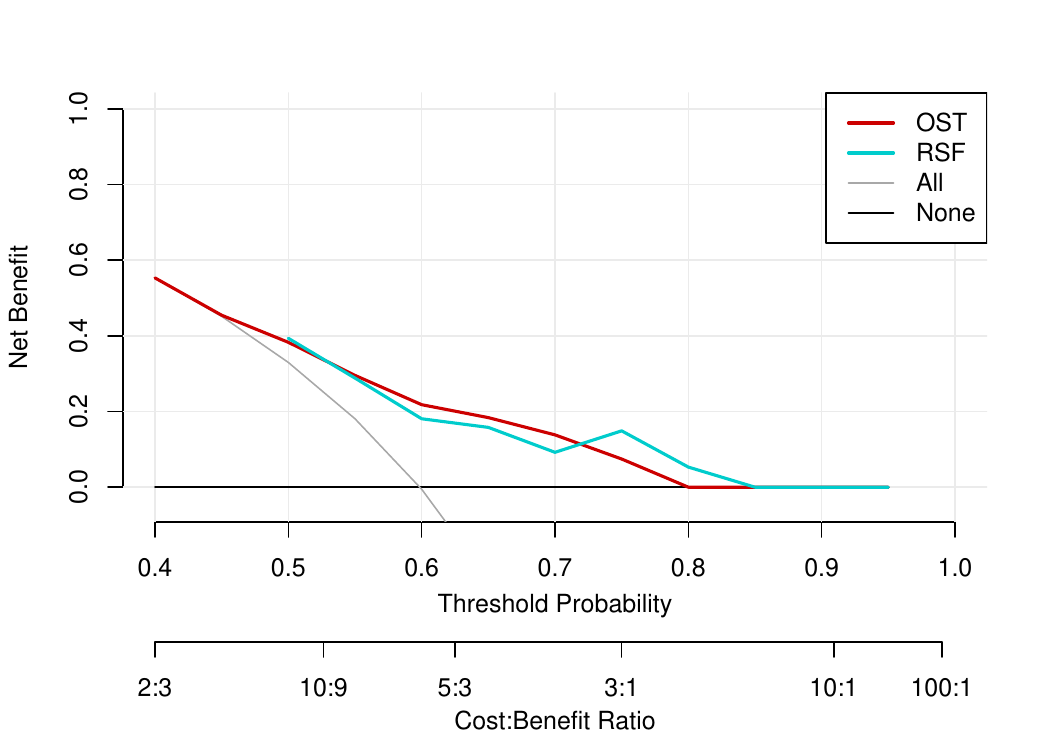}
        \caption*{\small B. Trained on YCU}
    \end{subfigure} ]\\
    \begin{subfigure}[b]{0.45\textwidth}
        \centering
          \vspace{-5mm}
         \includegraphics[width=\textwidth]{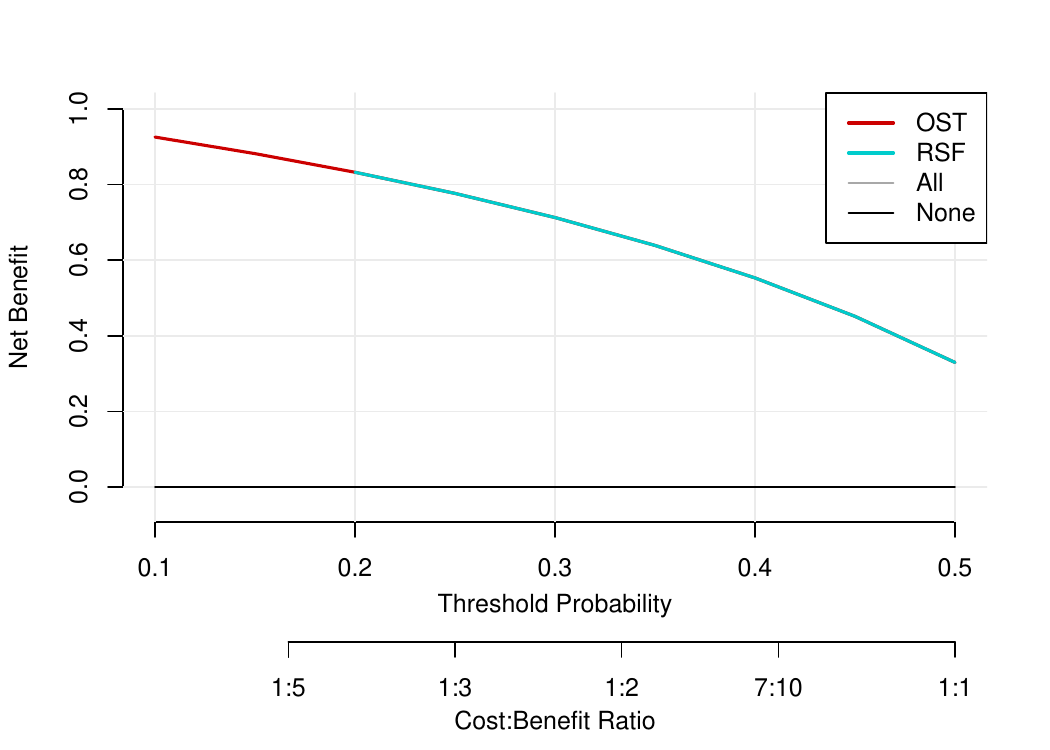}
        \caption*{\small  C. Trained on Graz}
    \end{subfigure}
    \begin{subfigure}[b]{0.45\textwidth}
        \centering
            \vspace{-5mm}
         \includegraphics[width=\textwidth]{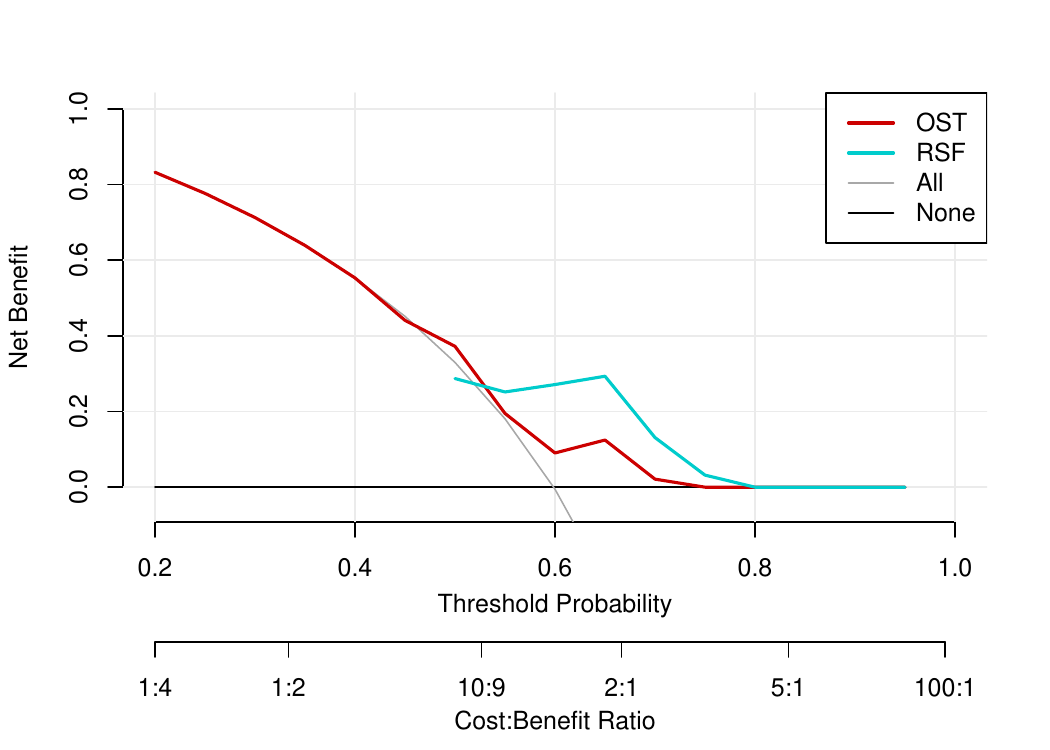}
         \caption*{\small D. Trained on UOR}
    \end{subfigure}
    \caption{Decision curve analysis showing the clinical utility of models trained in four different cohorts and externally validated in the JHH surgery-alone cohort.}
    \label{fig:DCA_untreated_jhh}
\end{figure}

\begin{figure}[ht!]

    \begin{subfigure}[b]{0.45\textwidth}
        \centering
        \vspace{-5mm}
         \includegraphics[width=\textwidth]{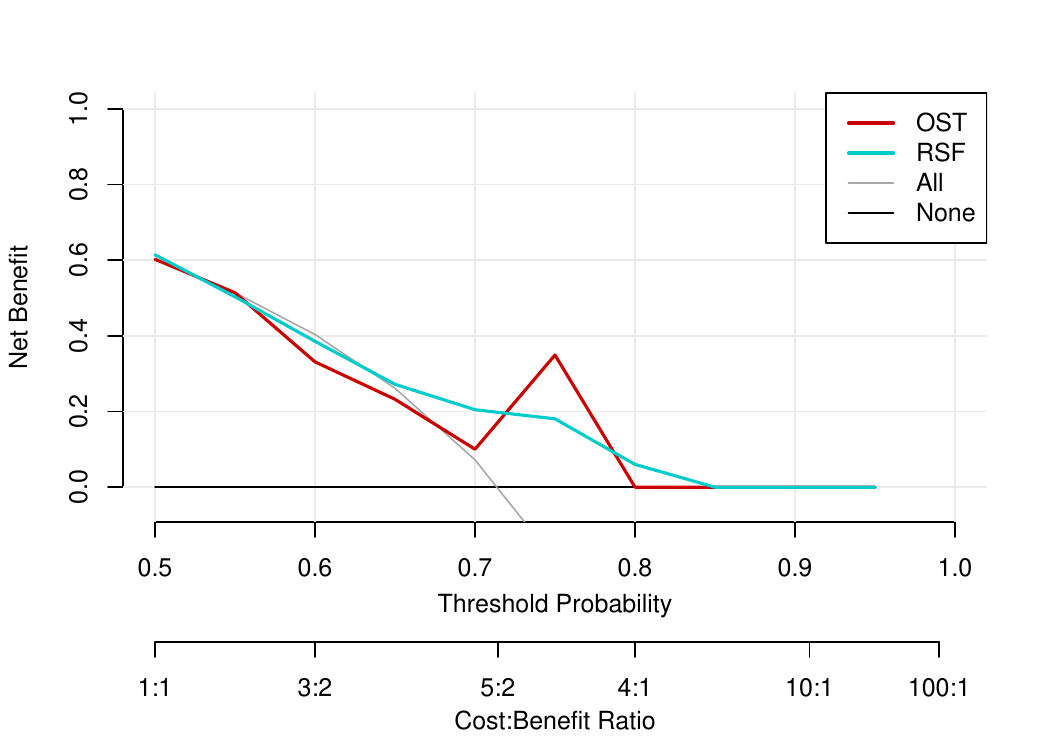}
        \caption*{\small A. Trained on JHH}
    \end{subfigure}
    \begin{subfigure}[b]{0.45\textwidth}
        \centering
        \vspace{-5mm}
         \includegraphics[width=\textwidth]{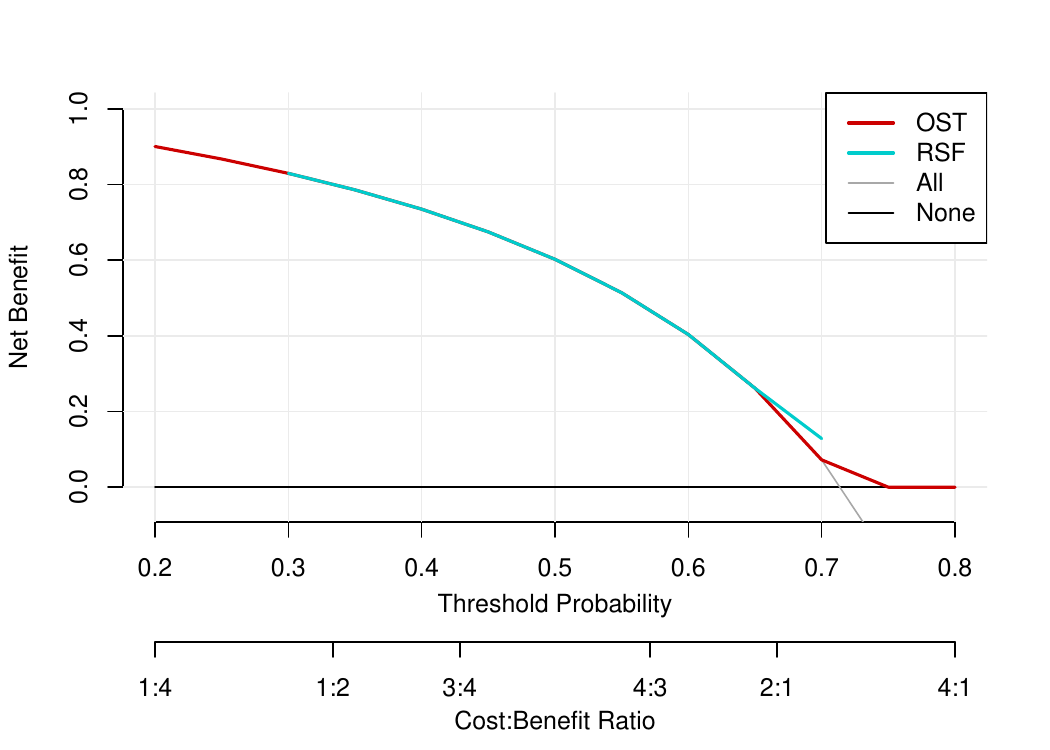}
        \caption*{\small  B. Trained on CCF}
    \end{subfigure} \\
    \begin{subfigure}[b]{0.45\textwidth}
        \centering
          \vspace{-5mm}
         \includegraphics[width=\textwidth]{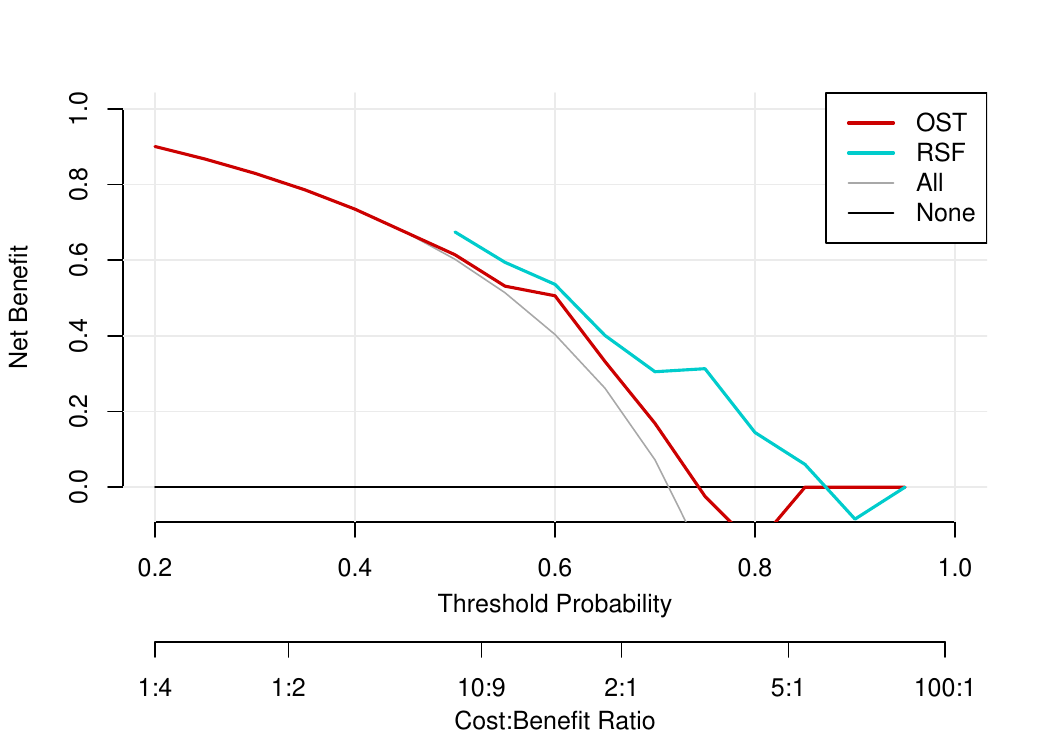}
        \caption*{\small C. Trained on UOR}
    \end{subfigure}
    \begin{subfigure}[b]{0.45\textwidth}
        \centering
          \vspace{-5mm}
         \includegraphics[width=\textwidth]{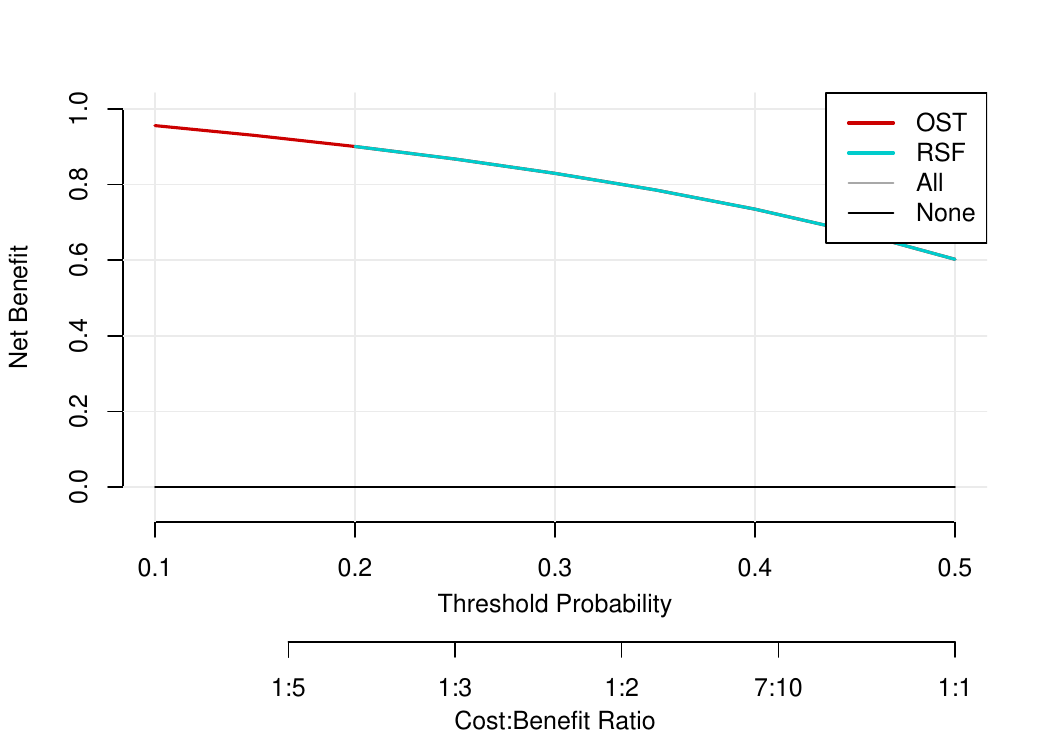}
        \caption*{\small  D. Trained on Graz}
    \end{subfigure}
    \caption{Decision curve analysis showing the clinical utility of models trained in four different cohorts and externally validated in the YCU surgery-alone cohort.}
    \label{fig:DCA_untreated_ycu}
\end{figure}

We then examined whether this improved cohort matching also translated into greater clinical utility. Decision curve analysis showed that models trained on cohorts with outcome profiles similar to the target cohort generally produced higher net benefit across clinically relevant threshold ranges than models trained on more dissimilar cohorts. For example, when patients from JHH who underwent surgery alone served as the target cohort, models trained on YCU and UOR—two cohorts with similarly high recurrence risk—outperformed both the “treat all” and “treat none” strategies over a meaningful range of thresholds, whereas models trained on the lower-risk CCF and Graz cohorts offered little or no clinical value beyond trivial strategies (Figure \ref{fig:DCA_untreated_jhh}). A similar pattern was observed when patients from YCU who underwent surgery alone were used as the target cohort (Figure \ref{fig:DCA_untreated_ycu}).

\begin{figure}[ht!]
    \begin{subfigure}[b]{0.45\textwidth}
        \centering
            \vspace{-5mm}
         \includegraphics[width=\textwidth]{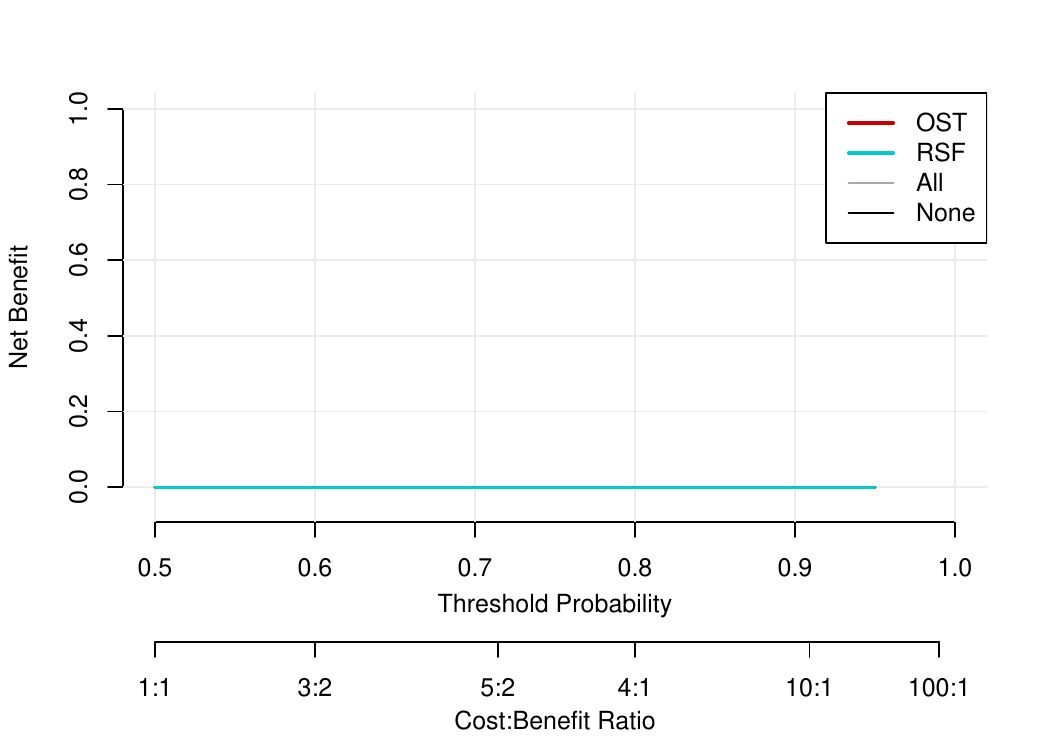}
        \caption*{\small A. Trained on JHH}
    \end{subfigure}
        \begin{subfigure}[b]{0.45\textwidth}
        \centering
            \vspace{-5mm}
         \includegraphics[width=\textwidth]{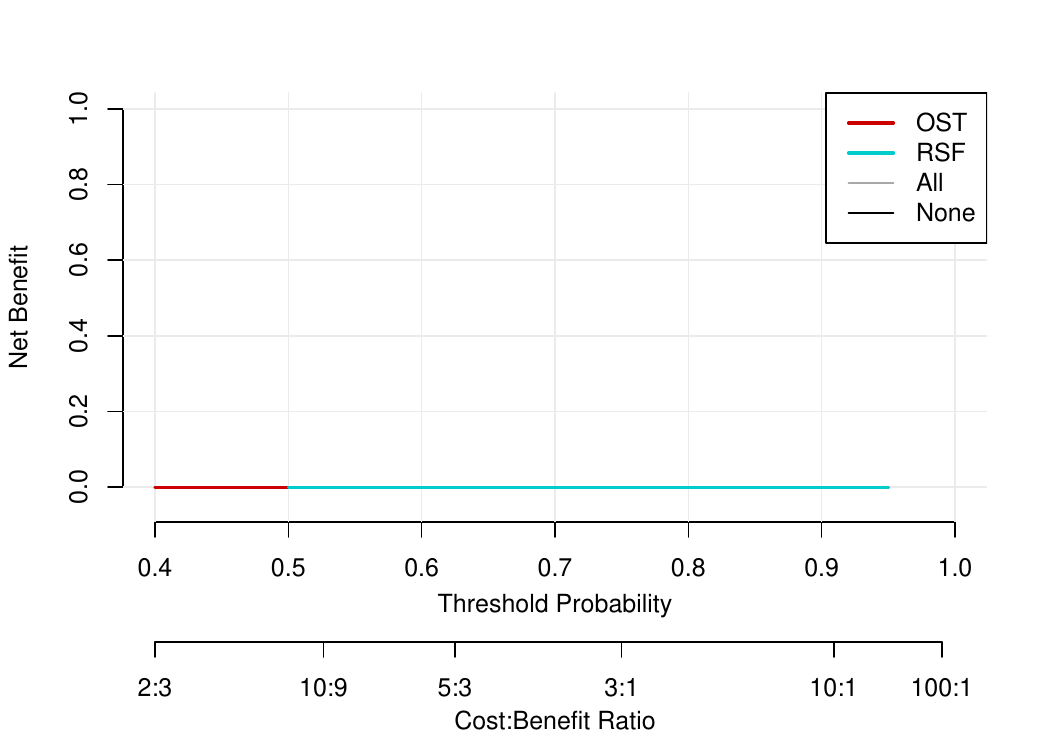}
        \caption*{\small B. Trained on YCU}
    \end{subfigure} \\
    \begin{subfigure}[b]{0.45\textwidth}
        \centering
         \vspace{-5mm}
         \includegraphics[width=\textwidth]{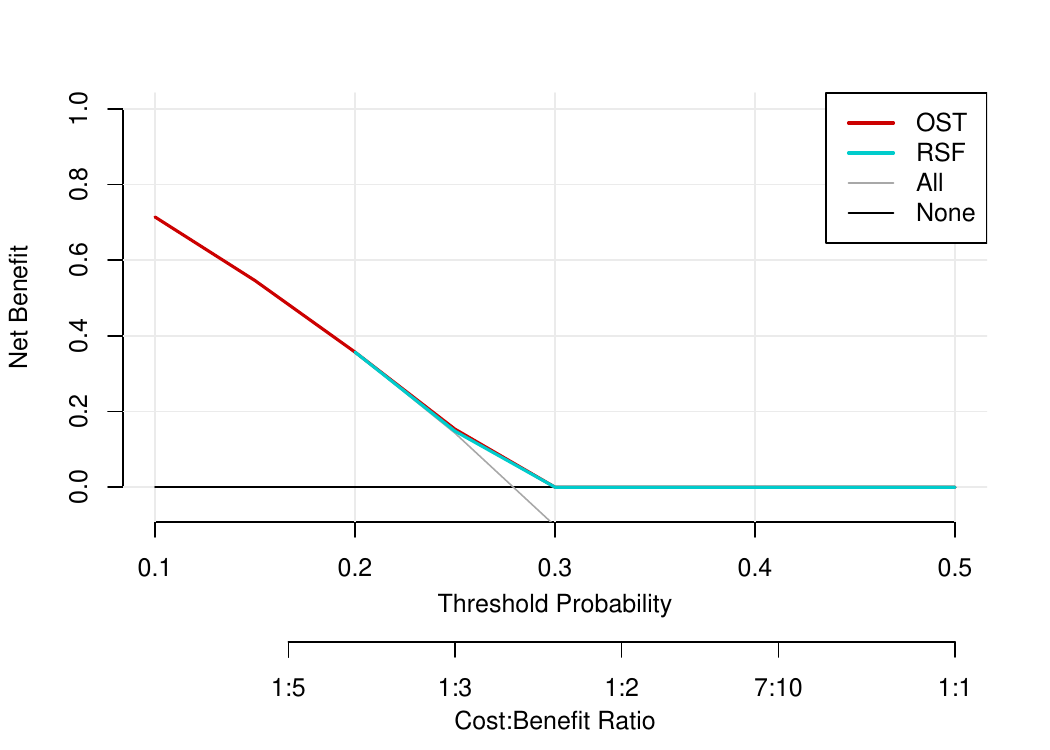}
         \caption*{\small C. Trained on Graz}
    \end{subfigure}
        \begin{subfigure}[b]{0.45\textwidth}
        \centering
         \vspace{-5mm}
         \includegraphics[width=\textwidth]{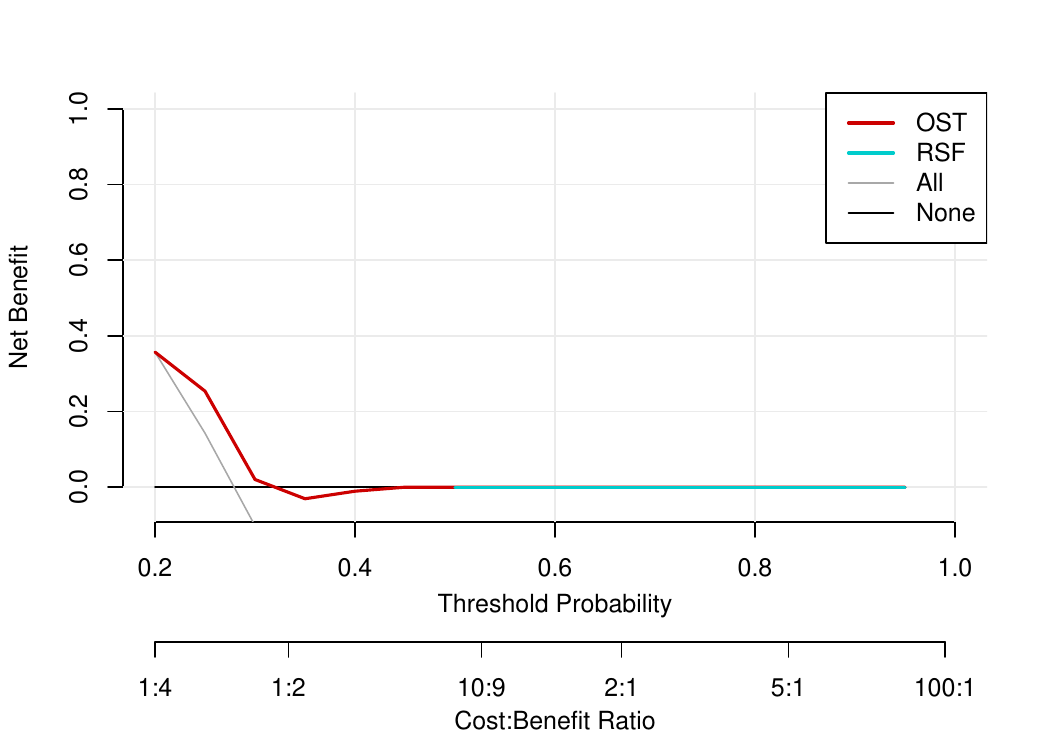}
         \caption*{\small  D. Trained on UOR}
    \end{subfigure}
    \caption{Decision curve analysis showing the clinical utility of models trained in four different cohorts and externally validated in the CCF surgery-alone cohort.}
    \label{fig:DCA_untreated_ccf}
\end{figure}

We next examined whether the same pattern held for the two lower-risk surgery-alone cohorts, CCF and Graz. As shown in Figure \ref{fig:DCA_untreated_ccf}, when CCF was used as the target cohort, the model trained on Graz patients who underwent surgery alone outperformed the other externally trained models, consistent with the fact that Graz was the most similar cohort to CCF in terms of risk of recurrence. However, despite this relative advantage, the corresponding DCA curve largely overlapped with the “treat all” strategy, indicating limited added clinical value.

\begin{figure}[ht!]
    \begin{subfigure}[b]{0.45\textwidth}
        \centering
         \includegraphics[width=\textwidth]{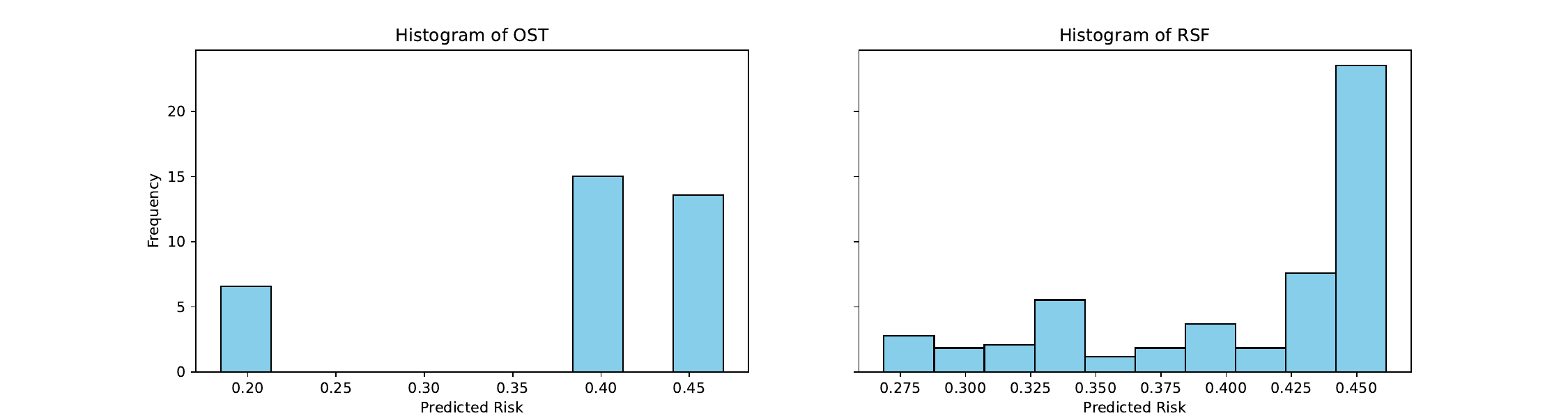}
        \caption*{\small A. Trained on Graz and Validated on CCF}
    \end{subfigure}
    \begin{subfigure}[b]{0.45\textwidth}
        \centering
         \includegraphics[width=\textwidth]{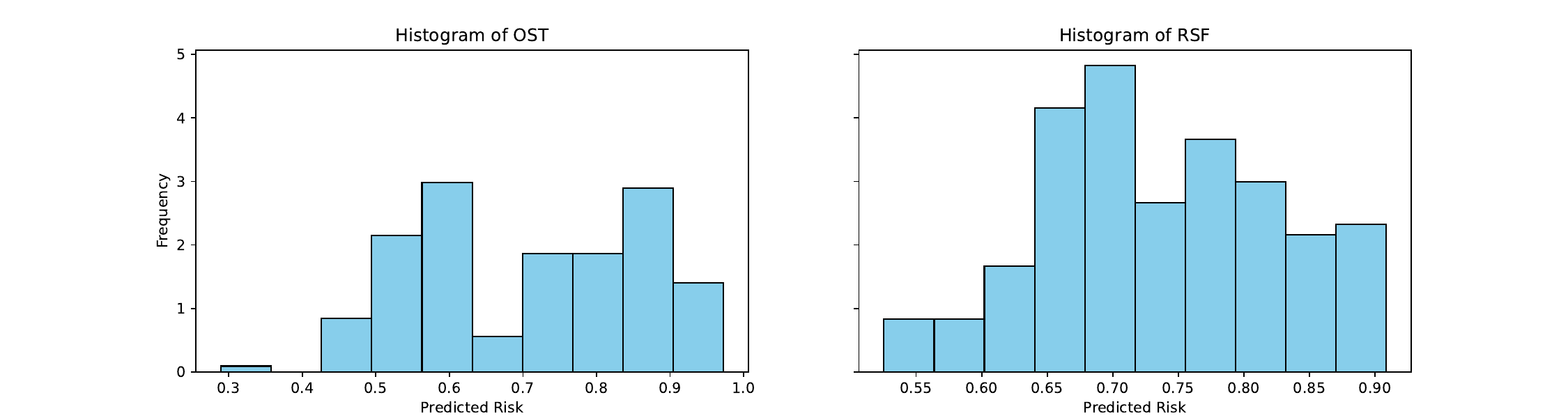}
        \caption*{\small B. Trained on UOR and Validated on JHH}
    \end{subfigure} \\
    \begin{subfigure}[b]{0.45\textwidth}
        \centering
         \includegraphics[width=\textwidth]{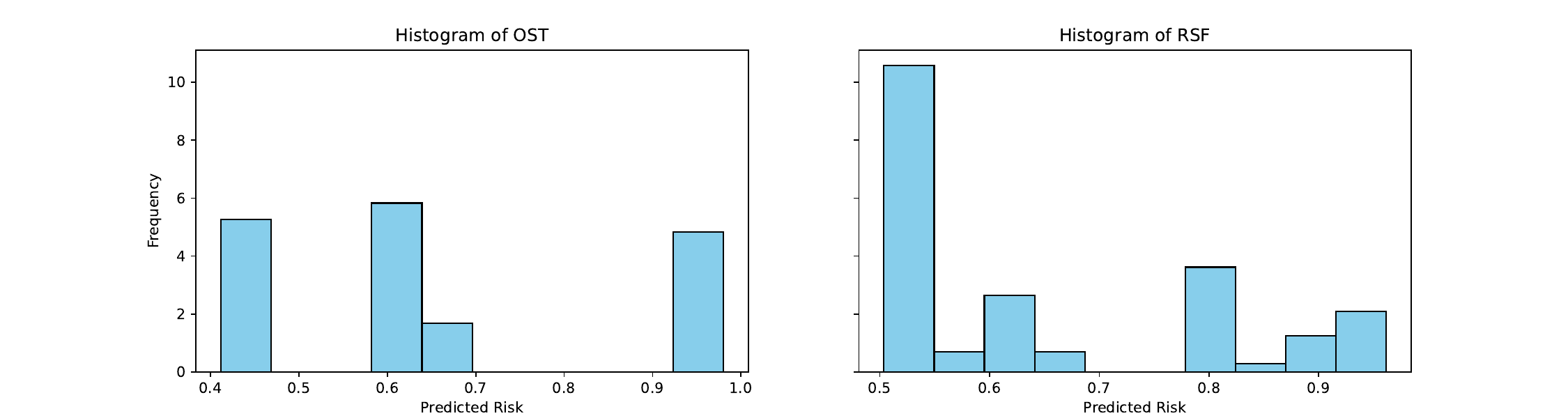}
        \caption*{\small C. Trained on YCU and Validated on JHH}
    \end{subfigure}    
    \caption{Histogram of Predicted Risks for Surgery-alone Cohorts }
    \label{fig:histogram_risk_untreated}
\end{figure}

To better understand this finding, we examined the distribution of predicted risks (Figure \ref{fig:histogram_risk_untreated}). In contrast to the better-performing models in the JHH target cohort—specifically, the models trained on YCU and UOR, which generated a broad range of predicted risks from approximately 0.3 to 0.9—the model trained on Graz produced a much narrower range of predictions, approximately 0.2 to 0.45, when applied to the CCF cohort. This suggests that the model assigned relatively low recurrence risk to most patients in the cohort. Because CCF itself is a predominantly low-risk cohort, such predictions can still yield favorable calibration and low ICI, as shown in Panel B of Figure \ref{fig:untreated_part3}. However, a model that predicts a narrow low-risk range for nearly all patients, even if reasonably calibrated on average, offers limited ability to distinguish patients with meaningfully different prognoses and therefore limited clinical utility.

\begin{figure}[ht!]
    \begin{subfigure}[b]{0.48\textwidth}
        \centering
        \vspace{-5mm}
        \includegraphics[width=\textwidth]{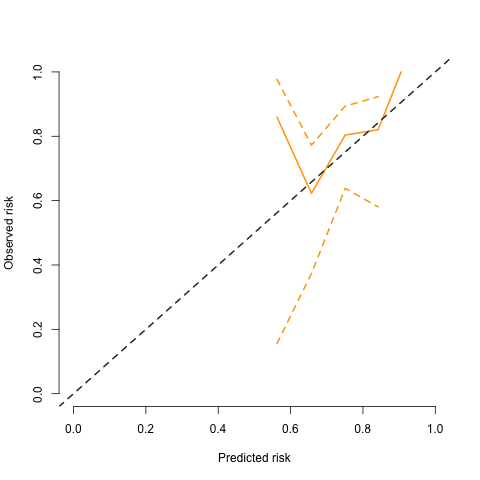}
        \caption*{\footnotesize A. Trained on UOR and Validated on JHH}
    \end{subfigure}
    \begin{subfigure}[b]{0.48\textwidth}
        \centering
        \vspace{-5mm}
        \includegraphics[width=\textwidth]{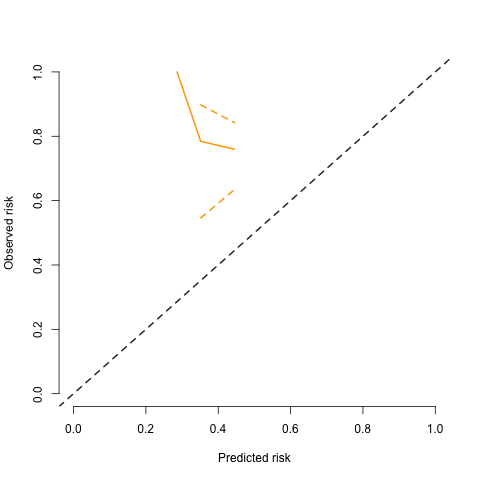}
        \caption*{\footnotesize B. Trained on Graz and Validated on JHH}
    \end{subfigure} \\
    \begin{subfigure}[b]{0.48\textwidth}
        \centering
        \includegraphics[width=\textwidth]{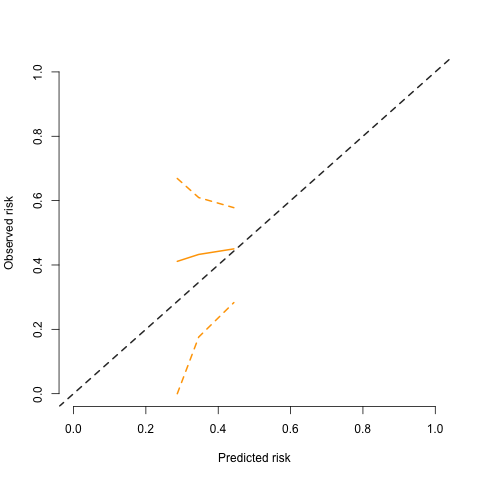}
        \caption*{\footnotesize C. Trained on Graz and Validated on CCF}
    \end{subfigure}
    \begin{subfigure}[b]{0.48\textwidth}
        \centering
        \includegraphics[width=\textwidth]{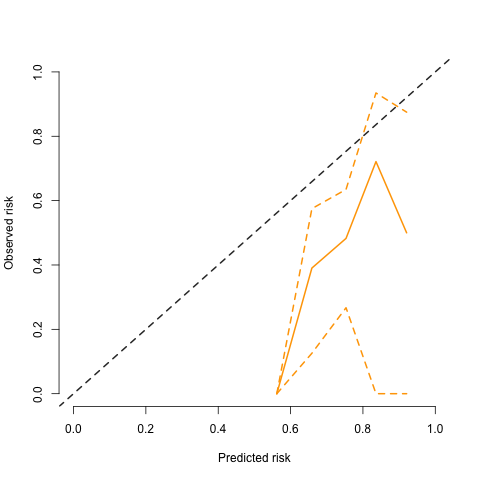}
        \caption*{\footnotesize D. Trained on UOR and Validated on CCF}
    \end{subfigure}
    \caption{Calibration plots for recurrence-free survival, showing external validation in the surgery-alone JHH and CCF cohorts of models developed in the surgery-alone UOR and Graz cohorts. Top left: UOR model validated in JHH. Top right: Graz model validated in JHH. Bottom left: Graz model validated in CCF. Bottom right: UOR model validated in CCF.}
    \label{fig:calibration_untreated}
\end{figure}

This interpretation is further supported by the calibration plots in Figure \ref{fig:calibration_untreated}, where we compare random survival forest models trained on the surgery-alone UOR and Graz cohorts. When validated on JHH patients who underwent surgery alone, the model trained on UOR patients—whose risk profile more closely resembles that of JHH—produced a calibration curve that closely followed the 45-degree line, indicating good agreement between predicted and observed risks. By contrast, the model trained on Graz patients underestimated risk in JHH, consistent with the lower-risk composition of the Graz cohort. Conversely, when validated on CCF patients who underwent surgery alone, the Graz-trained model produced a calibration curve close to the 45-degree line, whereas the UOR-trained model systematically overestimated risk because UOR consists of substantially higher-risk patients. Importantly, however, even in the setting where the Graz model was well calibrated in CCF, its predicted risks remained concentrated within a relatively narrow range. Thus, similarity between development and target cohorts may improve calibration, but clinical usefulness also depends on whether the model produces a sufficiently informative spread of predicted risks to support treatment decisions.

Similar patterns were observed in the surgery-plus-chemotherapy cohorts. For example, when patients from CCF who received adjuvant chemotherapy were used as the target cohort, models trained on the French and Graz surgery-plus-chemotherapy cohorts achieved greater net benefit than those trained on JHH and YCU, consistent with their closer similarity to the target population. Additional DCA plots and predicted-risk histograms are provided in the Supplementary Appendix.

\section{Discussion}
In this methodological study, we addressed a central challenge in prognostic modeling: that calibration is not an intrinsic property of a model, but depends in part on the population in which the model is applied. Across geographically distinct cohorts of patients with colorectal liver metastases, we found that external calibration varied substantially according to the degree of alignment between the training and target cohorts. More specifically, calibration deteriorated when cohorts differed either in the distribution of observed prognostic factors or in the relationship between those factors and outcomes. These findings provide empirical support, in a multicenter real-world setting, for the principle articulated by Andrew Vickers that calibration is a joint property of the model and the cohort to which it is applied.\cite{vickers2010everything}

This observation is particularly important in light of expert guidance on prognostic model evaluation, which has identified calibration as the most important property of a prognostic model.\cite{alba2017discrimination} Nonetheless, in the current literature, discrimination metrics such as the C-index are reported routinely, whereas calibration is often underreported despite its more direct relevance to many clinical decisions. In fact, a recent review of prognostic models for colorectal liver metastases found that fewer than half of the models assessed calibration, while nearly all reported discrimination.\cite{kokkinakis2024clinical} This imbalance is problematic because discrimination and calibration answer fundamentally different questions. The C-index evaluates whether a model ranks patients correctly, whereas calibration evaluates whether predicted probabilities correspond to actual outcome frequencies. In many clinical settings, particularly when the goal is to estimate absolute risk and guide management, this latter question may be more important. Importantly, a model may provide clinically meaningful estimates of absolute risk even if its ability to rank individual patients is only modest. For example, consider three CRLM patients for whom a model predicts 5-year recurrence risks of 40\%, 42\%, and 43\%. These estimates are close to the cohort’s empirical recurrence risk and are therefore well calibrated. Yet if the actual recurrence order of these patients were reversed, the model’s C-index would be poor This is especially relevant in complex diseases such as CRLM, where high discrimination may be difficult to achieve because important prognostic determinants remain unmeasured. By contrast, well-calibrated risk estimates can still support clinical decision-making even in the presence of incomplete biological knowledge.

The first methodological contribution of this study addressed this problem from the perspective of the model developer. We proposed a meta-analysis-informed importance-weighting strategy that shifts the effective training distribution toward a broader target population using only summary-level outcome and covariate information that s typically available in meta-analyses. Empirically, this approach improved average external calibration across cohorts without materially affecting discrimination. The gain was greatest when the original training cohort was most dissimilar to the external cohorts in which the model was evaluated, which is consistent with the underlying rationale of the method: the more idiosyncratic the derivation cohort, the more benefit there is in moving training toward a broader reference population. Conceptually, this improvement appears to arise primarily through addressing concept shift. In our setting, concept shift may reflect both differences in standards of care across centers and differences in unobserved prognostic factors across cohorts. In CRLM, for example, prognosis may depend not only on routinely recorded clinicopathologic variables, but also on tumor genomics, immune microenvironment, micrometastatic burden, and other latent biological features that are incompletely captured in structured datasets. By aligning the training process to outcome distributions observed in a broader external reference population, the proposed framework may partially account for these otherwise unmeasured differences and thereby yield better calibrated predictions across heterogeneous cohorts.

This methodological contribution should also be understood in the context of the broader literature on distribution shift. A large body of statistical work has studied importance weighting as a means of addressing covariate shift, typically by reweighting the training data so that its covariate distribution better matches that of the target population. Early work by Shimodaira et al introduced this idea in the context of maximum likelihood estimation under covariate shift,\cite{shimodaira2000improving} while Bickel and colleagues proposed to formulate classification tasks as an integrated optimization problem to account for train-test distribution mismatch without directly estimating each distribution.\cite{bickel2009discriminative}  
Other methods, such as kernel mean matching, were developed to align distributions in feature space by matching moments of the training and test sets.\cite{huang2006correcting,gretton2009covariate} Related work has also shown that standard cross-validation can become biased under covariate shift and proposed importance-weighted alternatives.\cite{sugiyama2007covariate} For a broader overview of this area, see the review by Nair and colleagues.\cite{nair2019covariate} In most of these approaches, however, the target-domain distribution is assumed to be directly observable during model development, either through patient-level target data or through explicit access to the target covariate distribution. That assumption is often unrealistic in clinical prognostic modeling, where models are commonly developed in a single institutional cohort and later applied to external centers for which individual-level data are unavailable at the time of training. Our approach differs in that it does not require direct access to the eventual target cohort. Instead, it uses summary-level information from the published literature to approximate a broader reference population and then shifts the effective training distribution toward that population.

Our framework is also related to the smaller literature on concept shift, which considers settings in which the relationship between predictors and outcomes differs across training and test populations. Nguyen et al., for example, showed that even ridge regression with infinite data may fail to generalize under strong concept shift\cite{nguyen2024generalization} while Tian et al. proposed using Geometric Sensitivity Decomposition to improve the detection of concept shift.\cite{tian2021exploring} Prior work in this area has focused mainly on characterizing or detecting such shifts. In medicine, however, concept shift may arise from differences in standards of care, case mix, follow-up intensity, or unmeasured biological factors across centers. Our goal was therefore more pragmatic: to improve calibration across heterogeneous real-world cohorts by using external summary information to partially account for both covariate and concept mismatch between training and validation sets. In this sense, our framework extends prior weighting approaches approaches to a setting in which neither the future deployment cohort nor its full patient-level distribution is known in advance.

The second methodological contribution of this study addressed the same problem from the perspective of the end user. Even if a model is optimized to perform well on average across external cohorts, the most generalizable model will not necessarily be the best model for a specific center. Clinicians typically face a different practical question: among the many prognostic models already available in the literature, which one is most likely to be well calibrated in their own patient population? We therefore proposed a simple model-selection strategy based on proximity in outcome space, using summary-level statistics such as KM-based recurrence-free survival to quantify similarity between the target cohort and candidate development cohorts. Across most cohort comparisons, the model trained on the most similar cohort also showed the best calibration in the target population. This suggests that even simple cohort-level outcome summaries may provide a useful and pragmatic guide for selecting among published prognostic models.

This strategy differs from conventional approaches that prioritize model reputation, prior external validation, or discrimination alone. It is also distinct from post-hoc recalibration approaches such as Platt scaling or isotonic regression. Those methods recalibrate predictions after model development and usually require data from the same population in which performance is later assessed, which may blur the distinction between model selection and model evaluation. By contrast, our approach uses external summary-level information to guide model choice before deployment and without recalibrating on the target data itself. In this sense, it is better aligned with the real-world problem faced by clinicians deciding which published model to adopt.

A particularly important finding of our study was that calibration and clinical utility, although related, are not identical. In many settings, models trained on cohorts similar to the target population achieved both lower ICI and greater net benefit on decision curve analysis. However, our results also revealed an important nuance: a model may be reasonably calibrated on average yet still provide limited clinical value if it predicts risks within a narrow range and therefore fails to separate patients into meaningfully different prognostic groups. This was illustrated by the low-risk Graz-to-CCF comparison, where the model trained on the Graz cohort was relatively well calibrated in the CCF cohort but generated only a narrow band of low predicted risks, leading to limited added value over trivial treatment strategies. By contrast, when models trained on YCU or UOR cohorts were applied to the JHH cohort, they generated a broader range of predicted risks and yielded greater clinical utility. These findings suggest that clinically useful prognostic models require not only good calibration, but also sufficient spread in predicted risk to support actionable stratification.

This study has limitations. First, the empirical analyses were conducted in colorectal liver metastases, and although this setting is clinically relevant, generalizability to other diseases remains to be established. Second, the meta-analysis-informed target distributions used in the weighting framework are themselves approximations and may not fully represent the true deployment population. Third, the outcome-space model-selection strategy intentionally relies on simple summary statistics for practicality, but richer summaries may further improve performance.

In conclusion, this study shows that calibration is strongly shaped by alignment between the development and target populations, and that poor calibration can be addressed from two directions: by training more generalizable models through meta-analysis-informed weighting, and by helping clinicians identify better-fitting models using simple outcome-based measures of cohort similarity. More broadly, our findings argue for a shift in how model generalizability and cohort-specific fit are assessed. This framework is highly practical for clinicians, both when training models intended to perform well on average across diverse settings and when selecting the model best suited to their own cohort for treatment-guiding predictions.

\newpage
\bibliography{reference.bib}

\pagebreak
\section*{Supplementary Appendix}
\begin{figure}[ht!]
    \begin{subfigure}[b]{0.45\textwidth}
        \centering
          \vspace{-5mm}
         \includegraphics[width=\textwidth]{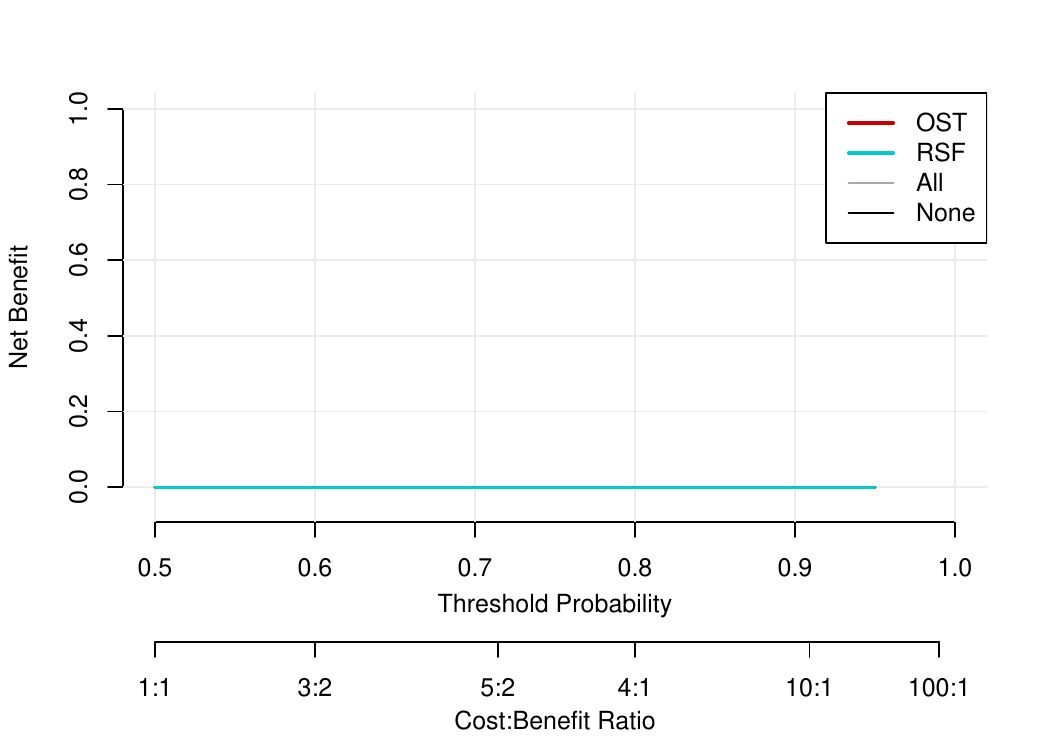}
        \caption*{\small  A. Trained on JHH}
    \end{subfigure}
    \begin{subfigure}[b]{0.45\textwidth}
        \centering
        \vspace{-5mm}
         \includegraphics[width=\textwidth]{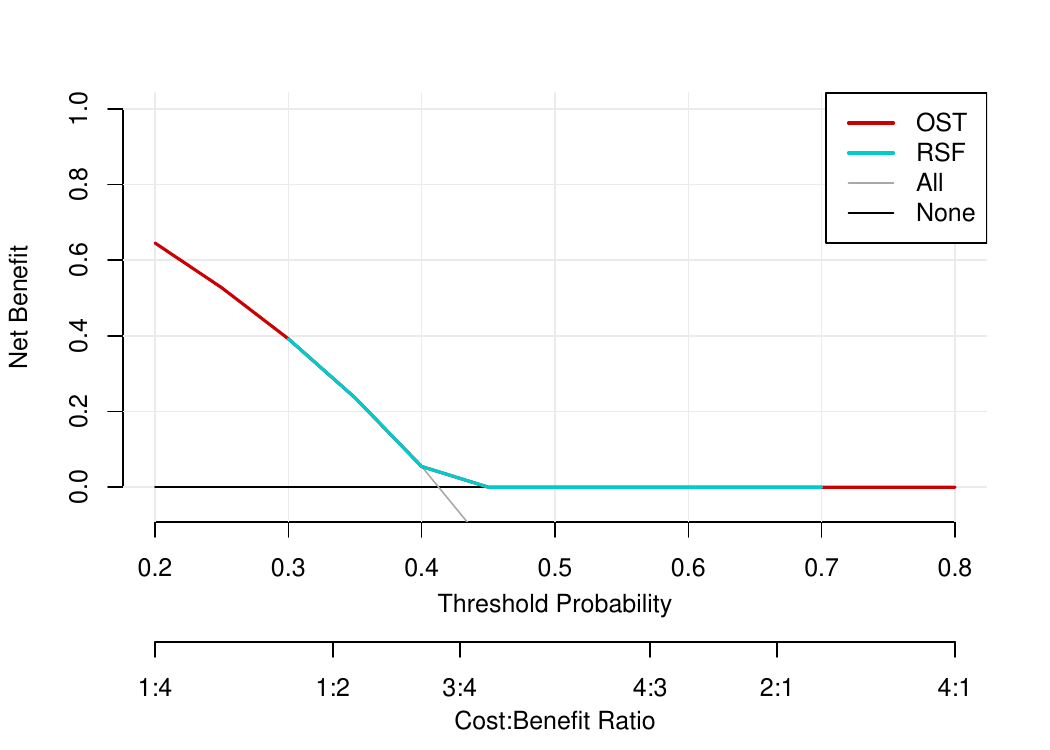}
        \caption*{\small B. Trained on CCF}
    \end{subfigure}\\
    \begin{subfigure}[b]{0.45\textwidth}
        \centering
          \vspace{-5mm}
         \includegraphics[width=\textwidth]{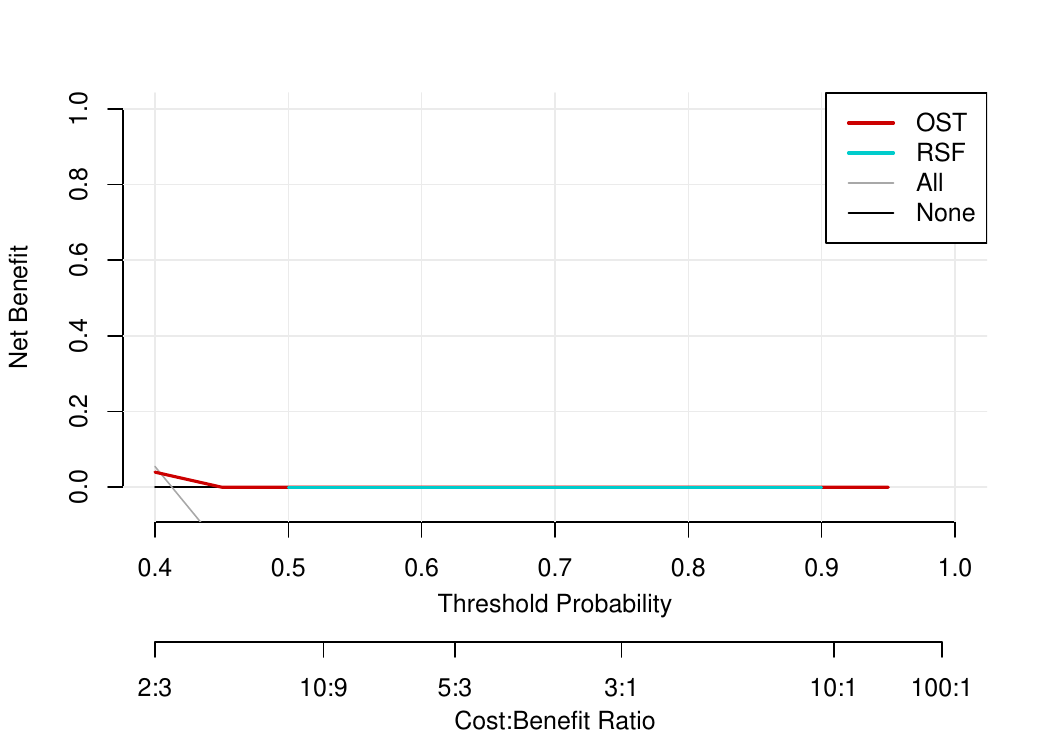}
        \caption*{\small C. Trained on YCU}
    \end{subfigure} 
    \begin{subfigure}[b]{0.45\textwidth}
        \centering
             \vspace{-5mm}
         \includegraphics[width=\textwidth]{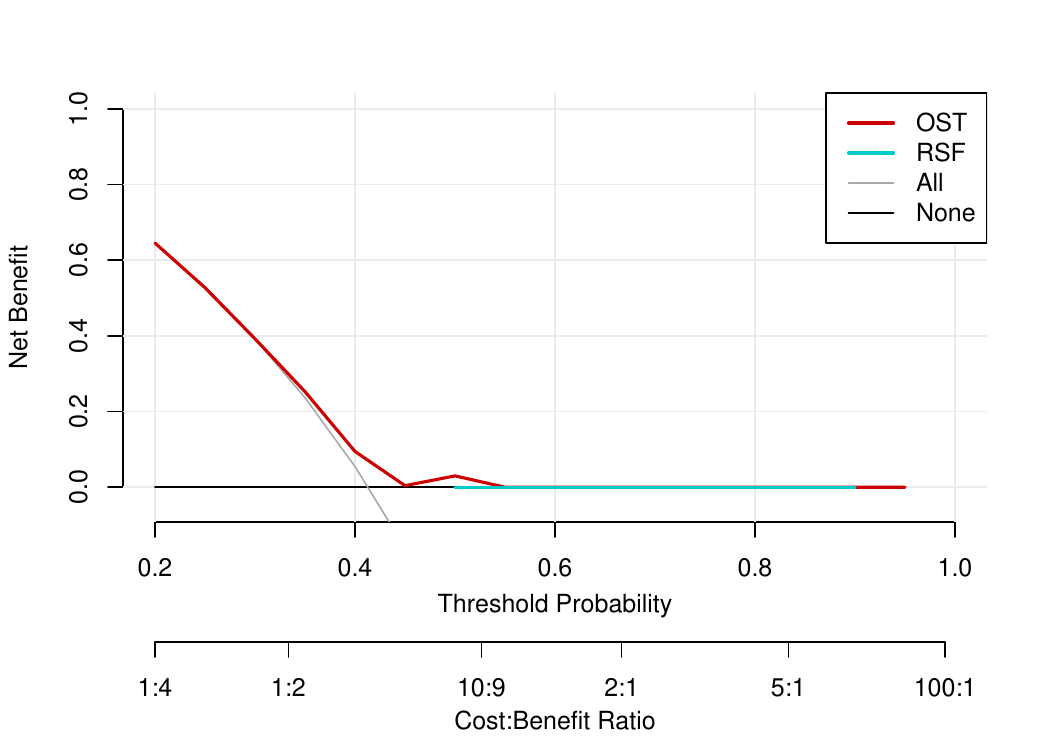}
        \caption*{\small D. Trained on UOR}
    \end{subfigure}
    \caption{Decision curve analysis showing the clinical utility of models trained in four different cohorts and externally validated in the Graz surgery-alone cohort.}
    \label{fig:DCA_untreated_graz}
\end{figure}

\begin{figure}[ht!]
    \begin{subfigure}[b]{0.45\textwidth}
        \centering
          \vspace{-5mm}
         \includegraphics[width=\textwidth]{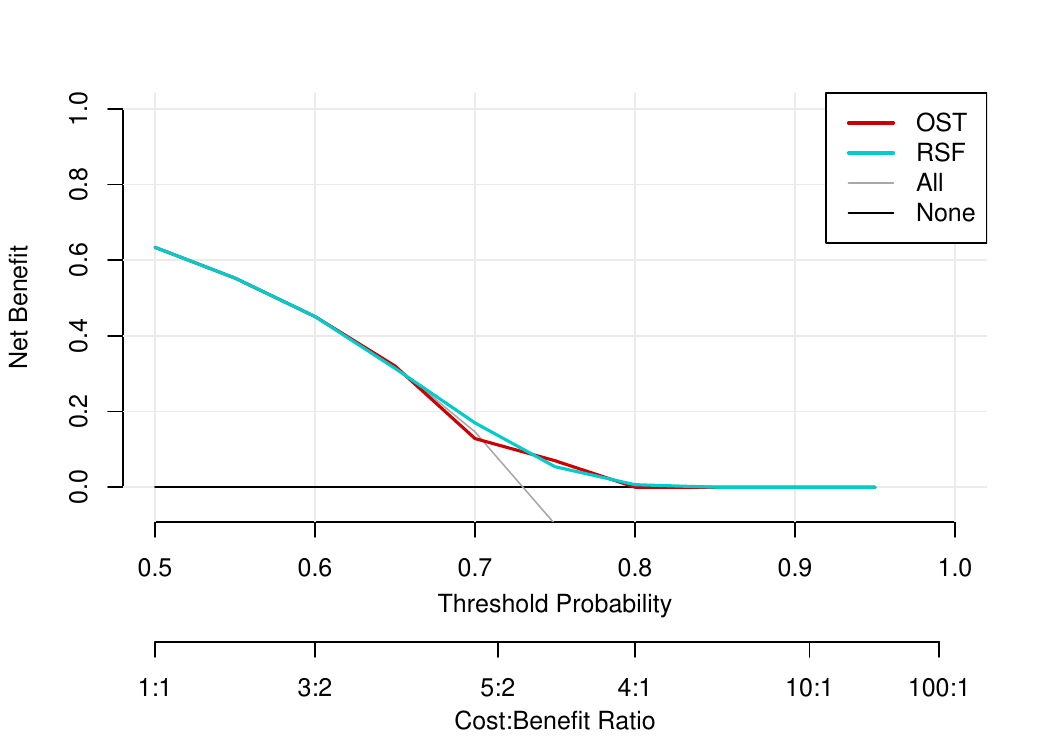}
        \caption*{\small  A. Trained on JHH}
    \end{subfigure}
    \begin{subfigure}[b]{0.45\textwidth}
        \centering
            \vspace{-5mm}
         \includegraphics[width=\textwidth]{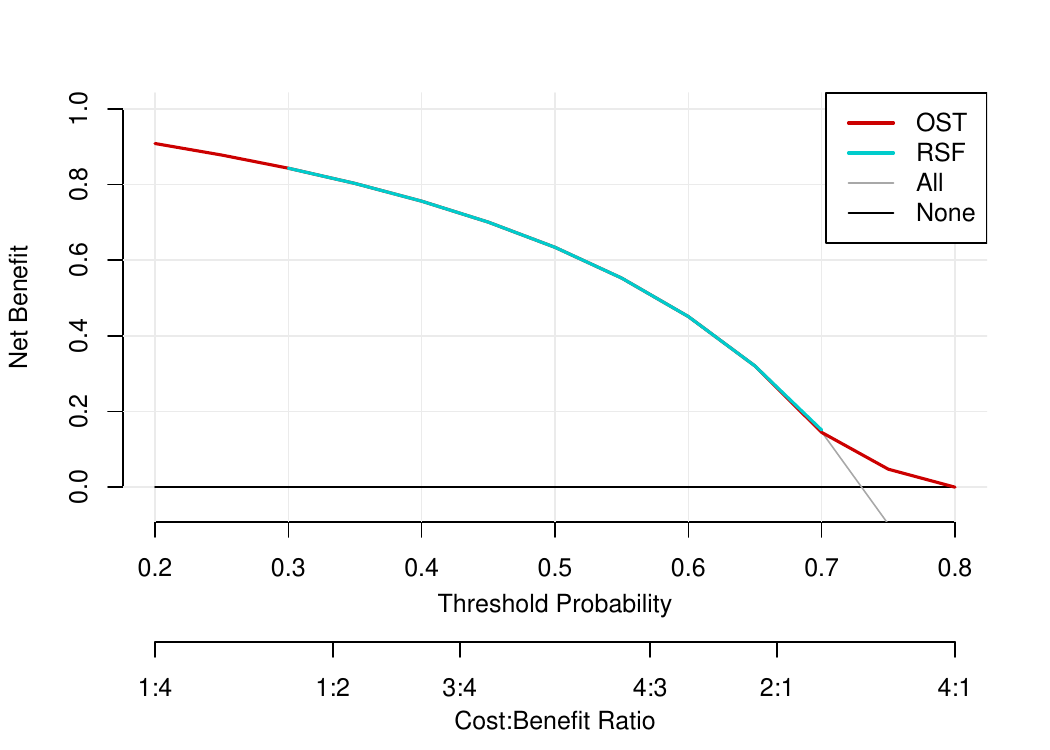}
        \caption*{\small B. Trained on CCF}
    \end{subfigure}\\
    \begin{subfigure}[b]{0.45\textwidth}
        \centering
          \vspace{-5mm}
         \includegraphics[width=\textwidth]{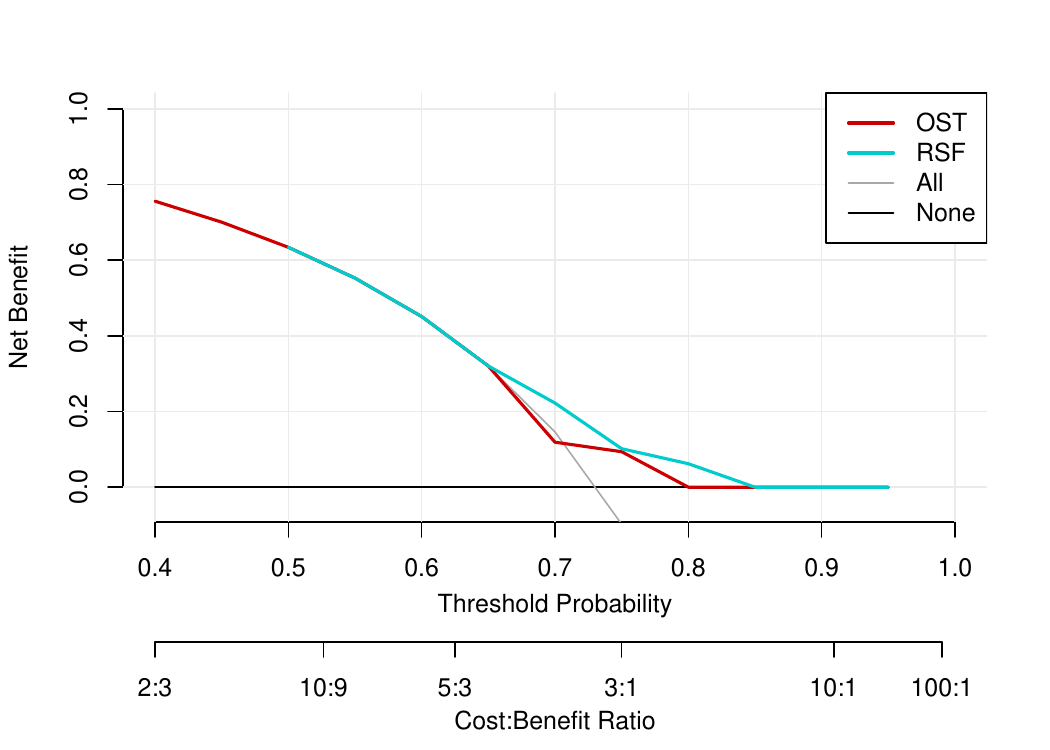}
        \caption*{\small C. Trained on YCU}
    \end{subfigure} 
    \begin{subfigure}[b]{0.45\textwidth}
        \centering
            \vspace{-5mm}
         \includegraphics[width=\textwidth]{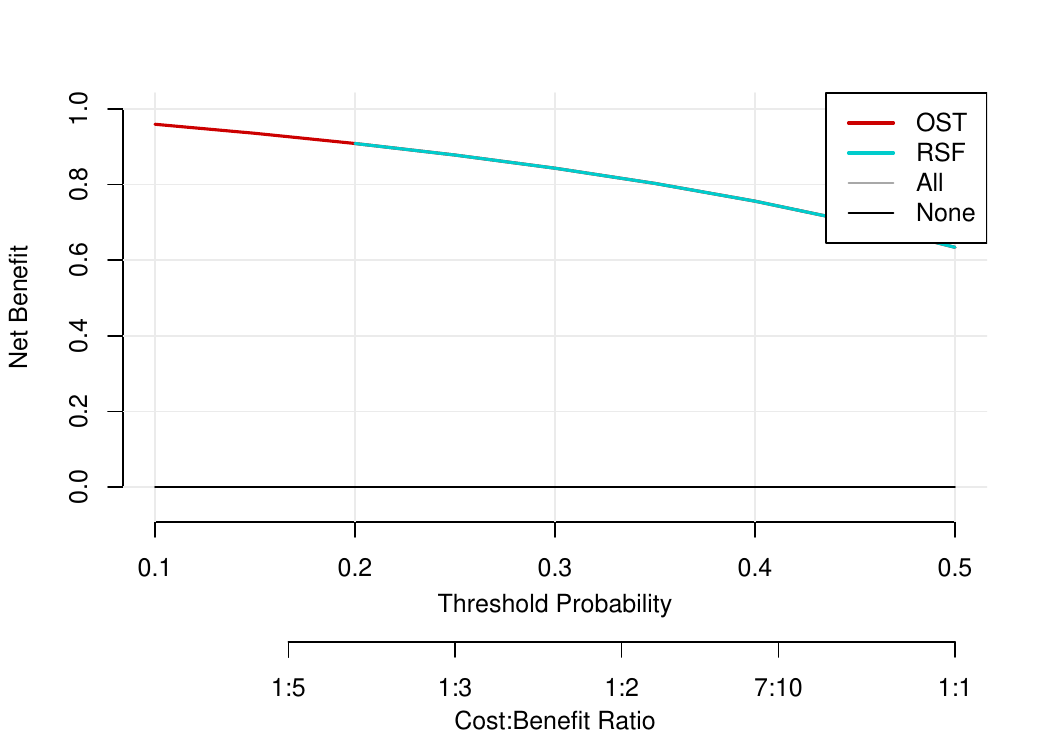}
        \caption*{\small D. Trained on Graz}
    \end{subfigure}
    \caption{Decision curve analysis showing the clinical utility of models trained in four different cohorts and externally validated in the UOR surgery-alone cohort.}
    \label{fig:DCA_untreated_uor}
\end{figure}

\begin{figure}
    \centering
    \includegraphics[width=0.75\linewidth]{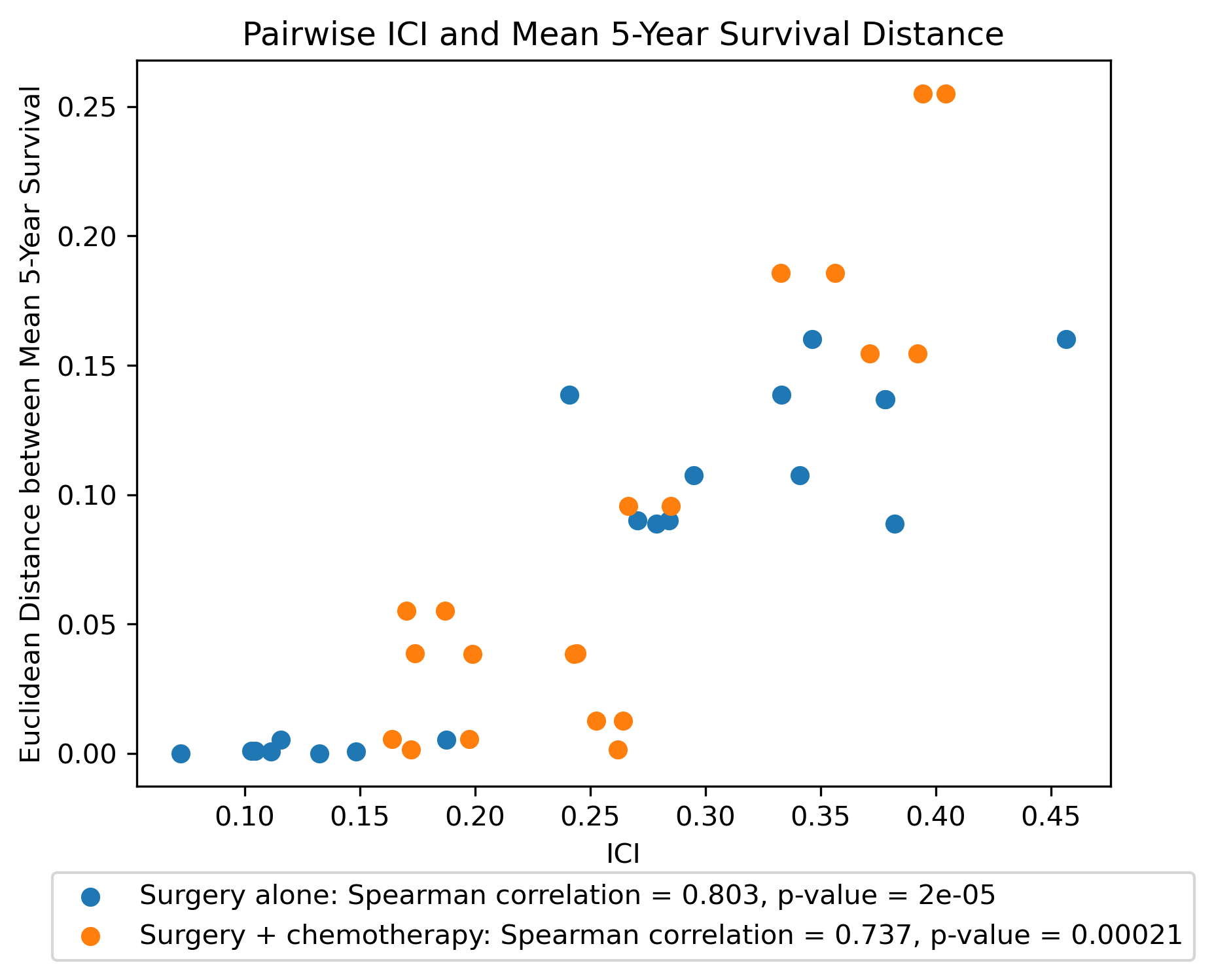}
    \caption{Correlation between ICI and Mean 5-Year Survival Distance}
    \label{fig:scatter_part3}
\end{figure}

\end{document}